\documentclass[aps,prl,reprint,superscriptaddress,longbibliography,floatfix,nofootinbib]{revtex4-2}
\usepackage{amsmath,amssymb,graphics,epsfig,epstopdf,color,verbatim,ulem,braket,tabularx}
\usepackage{multirow}
\usepackage[colorlinks,linkcolor=blue,citecolor=blue,urlcolor=blue,bookmarks=false]{hyperref}

\usepackage{listings}
\usepackage{cancel}
\usepackage{mathrsfs}
\usepackage{soul}
\usepackage{color}
\usepackage{url}

\usepackage{amsthm}
\usepackage{epsf}
\usepackage{fancyhdr}
\usepackage{diagbox}
\usepackage{booktabs}
\newcommand{\beginsupplement}{%
        \setcounter{table}{0}%
        \renewcommand{\thetable}{S\arabic{table}}%
        \renewcommand{\theHtable}{S\arabic{table}}%
        \setcounter{figure}{0}%
        \renewcommand{\thefigure}{S\arabic{figure}}%
        \renewcommand{\theHfigure}{S\arabic{figure}}%
        \setcounter{equation}{0}%
        \renewcommand{\theequation}{S\arabic{equation}}%
        \renewcommand{\theHequation}{S\arabic{equation}}%
}

\makeatletter
\def\ps@plain{%
  \let\@oddhead\@empty
  \let\@evenhead\@empty
  \def\@oddfoot{\hfil\thepage\hfil}%
  \let\@evenfoot\@oddfoot
}
\let\ps@headings\ps@plain
\let\ps@myheadings\ps@plain
\makeatother

\begin{document}
\pagestyle{plain}

\title{Competing Extended-$s$- and $d$-Wave Pairing from Distinct Spin-Fluctuation Channels in Stoichiometric $\mathrm{FeTe}$}

\author{Yichen Hua*}
\affiliation{Department of Physics, Southern University of Science and Technology, Shenzhen 518055, China}
\affiliation{Quantum Science Center of Guangdong-Hong Kong-Macao Greater Bay Area, Shenzhen 518045, China}
\thanks{These authors contributed equally to this work.}

\author{Wen-lin Yang*}
\affiliation{Department of Physics, Southern University of Science and Technology, Shenzhen 518055, China}
\thanks{These authors contributed equally to this work.}

\author{Jian-Jian Miao}

\affiliation{Quantum Science Center of Guangdong-Hong Kong-Macao Greater Bay Area, Shenzhen 518045, China}
\author{Hu Xu$^{\dag}$}
\affiliation{Department of Physics, Southern University of Science and Technology, Shenzhen 518055, China}
\author{Changming Yue$^{\dag}$}
\affiliation{State Key Laboratory of Quantum Functional Materials, Department of Physics, and Guangdong Basic Research Center of Excellence for Quantum Science, Southern University of Science and Technology (SUSTech), Shenzhen 518055, China}

\begin{abstract}
The recent observation of superconductivity in stoichiometric $\mathrm{FeTe}$ raises the question of how pairing develops in this tetragonal 11-type chalcogenide once interstitial Fe is removed. We construct an experimentally constrained five-orbital tight-binding model from first-principles calculations and treat electronic correlations and pairing within the fluctuation-exchange approximation. 
The linearized Eliashberg equation yields competing extended-$s$- and $d_{x^2-y^2}$-wave spin-singlet pairing instabilities. Near stoichiometric filling, spin fluctuations near $(\pi,0)$ and $(0,\pi)$ in the unfolded one-Fe Brillouin zone connect the $\Gamma/M$ hole pockets with the $X/Y$ electron pockets and favor an extended-$s$ gap that changes sign between the hole and electron sheets. Upon electron doping, depletion of the hole pockets shifts the dominant scattering toward the $X$--$Y$ channel near $(\pi,\pi)$, making the $d_{x^2-y^2}$-wave state the leading instability, with nodal lines that avoid most of the Fermi surface. The relative strengths of the two pairing channels vary with filling and interaction strength as the dominant spin-fluctuation channel changes. Under matched interaction strength, temperature, and filling, the extended-$s$ eigenvalue is larger in $\mathrm{FeTe}$ than in $\mathrm{FeSe}$ throughout the range considered, while the $d_{x^2-y^2}$-wave eigenvalue is also generally larger, particularly under electron doping. These results give concrete gap structures against which spectroscopic measurements of stoichiometric $\mathrm{FeTe}$ can be compared.
\end{abstract}

\maketitle

{\it Introduction. ---}
Iron-based superconductors form a major class of unconventional superconductors in which magnetism, nematicity, and superconductivity are closely intertwined~\cite{Kamihara2008,Chen2008SmFeAsO,Ren2008SmFeAsO,Stewart2011,Paglione2010,Johnston2010,Fernandes2022,Dai2015,Hosono2015KurokiReview,Kuroki2008DisconnectedFS}. They are commonly grouped into the 1111~\cite{Kamihara2008,Chen2008SmFeAsO,Ren2008SmFeAsO}, 122~\cite{Rotter2008,Sefat2008,Canfield2010}, and 11~\cite{Hsu2008,Fang2008,Wang2012FeSeSTO} families, among which the 11 family is represented by the iron chalcogenide $\text{FeSe}$. Under ambient pressure, stoichiometric $\text{FeSe}$ is an intrinsic superconductor~\cite{Kreisel2020,Coldea2018,Bohmer2018,Shibauchi2020,Hsu2008}. On cooling, $\mathrm{FeSe}$ undergoes a nematic phase transition well above $T_c$, lowers the rotational symmetry from $C_4$ to $C_2$. Its superconductivity is therefore believed to be closely intertwined with nematic order~\cite{Bohmer2018,McQueen2009,Shimojima2014,Nakayama2014,Baek2015,Bohmer2015,Watson2015,Sprau2017,Hashimoto2018}. 
$\text{FeTe}$ has a crystal structure similar to that of $\text{FeSe}$. However,  stoichiometric $\text{FeTe}$ has been difficult to realize experimentally, while Fe-rich $\text{FeTe}$ typically exhibits an antiferromagnetic ground state.  $\text{FeTe}$ has therefore long been regarded as an antiferromagnetic metal rather than a superconductor~\cite{Bao2009,Li2009,Ma2009,Rodriguez2011,Stock2011,Zaliznyak2012,Gronvold1954,Ward1979,Han2010}. 

Chang \textit{et al.} recently grew 40-unit-cell $\mathrm{FeTe}$ films on SrTiO$_3$ substrates by molecular-beam epitaxy and performed post-growth annealing under a Te flux to remove interstitial Fe~\cite{Yan2026FeTe}. The resulting stoichiometric films are superconducting, with $T_c$ up to 13.5 K, higher than that of bulk $\mathrm{FeSe}$ at ambient pressure. This observation changes the usual view of $\mathrm{FeTe}$ as the nonsuperconducting antiferromagnetic member of the 11 family. The reported stoichiometric films show no antiferromagnetic order, and no nematic order has yet been reported.
These characteristics establish stoichiometric FeTe as a distinct platform for investigating the interplay between electronic structure, magnetism, and superconducting pairing in iron-based superconductors.
\begin{figure*}[htp]     \includegraphics[clip,width=0.85\paperwidth,angle=0]{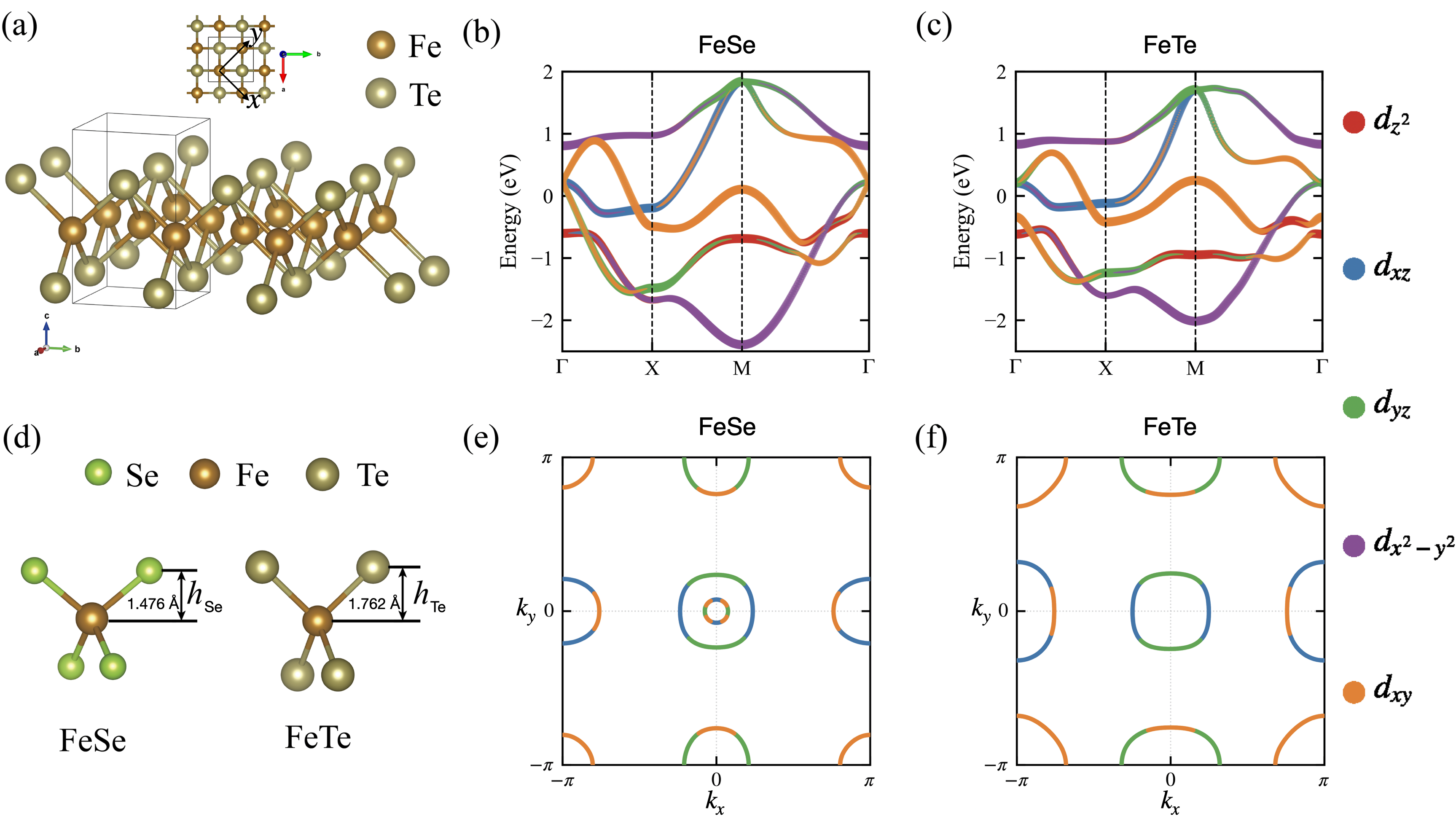}
    \caption{(a) Crystal structure of  $\text{FeTe}$. The local coordinate axes are also shown. (b-c) Unfolded low-energy bands  of the five-orbital tight-binding models for (b) $\text{FeSe}$ and (c) $\text{FeTe}$. Marker colors denote Fe-$3d$ orbital weights. (d) Local
tetrahedral coordination and anion heights in $\text{FeSe}$ and $\text{FeTe}$. (e-f) Fermi surfaces in the unfolded Brillouin zone for (e) $\text{FeSe}$ and (f) $\text{FeTe}$, with orbital character indicated by color.}
    \label{fig1}
\end{figure*}

The momentum-resolved spectral function, Fermi-surface topology, and gap symmetry of these films have not yet been measured, leaving the microscopic origin of their superconductivity unresolved. Earlier theoretical work addressed primarily the electronic and magnetic structure of $\mathrm{FeTe}$ or the broader $\mathrm{FeSe}_{1-x}\mathrm{Te}_x$ series ~\cite{Subedi2008FeChalcogenides,Ma2009,Fang2009MagneticOrder,Turner2009FeTe,Moon2010ChalcogenHeight,Ducatman2012Chalcogenides,Ducatman2014FeTe,Glasbrenner2015IronChalcogenides}. Superconductivity or orbital-dependent correlation in strained $\mathrm{FeTe}$ has also been studied theoretically~\cite{Ciechan2014StrainedFeTe} or  experimentally~\cite{Han2010,Xu2026StrainedFeTe,Changyoung-PRB-2026}. A microscopic pairing calculation tied to the structure of the newly reported stoichiometric films is still missing.

Here, we investigate a spin-fluctuation pairing scenario for stoichiometric $\mathrm{FeTe}$ using an experimentally constrained five-orbital tight-binding (TB) model, with multiorbital interactions treated within the fluctuation-exchange approximation. We find competing spin-singlet extended-$s$- and $d_{x^2-y^2}$-wave states selected by distinct spin-fluctuation channels: near stoichiometric filling, hole--electron scattering near $(\pi,0)$ and $(0,\pi)$ favors extended-$s$ pairing, whereas electron doping suppresses the hole pockets and shifts the dominant scattering to the $X$--$Y$ electron-pocket channel near $(\pi,\pi)$, favoring $d_{x^2-y^2}$ pairing. The competition between these two states can be tuned by filling and interaction strength, directly linking Fermi-surface topology and magnetic response to the superconducting gap symmetry. 

{\it Method. ---}
Density-functional calculations for tetragonal $\mathrm{FeTe}$ (space group $P4/nmm$) were performed with 
VASP~\cite{Kresse1993VASP,Kresse1996VASP}, using the projector augmented-wave method~\cite{PhysRevB.50.17953,Kresse1999PAW} and the Perdew--Burke--Ernzerhof generalized-gradient approximation~\cite{PhysRevLett.77.3865}. We used the experimental lattice parameters $a=3.862$~\AA, $c=6.262$~\AA, and an $\text{Fe-Te}$ height of $1.762$~\AA ~\cite{Yan2026FeTe}. The low-energy Fe-$3d$ bands were represented using maximally localized Wannier functions (MLWFs) of $d_{3z^2-r^2}$, $d_{xz}$, $d_{yz}$, $d_{x^2-y^2}$, and $d_{xy}$ character with the Wannier90 package ~\cite{MOSTOFI2008685,RevModPhys.84.1419}. Using a band-unfolding procedure adapted from the standard approach for iron-based superconductors, we projected the electronic structure from the folded 2-$\text{Fe}$ Brillouin zone into the effective 1-$\text{Fe}$ Brillouin zone, yielding a five-orbital noninteracting TB Hamiltonian~\cite{Kuroki2008DisconnectedFS}. The interlayer hoppings along the $z$-direction are much smaller than the in-plane ones and are therefore neglected. The effective 2D Hamiltonian for FeTe is expressed as:
\begin{equation}
\begin{aligned}
H_{0}&=\sum_{ij,\alpha\beta,\sigma}[t_{ij}^{\alpha\beta}c_{i\alpha\sigma}^{\dagger}c_{j\beta\sigma}+h.c.
]+\sum_{i\alpha\sigma}\varepsilon_{\alpha}n_{i\alpha\sigma},
\end{aligned}
\end{equation}
where $c^{\dagger}_{i\alpha\sigma}$ creates an electron with spin $\sigma$ in orbital $\alpha$ at site $i$, and $n_{i\alpha\sigma}=c^{\dagger}_{i\alpha\sigma}c_{i\alpha\sigma}$. Here $\alpha$,$\beta$=$1,\ldots,5$ label the five Fe (3d) orbitals. The hopping integrals $t_{ij}^{\alpha\beta}$ and on-site energies $\varepsilon_\alpha$ 
are tabulated in Section A of the Supplemental Material (SM).

We consider a local Hubbard--Kanamori interaction Hamiltonian for the five Fe-3$d$ orbitals~\cite{Kanamori1963TransitionMetals}, written as  
\begin{equation}
\begin{aligned}
H_{\mathrm{int}}={}&  
U \sum_{i,\alpha}  
n_{i\alpha\uparrow} n_{i\alpha\downarrow}
+ U' \sum_{i,\alpha<\beta}\sum_{\sigma,\sigma'}  
n_{i\alpha\sigma} n_{i\beta\sigma'} \  
\\
&-J_{\mathrm H}  
\sum_{i,\alpha<\beta}\sum_{\sigma,\sigma'}  
c^{\dagger}_{i\alpha\sigma}  
c^{\dagger}_{i\beta\sigma'}  
c_{i\alpha\sigma'}  
c_{i\beta\sigma}  
\\
&+ J_{\mathrm H}  
\sum_{i,\alpha<\beta}  
\left(  
c^{\dagger}_{i\alpha\uparrow}  
c^{\dagger}_{i\alpha\downarrow}  
c_{i\beta\downarrow}  
c_{i\beta\uparrow}  
+h.c.  
\right),  
\end{aligned}
\end{equation}
where $U$, $U'$, and $J_{\mathrm H}$ denote the intra-orbital Hubbard repulsion, inter-orbital Coulomb repulsion, and Hund’s coupling, respectively. For a rotationally invariant interaction, $U' = U - 2J_{\mathrm H}$, and we fix $J_{\mathrm H}=U/6$ throughout this work. 

The fluctuation-exchange (FLEX) approximation  ~\cite{Bickers1989FLEX,PhysRevB.43.8044,PhysRevB.55.2122} is employed to study spin fluctuations and superconducting pairing tendencies ~\cite{Ikeda2010GapAnisotropy,PhysRevB.85.134507,Usui2012SweetSpot,Rademaker2021FeSeSTO,Kuroki-2024-PRL-327} in $\text{FeTe}$. In this work, we use a modified version of the FLEX code originally implemented by Witt~\cite {Witt2021EfficientFLEX} with feature of data compression provided by sparse-ir \cite{Shinaoka2017IR,Wallerberger2023SparseIR}. 
The spin and charge susceptibilities are defined as
\begin{equation}
\hat{\chi}^{C/S}(q)
=
\left[
\hat{1}\pm \hat{\chi}^{0}(q)\hat{U}^{C/S}
\right]^{-1}
\hat{\chi}^{0}(q),
\end{equation}
where $q\equiv(\mathbf{q},i\nu_m)$ denotes the momentum-frequency (bosonic) variable, and $\hat{\chi}^{0}(q)$ is the irreducible bare susceptibility constructed from the renormalized Green's functions $\hat{G}(k)$ [$k\equiv(\mathbf{k},i\omega_n)$]. The matrices $\hat{U}^{S}$ and $\hat{U}^{C}$ are the interaction vertices in the spin and charge channels, respectively. We focus on \(U\leq1.5\) eV at temperature $T=0.001~\mathrm{eV}$, where the FLEX calculations remain numerically stable and allow us to track the evolution of the leading pairing instabilities.
We use $A(\mathbf{k},0)\approx -\mathrm{Im\ Tr}G({\mathbf{k},i\pi T})/\pi$ as a low-energy spectral-weight proxy for the Fermi surface, thereby avoiding analytic continuation.
To investigate the superconducting pairing symmetry, we diagonalize the linearized Eliashberg equation in the spin-singlet channel and retain the four largest eigenpairs $(\lambda,\Delta)$. Here, $\lambda$ measures the pairing tendency, while $\Delta(k)$ at the lowest Matsubara frequency represents the gap function. More technical details about FLEX and the gap function can be found in Section B of the SM.

\begin{figure}[htp]
\includegraphics[clip,width=\columnwidth,angle=0]{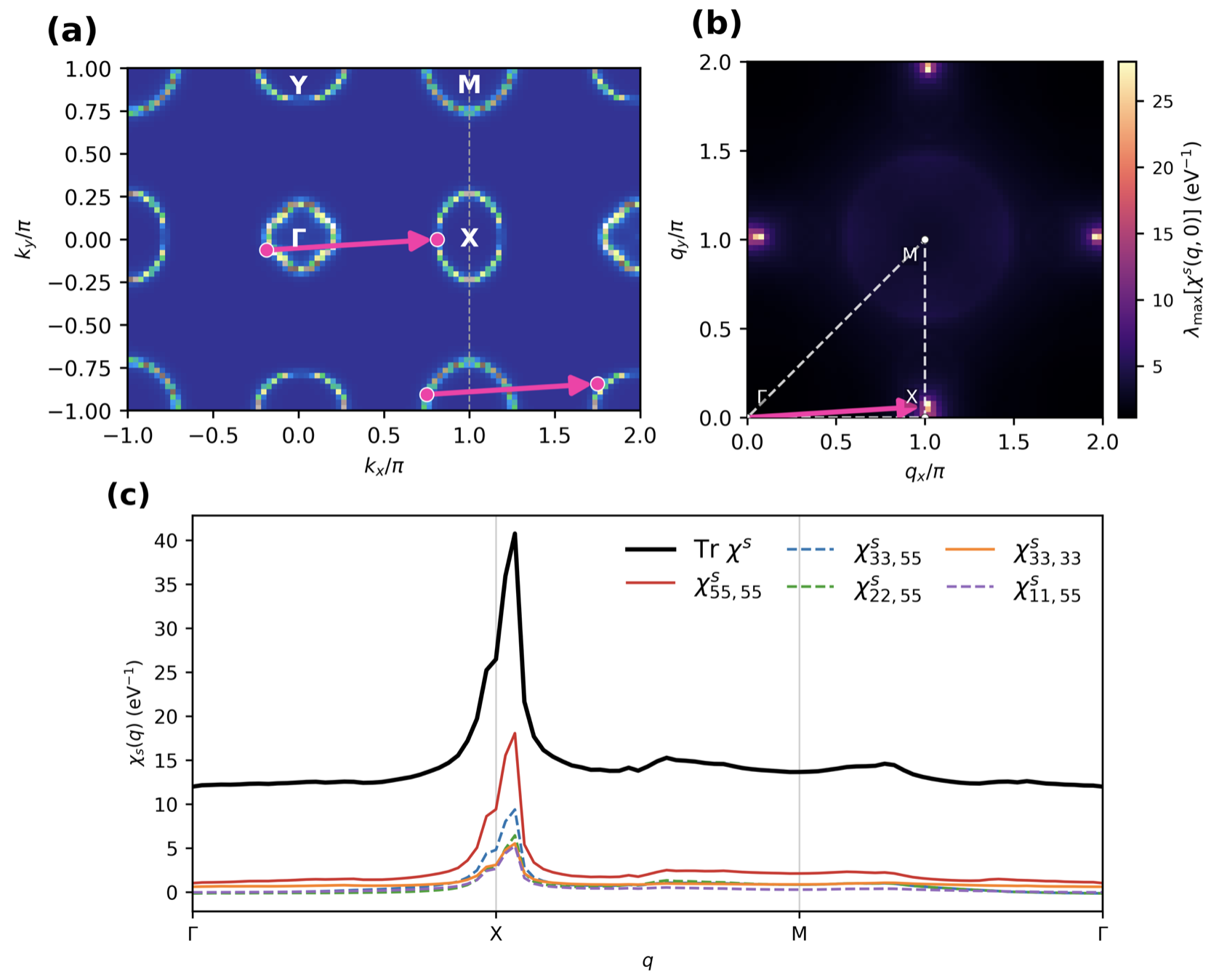}
    \caption{Fermi surface and spin susceptibility of stoichiometric $\text{FeTe}$ (n=1.20 electrons per orbital). Here, \(U=1.5\), and \(T=0.001\).
(a) \(A(\mathbf{k}, i\pi T)\) shown in an extended Brillouin-zone
representation. The pink arrows mark two equivalent translations by the same nesting wave vector
\(\mathbf{Q}/\pi=(1,0.0625)\).
(b) Largest eigenvalue of the spin susceptibility matrix \(\chi^s(\mathbf{q},0)\). The dashed white line marks
the high-symmetry ${\bf q}$-path \(\Gamma-X-M-\Gamma\), and the pink arrow indicates the same \(\mathbf{Q}\) vector as in panel (a). The center of panel (b) corresponds to the $M$ point rather than the $\Gamma$ point.
(c) Static spin susceptibility along the ${\bf q}$-path. The black curve is \(\mathrm{Tr}\,\chi^s(\mathbf{q})\),
while the colored curves show the dominant matrix elements \(\chi^s_{ij,kl}(\mathbf{q})\) 
. The orbital indices are \(1\equiv d_{3z^2-r^2}\), \(2\equiv d_{xz}\), \(3\equiv d_{yz}\), \(4\equiv d_{x^2-y^2}\), and \(5\equiv d_{xy}\).
}
    \label{fig2}
\end{figure}

{\it Ab initio calculations. ---} 
The anion height $h_X$, defined as the distance between the anion and the nearest Fe layer, affects the electronic structure and $T_c$ in iron-based superconductors ~\cite{PhysRevB.81.205119,Mizuguchi_2010,KUROKI2011307}. It is therefore useful to compare the electronic structures of FeTe and FeSe. As shown in Figs.~\ref{fig1}a and \ref{fig1}d, the anion height $h_{\text{Te}}$ in FeTe is larger than $h_{\text{Se}}$ in FeSe, with $h_{\text{Te}}=1.762$ $\AA$ ~\cite{Yan2026FeTe} and $h_{\text{Se}}=1.476$ $\AA$ ~\cite{PhysRevB.79.014522}.  
Apparently, the value of $h_X$ affects the band structure and Fermi surfaces, as shown in Figs.~\ref{fig1}(c-f).
Compared with $\mathrm{FeSe}$, $\mathrm{FeTe}$ exhibits markedly enlarged pockets at the $M$ and $X$ points via two distinct pathways. For the in-plane $d_{xy}$ orbital, a larger $h_{\text{X}}$ lowers its orbital energy (from $-0.069$~eV in FeSe to $-0.184$~eV in FeTe, as detailed in Tables~S1 and~S2 of SM) and submerges the $\Gamma$-centered hole pocket below the Fermi level. At fixed total filling, the accompanying chemical potential adjustment enlarges the $M$-centered hole pockets. Concurrently, the $X$ ($Y$) electron pockets, dominated by the $d_{xz}$ ($d_{yz}$) orbitals, are sensitive to anion-mediated indirect hopping. In the $\mathrm{FeTe}$ model, the dominant next-nearest-neighbor hopping $t^{[1,1]}_{2,2}$ decreases from $0.221$~eV in $\mathrm{FeSe}$ to $0.196$~eV in $\mathrm{FeTe}$ (see SM), flattening the conduction band dispersion and widening the electron pockets at the $X$ ($Y$) points.

{\it Spin susceptibility and nesting in $\text{FeTe}$. ---}
We first consider the nominally stoichiometric system, corresponding to an average orbital occupation of $n=1.2$ electrons. At $U=1.5~\mathrm{eV}$, the Fermi surface renormalized within FLEX is shown in Fig.~\ref{fig2}a, while Fig.~\ref{fig2}b presents the corresponding maximum eigenvalue of $\chi^S({\bf q},0)$, which exhibits pronounced peaks near ${\bf Q}=(\pi,0)$ and $(0,\pi)$. 
By resolving the Green's-function contributions to $\chi^{S}$ over the Brillouin zone, we find that the dominant contributions to these peaks of $\chi^S({\bf q},0)$ arise from the nesting between the $\Gamma$ pocket and the $X/Y$ pockets, as well as between the $M$ pocket and the $X/Y$ pockets, as indicated by the pink arrows in Fig.~\ref{fig2}a. Fig.~\ref{fig2}c further shows the dominant orbital contributions to $\chi^S$, with the leading contributions involving the $d_{xy}$, $d_{yz}$, and $d_{xz}$ orbitals.

\begin{figure}[htp]   \includegraphics[clip,width=\columnwidth,angle=0]{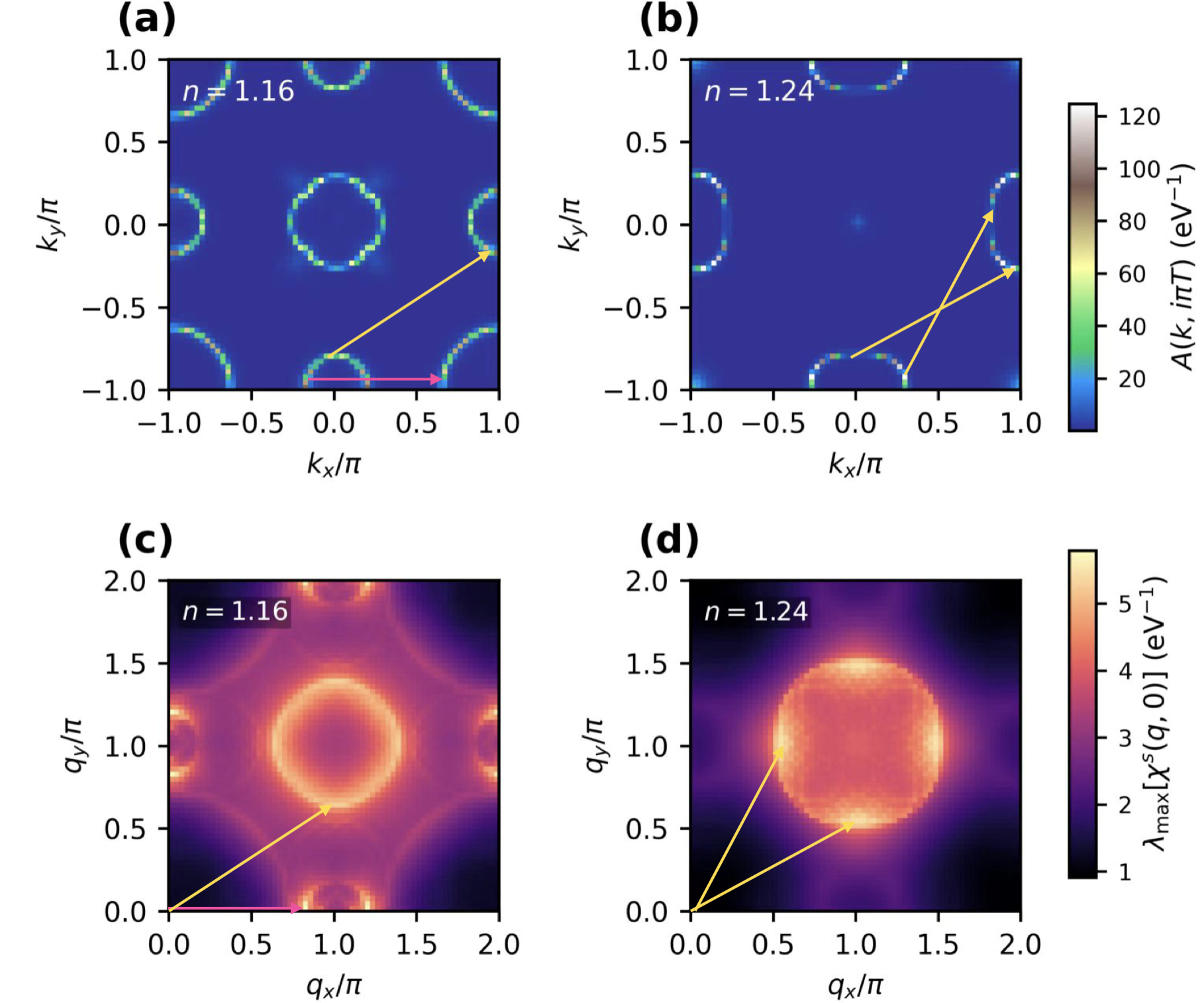}
    \caption{Fermi surfaces and spin susceptibilities of hole- and electron-doped $\text{FeTe}$. Here,  $U=1.5$, $T=0.001$. The first row shows \(A(\mathbf{k}, i\pi T)\) for $\text{FeTe}$  with fillings (a) $n=1.16$ and (b) $n=1.24$. 
The second row shows the corresponding largest eigenvalue of the spin susceptibility for (c) $n=1.16$ and (d) $n=1.24$. The common color scale in panels (c,d) allows direct comparison of the susceptibility strength between
the two fillings. The pink arrows indicate $(0,\pi)$-type nesting, while the yellow arrows denote $(\pi,\pi)$-type nesting.}
    \label{fig3}
\end{figure}

\begin{figure}[htp]
\includegraphics[clip,width=\columnwidth,angle=0]{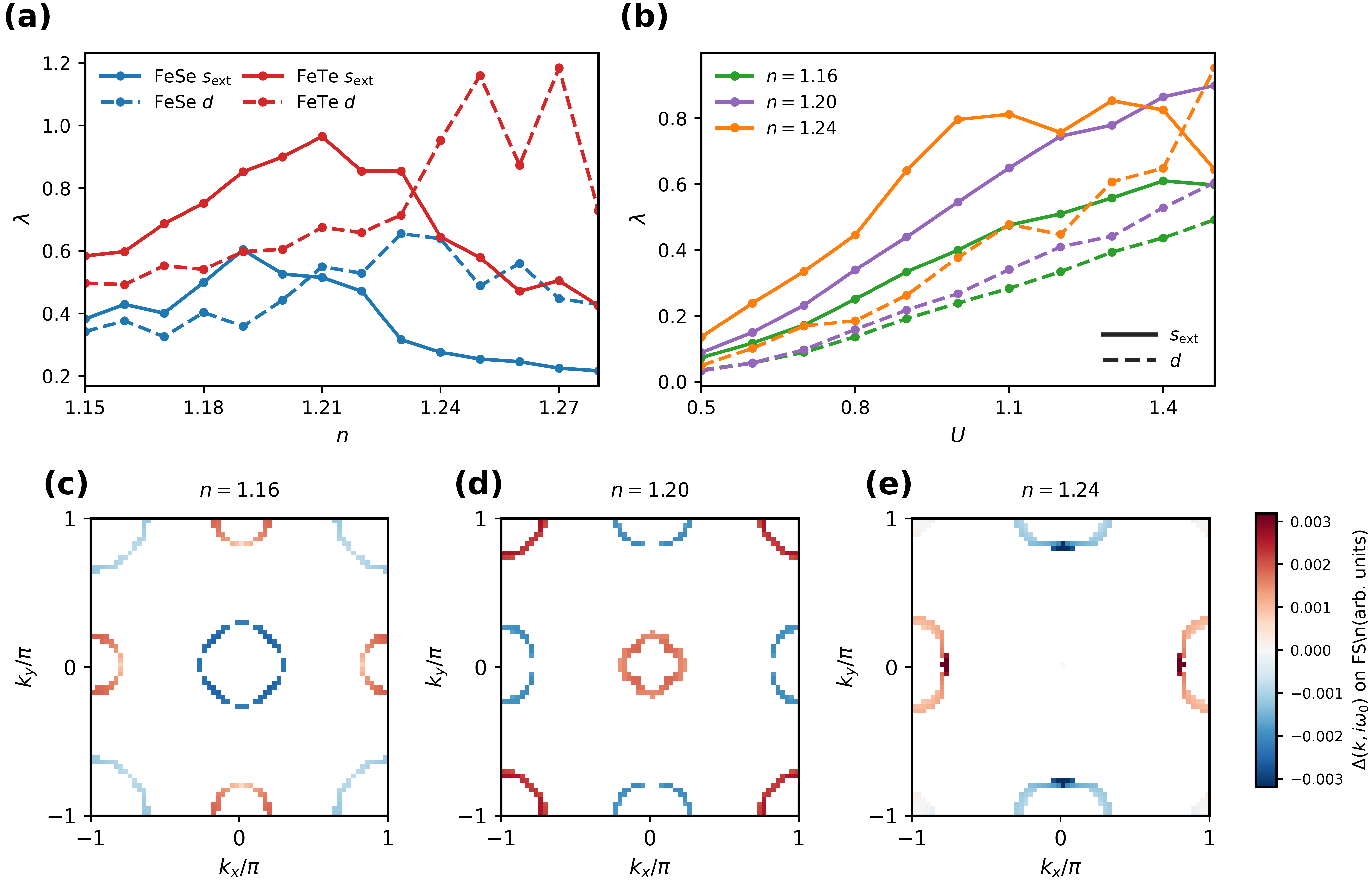}
    \caption{Pairing eigenvalues and leading gap structures.
(a) Filling dependence of the leading $s_{\mathrm{ext}}$- and $d$-wave eigenvalues $\lambda$ for $\text{FeSe}$ and $\text{FeTe}$ at $U=1.5$ and $T=0.001$.
(b) Interaction dependence of $\lambda$ for $\text{FeTe}$ at $n=1.16$, $1.20$, and $1.24$; solid and dashed lines denote the $s_{\mathrm{ext}}$- and $d$-wave channels, respectively.
(c)--(e) Fermi-surface projections of the leading gap function for $\text{FeTe}$ at $U=1.5$, $T=0.001$, and fillings $n=1.16$, $1.20$, and $1.24$, respectively.
The color scale indicates the sign and relative amplitude of the gap function on the Fermi surface.}
    \label{fig4}
\end{figure}

Fig.~\ref{fig3} shows the hole- and electron-doped cases at $n=1.16$ and $n=1.24$, respectively, with the corresponding Fermi surfaces in panels (a) and (b), and the maximum eigenvalue of $\chi^S({\bf q},0)$ in panel (c) and (d). With increasing $n$, the $\Gamma$- and $M$-centered hole pockets shrink and disappear, while the $X/Y$ electron pockets expand. At $n=1.24$, the $\Gamma$ and $M$ hole pockets are nearly absent, and the $d_{xy}$ orbital no longer contributes to the Fermi surface. In the hole-doped case (Fig.~\ref{fig3}c), $\chi^S$ exhibits two branches of maxima. The first branch is located near $(0,\pi)$ and $(\pi,0)$, similar to the stoichiometric case, while the second branch appears near $(\pi,\pi)$. The first branch originates from the same nesting channels as in the stoichiometric case, whereas the second branch corresponds to scattering between the $X$ and $Y$ electron pockets. In the electron-doped case (Fig.~\ref{fig3}d), where the hole pockets are absent, the dominant peak of $\chi^S$ is associated with scattering between the $X$ and $Y$ electron pockets and therefore appears near $(\pi,\pi)$.
Comparing the largest eigenvalues of $\chi^S$ in Figs.~\ref{fig2} and \ref{fig3} shows that the peak near $(\pi,\pi)$ is present throughout the fillings considered. Near stoichiometric filling, however, the nesting associated with the $(0,\pi)$ and $(\pi,0)$ channels is enhanced, making the $(\pi,\pi)$ peak relatively less prominent.

\begin{figure}    \includegraphics[clip,width=3in,angle=0]{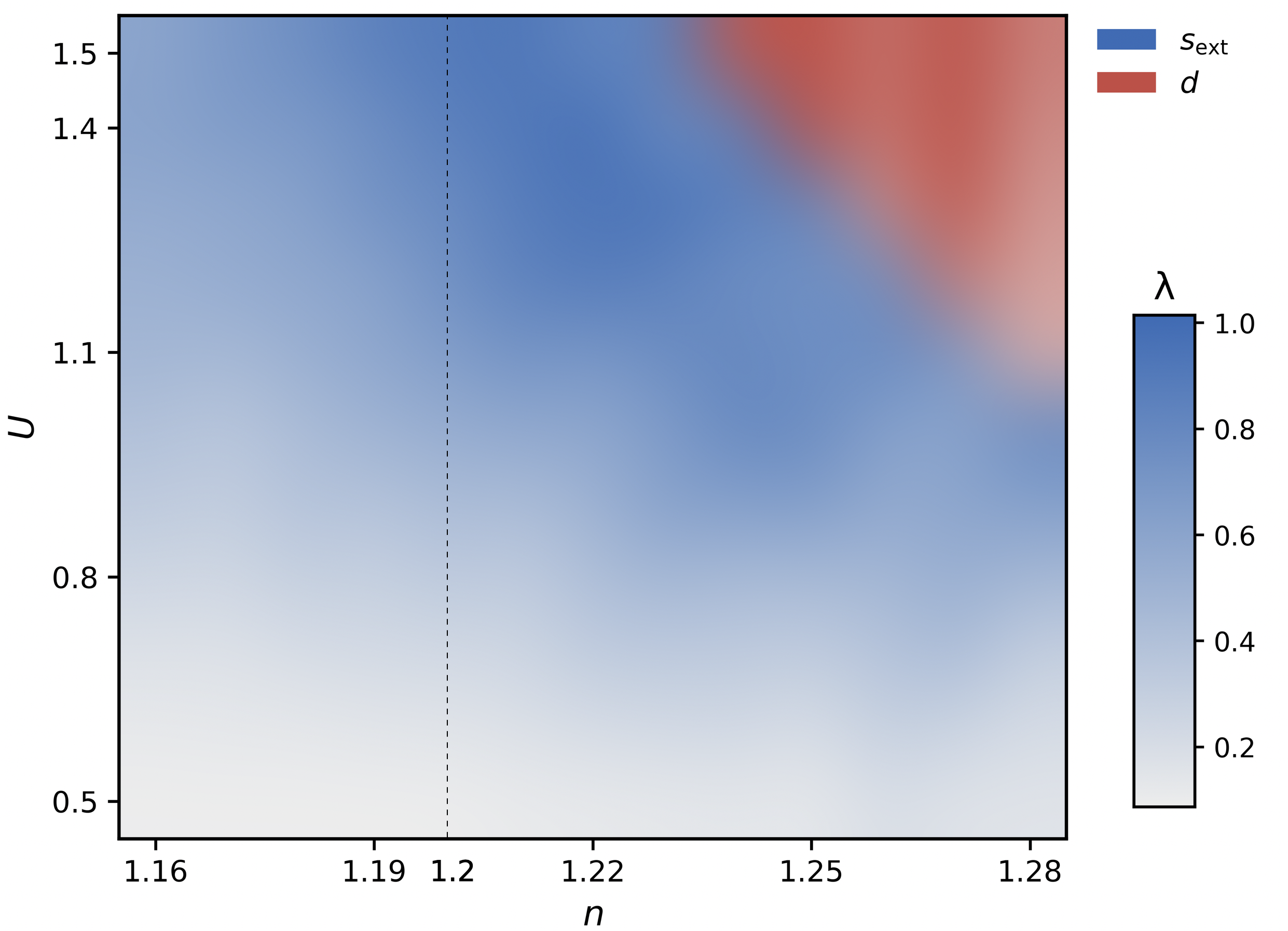}
    \caption{Pairing-symmetry phase diagram of $\text{FeTe}$ in the $U$--$n$ plane at $T=0.001$. The color indicates the leading pairing channel obtained from the largest Eliashberg eigenvalue: extended $s$-wave or $d$-wave. The vertical dashed line indicates the filling $n=$1.2 of the stoichiometric $\text{FeTe}$.
}
    \label{fig5}
\end{figure}

{\it Pairing symmetry and doping phase diagram. ---}
Within the parameter regime considered here, the two competing pairing symmetries are the extended $s$-wave and the $d_{x^2-y^2}$-wave states.
Figure~\ref{fig4}a shows the evolution of the eigenvalues $\lambda$ associated with the competing extended $s$-wave and $d$-wave pairing states in $\text{FeSe}$ and $\text{FeTe}$ as a function of $n$. In the hole-doped regime, the extended $s$-wave pairing is more favorable, and $\lambda_{s_{\mathrm{ext}}}$ exhibits a dome-like behavior in both $\text{FeSe}$ and $\text{FeTe}$. With increasing $n$, however, the $d$-wave state eventually overtakes the extended $s$-wave state and becomes dominant, which can be attributed to the shrinking of the hole pockets and the weakening of the nesting near $(0,\pi)$ and $(\pi,0)$. 
Fig.~\ref{fig4}a further shows that the extended $s$-wave eigenvalue is substantially larger in $\text{FeTe}$ than in $\text{FeSe}$ throughout the filling range considered. The $d_{x^2-y^2}$-wave eigenvalue is also generally larger in $\text{FeTe}$, with the enhancement particularly pronounced on the electron-doped side. Thus, within the common parameter set used here, $\text{FeTe}$ has a stronger pairing tendency in both channels, while retaining the same filling-driven competition between them.
Previous theoretical studies of $\mathrm{FeSe}$ have also found competing $s$- and $d$-wave pairing channels. Their relative strengths are affected by nematic order and orbital-selective quasiparticle coherence~\cite{Kang2018Nematic,Kreisel2017Orbital}, while incipient bands can contribute to pairing in electron-doped $\mathrm{FeSe}$-derived systems~\cite{Linscheid2016Incipient,Gao2018QPI}. These results are consistent with the $s$-to-$d$ crossover shown in Fig.~\ref{fig4}a.
We also find pronounced orbital-selective correlations in FeTe. The mass enhancement $z$ increases rapidly as $U$ increases. At $U = 1.5$ eV, the mass enhancement $z$ for the $d_{xy}$ orbital approaches 5 whereas the other orbitals has $z\sim$ (1.5-3). Besides, FeTe exhibits substantially stronger orbital-dependent renormalization than FeSe, indicating a more pronounced orbital-selective correlation effect (see Section C of the SM). 

Figure~\ref{fig4}b shows the evolution of the pairing symmetry in $\text{FeTe}$ as a function of $U$ at different fillings. 
We find that, within the interaction range considered here, the extended $s$-wave state remains the leading pairing instability at the hole-doped and stoichiometric fillings, although the $d$-wave channel becomes increasingly competitive as $U$ increases. On the electron-doped side, by contrast, the $d_{x^2-y^2}$-wave eigenvalue grows sufficiently rapidly to overtake the extended $s$-wave one at large $U$. This trend is consistent with the evolution of the spin fluctuations shown in Fig.~\ref{fig2} and Fig.~\ref{fig3}. In the hole-doped case, the spin susceptibility contains appreciable contributions near $(\pi,\pi)$, in addition to the peaks near $(\pi,0)$ and $(0,\pi)$, but the hole–electron scattering channels remain sufficiently strong to favor extended $s$-wave pairing over the interaction range considered. Upon electron doping, the depletion of the $\Gamma$ and M hole pockets suppresses these hole–electron scattering channels and enhances the relative importance of the scattering near $(\pi,\pi)$, thereby favoring $d$-wave pairing.
Figure~\ref{fig4}c-Fig.~\ref{fig4}e show the leading gap functions obtained at three representative fillings. The extended $s$-wave state is the leading pairing symmetry at $n=1.16$ and $n=1.20$. 
The sign reversal appears in two distinct places: one involves the hole pocket around $\Gamma$ and the electron pockets around $X/Y$, and the other involves the hole pocket around $M$ and those same $X/Y$ electron pockets.
Both sign reversals are associated with the $(0,\pi)$- and $(\pi,0)$-type antiferromagnetic spin fluctuations. When the $d$-wave state becomes the leading instability, the hole pockets are nearly removed from the Fermi surface by electron doping. As a result, the sign change mainly occurs within the $X/Y$ electron-pocket sector, corresponding to the $(\pi,\pi)$-type spin fluctuations. Moreover, since the nodal lines barely intersect the Fermi surface, the resulting state behaves a $d$-wave state that is nearly nodeless. 

Figure~\ref{fig5} shows the pairing-symmetry diagram of $\text{FeTe}$ at $T=0.001~\mathrm{eV}$ within the FLEX approximation. The blue region indicates that the extended $s$-wave state has the leading pairing symmetry, while the red region denotes the regime where the $d_{x^2-y^2}$-wave pairing state is dominant. This phase diagram shows that superconductivity in $\text{FeTe}$ is highly sensitive to both doping and interaction strength. In general, larger $U$ and higher electron doping drive the system toward the $d$-wave state, whereas the extended $s$-wave state, together with its associated $(0,\pi)$ component, dominates in regions close to the stoichiometric filling at large $U$.

{\it Conclusion. ---}
Starting from the experimentally determined structure of stoichiometric $\text{FeTe}$, we construct a five-orbital tight-binding model and treat spin fluctuations and pairing within the FLEX approach. Near stoichiometric filling, hole--electron scattering near $(\pi,0)$ and $(0,\pi)$ favors an extended $s$-wave state. Electron doping depletes the hole pockets and shifts the dominant scattering toward the $X$--$Y$ channel near $(\pi,\pi)$, favoring a $d_{x^2-y^2}$-wave state whose nodal lines largely avoid the remaining Fermi surfaces. The $U$--$n$ phase diagram shows that filling and interaction strength jointly tune the competition between these two pairing states. For matched parameters, the extended $s$-wave eigenvalue is larger in $\text{FeTe}$ than in $\text{FeSe}$ across the filling range considered, while the $d_{x^2-y^2}$-wave eigenvalue is also generally larger, particularly on the electron-doped side. These results provide a microscopic spin-fluctuation pairing scenario for stoichiometric $\text{FeTe}$ and predict a doping-driven change in gap symmetry that can be tested by momentum-resolved and phase-sensitive measurements.

{\it Acknowledgements. ---}
C.Y. acknowledges support from the Guangdong Major Project of Basic Research (Grant No. 2025B0303000004), the Guangdong Provincial Quantum Science Strategic Initiative (Grant Nos. GDZX2401004 and GDZX2501001), and the National Natural Science Foundation of China (Grant No. 12474231).  J.-J.M is supported by NSFC (No. 12404171) and the Guangdong Project (Grant No. 2024QN11X176). 

\clearpage

\clearpage
\beginsupplement
\begin{widetext}

\begin{center}
{\large\bf Supplementary Material for\\[0.5ex]
Competing Extended-$s$- and $d$-Wave Pairing from Distinct Spin-Fluctuation Channels in Stoichiometric $\mathrm{FeTe}$}
\end{center}
\vspace{1em}

\subsection{A. DFT Calculation}

In this section of the Supplemental Material, we provide the detailed methodology of our first-principles calculations, the band unfolding procedure enabled by the reduction of the 3D glide mirror plane to a pure translation in the 2D limit and the explicit construction of the tight-binding (TB) effective Hamiltonians for bulk FeTe.
\subsubsection{First-Principles Methodology}
To accurately capture the low-energy electronic structure of bulk FeTe, first-principles density functional theory (DFT) calculations were performed using the projector augmented-wave (PAW)\cite{blochl1994projector} method as implemented in the Vienna Ab initio Simulation Package (VASP)\cite{kresse1996efficient}. The exchange-correlation interaction was treated within the generalized gradient approximation (GGA) using the Perdew-Burke-Ernzerhof (PBE) functional\cite{perdew1996generalized}. The plane-wave cutoff energy was set to 400 eV. A dense $\Gamma$-centered $k$-point mesh of $10 \times 10 \times 7$ was employed to sample the Brillouin zone of the bulk primitive cell. The energy convergence criterion was set to $10^{-7}$ eV. The experimental bulk lattice parameters were used as the starting point for the structural relaxation (a = 3.862 \AA, c = 6.262 \AA, and an Fe-Te separation of 1.762 \AA).

\begin{figure*}[htp]
\centering
\includegraphics[width=0.5\textwidth]{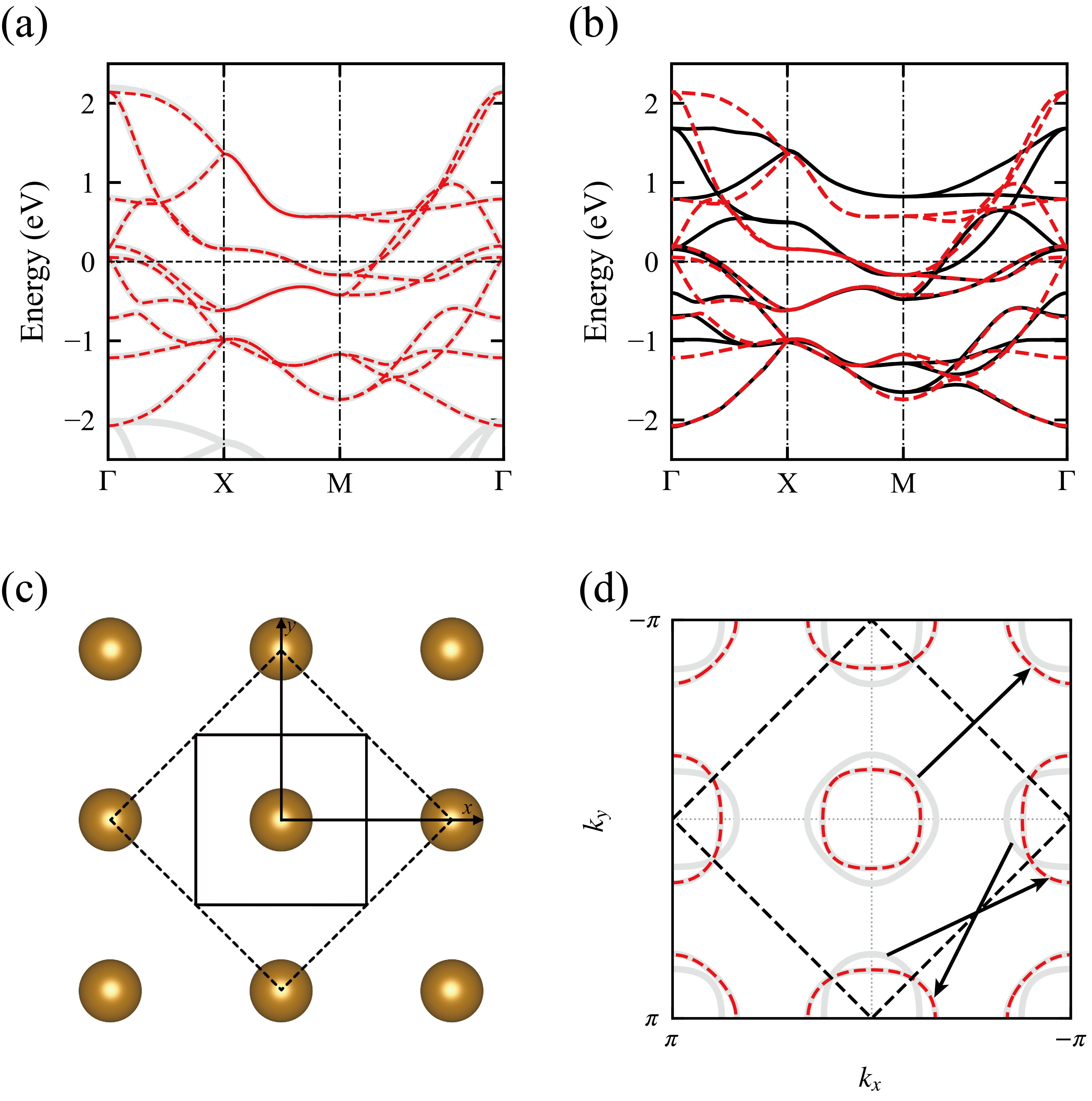}
    \caption{
     (a) Electronic band structure of the 10-band TB model (red dashed lines) compared with the DFT band structure (gray solid lines) of bulk FeTe. 
     (b) Comparison of the 10-band TB model with (red solid lines) and without (black solid lines) the interlayer $z$-direction hopping terms ($t_z$). 
     (c) Real-space crystal structure of FeTe: the black solid lines indicate the primitive 1-Fe unit cell, while the black dashed lines denote the structural 2-Fe unit cell. 
     (d) Fermi surface evolution and band unfolding. The gray solid lines represent the Fermi surface of the 2-Fe unit cell, while the red dashed lines show the Fermi surface in the folded Brillouin zone of the 2-Fe unit cell (indicated by black dashed lines). The black arrows indicate the mapping process of the band unfolding procedure from the 2-Fe unit cell to the primitive Brillouin zone.
    }
    \label{s1}
\end{figure*}

\subsubsection{The 10-Band TB Model}
To construct the low-energy effective tight-binding Hamiltonian, we mapped the first-principles DFT results onto a realistic TB model using the Wannier90 code\cite{mostofi2008wannier90}. By generating maximally localized Wannier functions (MLWFs) spanning the five Fe-$3d$ orbitals ($d_{3z^2-r^2}$, $d_{xz}$, $d_{yz}$, $d_{x^2-y^2}$, and $d_{xy}$), we obtained a highly accurate 10-band TB model.As illustrated in Fig.~\ref{s1}(c), the checkerboard arrangement of the Te atoms necessitates the use of a 2-Fe supercell in real space, as opposed to the primitive 1-Fe unit cell. The calculated 10-band TB band structure (red dashed lines) shows excellent agreement with the DFT dispersion (gray solid lines) within the energy window of $[-2.0, 2.5]$ eV around the Fermi level, as presented in Fig.~\ref{s1}(a). Furthermore, to quantify the contribution of interlayer coupling to the electronic structure, Fig.~\ref{s1}(b) provides a comparison between the full 10-band model and a modified version where the inter-layer $z$-direction hopping terms ($t_z$) are artificially omitted. This comparison confirms that our model accurately captures the physical effects of interlayer interactions on the electronic states near the Fermi level.

\subsubsection{Band Unfolding and the 5-Band Effective Model}
\begin{figure}
\centering
\includegraphics[width=0.85\textwidth]{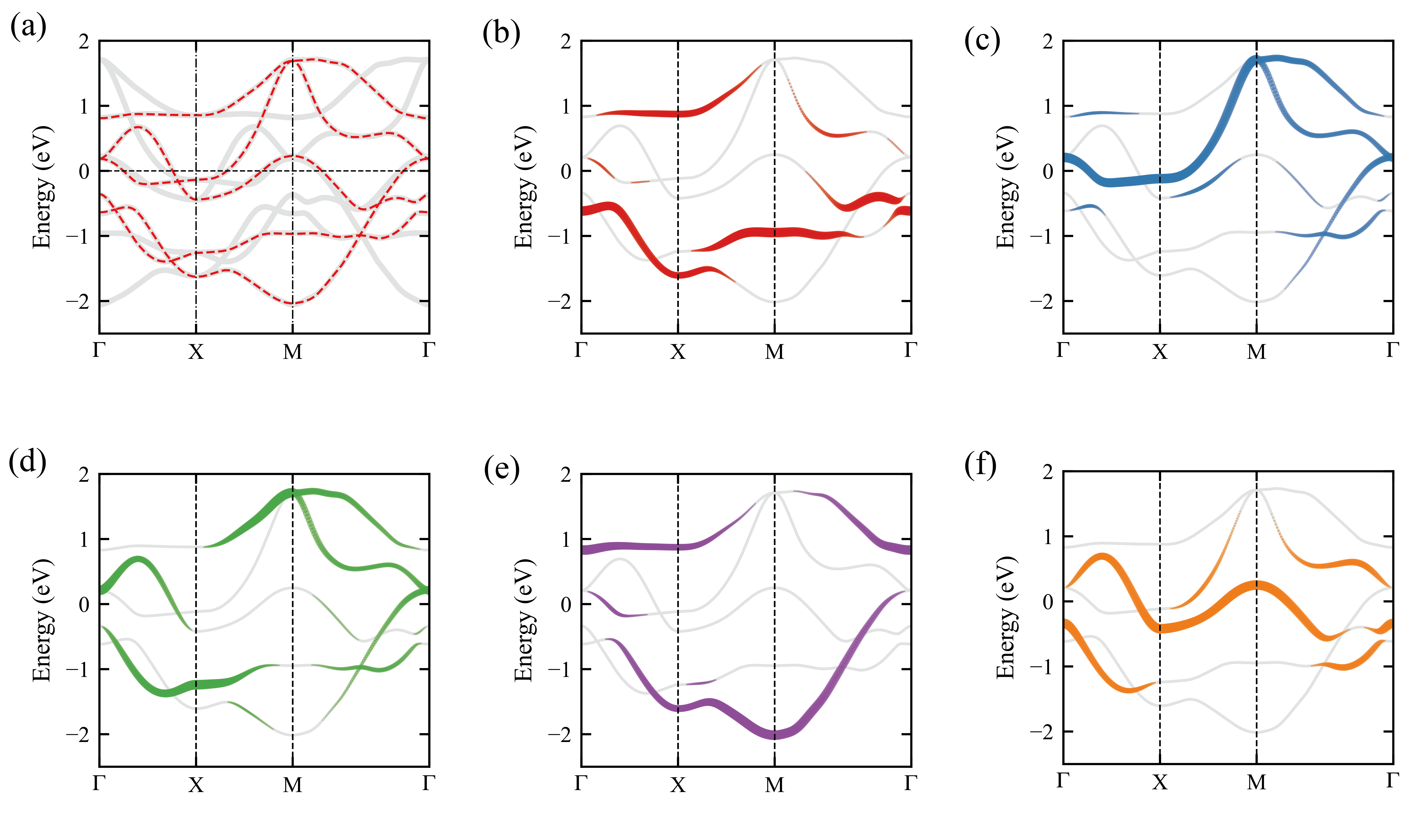}
    \caption{
    (a) Energy band structures of the 10-band TB model (gray solid lines) and the reduced 5-band effective model (red dashed lines) in the 1-Fe Brillouin zone. 
    (b)-(f) Wannier-orbital-projected band structures of the 5-band effective model, illustrating the dominant weight of the Fe $3d$ orbitals: (b) $d_{z^2}$, (c) $d_{xz}$, (d) $d_{yz}$, (e) $d_{x^2-y^2}$, and (f) $d_{xy}$.The color scale indicates the orbital weight.
    }
    \label{s2}
\end{figure}
Because the 2-Fe supercell leads to a folded Brillouin zone, we employed the band unfolding to derive a 5-band effective model for the 1-Fe Brillouin zone\cite{kuroki2008unconventional}. This procedure is physically justified by the $P4/nmm$ glide mirror symmetry, which maps onto a pure in-plane translation in the 2D limit. Fig.~\ref{s1}(d) illustrates the mapping process: the gray lines represent the primitive Fermi surface, and the red dashed lines show the folded surface in the supercell Brillouin zone (black dashed lines), with black arrows indicating the unfolding transformation. Utilizing this, the 10-band model is reduced to the 5-band representation. Fig.~\ref{s2}(a) demonstrates that the dispersion of the reduced 5-band model perfectly overlaps with the low-energy states of the 10-band model. Fig.~\ref{s2}(b)-(f) further present the orbital-projected band structure, delineating the dominant contributions of the five Fe $3d$ orbitals ($d_{z^2}, d_{xz}, d_{yz}, d_{x^2-y^2}, d_{xy}$).The explicit on-site energies and all hopping parameters $t_{ij}^{\alpha\beta} > 0.01$ eV are meticulously tabulated in Table S1.

\begin{table}[htbp]
\centering
\caption{\textbf{Tight-binding parameters for FeTe.}Hopping integrals $t_{ij}^{\alpha\beta}$ in units of 1 eV. $[i,j]$ denotes the in-plane hopping vector, and $(\alpha,\beta)$ the orbitals. The on-site energies are $(\varepsilon_1,\varepsilon_2,\varepsilon_3,\varepsilon_4,\varepsilon_5) = (\text{-0.469, 0.078, 0.078, -0.424, -0.184})$ eV. $\sigma_y$, $I$, and $\sigma_d$ correspond to $t_{i,-j}^{\alpha\beta}$, $t_{-i,-j}^{\alpha\beta}$, and $t_{j,i}^{\alpha\beta}$, respectively, where ``$\pm$'' and ``$\pm(\alpha',\beta')$'' in the $(\alpha,\beta)$ row mean that the corresponding hopping is equal to $\pm t_{ij}^{\alpha\beta}$ and $\pm t_{ij}^{\alpha'\beta'}$, respectively. This table, combined with the relation $t_{ij}^{\alpha\beta} = t_{ji}^{\beta\alpha}$,gives all the in-plane hoppings $\ge 0.01$ eV.}
\label{tab:hopping_fete}
\renewcommand{\arraystretch}{1.2}
\begin{tabular*}{\textwidth}{@{\extracolsep{\fill}} lccccccccc}
\hline\hline
\diagbox[width=6.5em, height=8.5ex]{$(\alpha,\beta)$}{$[i, j]$} & [1,0] & [1,1] & [2,0] & [2,1] & [2,2] & $\sigma_y$ & $I$ & $\sigma_d$  \\
\hline 
(1, 1) & 0.060&-0.046&      &-0.005&-0.017& $+$ & $+$ & $+$ \\
(1, 2) &      & 0.135&      & 0.007& 0.017& - & - & $+(1,3)$\\
(1, 3) & 0.165&-0.135& 0.023&-0.015&-0.017& + & - & -(1,2) \\
(1, 4) & 0.315&      & 0.031& 0.007&      & + & + & - \\
(1, 5) &      &-0.038&      &      & 0.008& - & + & \\
(2, 2) &-0.065& 0.196& 0.012&-0.037& 0.030& + & + & +(3,3) \\
(2, 3) &      &-0.097&      & 0.026&-0.042& - & + & + \\
(2, 4) &      & 0.068&      & 0.027&-0.005& - & - & +(3,4) \\
(2, 5) & 0.244& 0.088&      & 0.009&      & + & - & -(3,5) \\
(3, 3) &-0.251& 0.196&-0.044& 0.013& 0.030& + & + & +(2,2) \\
(3, 4) &-0.315& 0.068&-0.008&-0.010&-0.005& + & - & +(2,4) \\
(3, 5) &      &-0.088&      &-0.021&      & - & - & -(2,5) \\
(4, 4) & 0.363&-0.020&-0.035&      & 0.018& + & + & + \\
(4, 5) &      &      &      & 0.012&      & - & + & - \\
(5, 5) &-0.043& 0.056& 0.007&-0.011&-0.019& + & + & + \\
\hline\hline 
\end{tabular*}
\end{table}
\begin{table}[htbp]
\centering
\caption{\textbf{Tight-binding parameters for FeSe.}Hopping integrals $t_{ij}^{\alpha\beta}$ in units of 1 eV. $[i,j]$ denotes the in-plane hopping vector, and $(\alpha,\beta)$ the orbitals. The on-site energies are $(\varepsilon_1,\varepsilon_2,\varepsilon_3,\varepsilon_4,\varepsilon_5) = (\text{-0.424, 0.051, 0.051, -0.528, -0.069})$ eV. $\sigma_y$, $I$, and $\sigma_d$ correspond to $t_{i,-j}^{\alpha\beta}$, $t_{-i,-j}^{\alpha\beta}$, and $t_{j,i}^{\alpha\beta}$, respectively, where $\pm$'' and $\pm(\alpha',\beta')$'' in the $(\alpha,\beta)$ row mean that the corresponding hopping is equal to $\pm t_{ij}^{\alpha\beta}$ and $\pm t_{ij}^{\alpha'\beta'}$, respectively. This table, combined with the relation $t_{ij}^{\alpha\beta} = t_{ji}^{\beta\alpha}$,gives all the in-plane hoppings $\ge 0.01$ eV.}
\label{tab:hopping_fese}
\renewcommand{\arraystretch}{1.2}
\begin{tabular*}{\textwidth}{@{\extracolsep{\fill}} lccccccccc}
\hline\hline
\diagbox[width=6.5em, height=8.5ex]{$(\alpha,\beta)$}{$[i, j]$} & [1,0] & [1,1] & [2,0] & [2,1] & [2,2] & $\sigma_y$ & $I$ & $\sigma_d$  \\
\hline
(1, 1) & & -0.011 & -0.016 & & -0.019 & $+$ & $+$ & $+$ \\
(1, 2) & & 0.145 & & -0.013 & 0.016 & $-$ & $-$ & $+(1,3)$ \\
(1, 3) & 0.180 & -0.145 & 0.010 & -0.013 & -0.016 & $+$ & $-$ & $-(1,2)$ \\
(1, 4) & 0.336 & & 0.028 & & & $+$ & $+$ & $-$ \\
(1, 5) & & -0.111 & & & 0.012 & $-$ & $+$ & \\
(2, 2) & -0.077 & 0.221 & & -0.017 & 0.027 & $+$ & $+$ & $+(3,3)$ \\
(2, 3) & & -0.098 & & -0.014 & -0.034 & $-$ & $+$ & $+$ \\
(2, 4) & & 0.088 & & 0.022 & & $-$ & $-$ & $+(3,4)$ \\
(2, 5) & 0.305 & 0.058 & & & & $+$ & $-$ & $-(3,5)$ \\
(3, 3) & -0.317 & 0.221 & -0.050 & 0.021 & 0.027 & $+$ & $+$ & $+(2,2)$ \\
(3, 4) & -0.363 & 0.088 & -0.016 & & & $+$ & $-$ & $+(2,4)$ \\
(3, 5) & & -0.058 & & -0.011 & & $-$ & $-$ & $-(2,5)$ \\
(4, 4) & 0.407 & -0.065 & & & 0.016 & $+$ & $+$ & $+$ \\
(4, 5) & & & & 0.014 & & $-$ & $+$ & $-$ \\
(5, 5) & 0.061 & 0.092 & & & -0.028 & $+$ & $+$ & $+$ \\
\hline\hline
\end{tabular*}
\end{table}

\begin{table}[htbp]
\centering
\caption{\textbf{The most significant interlayer hopping integrals neglected in the 2D projection.} Listed here are the top five largest out-of-plane hopping paths for both FeSe and FeTe. The hopping vector is denoted by $[R_x, R_y, R_z]$ (with $R_z \neq 0$), and $(n \to m)$ represents the corresponding orbital indices in the 10-orbital (2-Fe) unit cell, where orbitals 1--5 and 6--10 belong to the two equivalent Fe atoms, respectively. The magnitudes $|t|$ are given in units of 1 eV. As explicitly shown, the maximum out-of-plane couplings ($\sim 0.031$~eV for FeSe and $\sim 0.062$~eV for FeTe) are an order of magnitude weaker than the dominant in-plane hoppings, thereby firmly validating the two-dimensional effective models.}
\label{tab:z_hopping}
\renewcommand{\arraystretch}{1.2}
\begin{tabular*}{\textwidth}{@{\extracolsep{\fill}} lcccc}
\hline\hline
Material & $[R_x, R_y, R_z]$ & $(n \to m)$ & Hopping Integral $t$ (eV) & $|t|$ (eV) \\
\hline 
FeSe & $[-2,  0, -1]$, etc. & (3 $\to$ 10) , (5 $\to$  8) & -0.0311 + 0.0000j & 0.0311 \\
     & $[ 0, -1,  1]$, etc. & (2 $\to$ 10) , (5 $\to$  7) &  0.0311 + 0.0000j & 0.0311 \\
     & $[ 0,  0, \pm1]$     & (5 $\to$  5) , (10 $\to$ 10) & -0.0306 + 0.0000j & 0.0306 \\
     & $[-1, -1, -1]$, etc. & (3 $\to$  5) , (8 $\to$ 10) &  0.0295 + 0.0000j & 0.0295 \\
     & $[-1, -1,  1]$, etc. & (3 $\to$  5) , (8 $\to$ 10) & -0.0295 + 0.0000j & 0.0295 \\
\hline
FeTe & $[ 0,  0, \pm1]$     & (1 $\to$  1) , (6 $\to$  6) & -0.0621 + 0.0000j & 0.0621 \\ 
     & $[-1, -1,  1]$, etc. & (3 $\to$  8) , (8 $\to$  3) & -0.0370 + 0.0000j & 0.0370 \\ 
     & $[ 0, -1, -1]$, etc. & (1 $\to$  5) , (6 $\to$ 10) &  0.0359 + 0.0000j & 0.0359 \\ 
     & $[-1,  0, -1]$, etc. & (1 $\to$  5) , (6 $\to$ 10) & -0.0359 + 0.0000j & 0.0359 \\ 
     & $[ 0,  0, \pm1]$     & (2 $\to$  7) , (7 $\to$  2) & -0.0341 + 0.0000j & 0.0341 \\ 
\hline\hline 
\end{tabular*}
\end{table}

\newpage

\subsection{B. FLEX Calculation}

\subsubsection{FLEX band renormalization}

FLEX is a self-consistent field-theoretical method formulated within the Baym--Kadanoff framework \cite{bickers1989conserving,bickers1991conserving,yanase2003theory,kuroki2008unconventional,witt2021efficient}. It self-consistently incorporates the electronic self-energy corrections arising from infinite series of bubble and ladder diagrams in the weak-to-intermediate correlation regime, and yields the corresponding spin and charge susceptibilities, as discussed in the main text. FLEX has played an important role in studies of pairing mechanisms and pairing symmetries in iron-based superconductors. In this part, we present the detailed procedure of the FLEX calculation:

We start from the bare Green's function, whose input parameters are provided in Table S1.
\begin{equation}
\hat{G}_0(k) = \left[ (i\omega_n+\mu)\hat{1} -\hat{H}_0({\bf k}) \right]^{-1},
\end{equation}
where $k\equiv(\mathbf{k},i\nu_m)$ denotes the  momentum-Matsubara frequency variable. Hatted quantities are matrices defined in the orbital space. 
$\hat{G}^0$ is related to the interacting Green's function $\hat{G}$ by the Dyson equation 
\begin{equation}
\hat{G}(k)^{-1} = \hat{G}_0(k)^{-1}-\Sigma(k), 
\end{equation}
with $\Sigma(k)$ the self-energy.
The irreducible bare susceptibility constructed from the Green's function is 
\begin{equation}
\chi^{0}_{ab,cd}(q) = -\frac{T}{N_k} \sum_k G_{ac}(k+q) G_{db}(k). 
\end{equation}
In the spin and charge channels, the interaction tensors are defined by
\begin{equation}
U^{S}_{abcd} =
\begin{cases}
U  \text{ ,} & a=b=c=d  \\
U' \text{, }  & a=c\neq b=d \\
J  \text{ ,} & a=b\neq c=d \\
J' \text{, }  & a=d \neq b=c
\end{cases}
\end{equation}
\begin{equation}
U^{C}_{abcd} =
\begin{cases}
U  \text{\ ,} & a=b=c=d  \\
-U'+J \text{, }  & a=c\neq b=d \\
2U'-J  \text{\ ,} & a=b\neq c=d  \\
J' \text{, }  & a=d \neq b=c
\end{cases}
\end{equation}
The effective interaction for this model is defined as
\begin{equation}
\hat{V}_{\rm eff}(q) = \frac{3}{2} \hat{U}^{S} \left[ \hat{\chi}^{S}(q)-\frac{1}{2}\hat{\chi}^{0}(q) \right] \hat{U}^{S} + \frac{1}{2} \hat{U}^{C} 
\left[ \hat{\chi}^{C}(q)-\frac{1}{2}\hat{\chi}^{0}(q) \right] \hat{U}^{C} + \frac{3}{2}\hat{U}^{S} - \frac{1}{2}\hat{U}^{C}. 
 \end{equation}
The normal self-energy can then be written compactly as 
\begin{equation}
\Sigma_{ab}(k) = \frac{T}{N_k} \sum_q \sum_{c,d} \left[ V_{\rm eff}(q) \right]_{ab,cd} G_{cd}(k-q). 
 \end{equation}
The resulting self-energy is then substituted into the Dyson equation to obtain the updated Green's function. The above procedure is repeated until both the self-energy and the Green's function no longer change during the self-consistency loop. In this way, we obtain the converged interaction-dressed Green's function, susceptibilities, and other physical quantities. In each self-consistency loop, we fix the electron filling and determine the chemical potential by a bisection search
\begin{equation}
n_{\rm fill} = 2\left[ 1+ {\rm Re} \sum_n w_n^{(0)} \frac{1}{N_kN_{\rm orb}} \sum_{\bf k} {\rm Tr}\,\hat{G}({\bf k},i\omega_n) \right],
\end{equation}
where $w_n^{(0)}$ denotes the numerical transformation weight from Matsubara frequency to $\tau=0^{-}$ used in the IR basis \cite{Shinaoka2017IRSM,wallerberger2023sparse}. In the calculations, we use Fourier transforms and the IR basis to optimize the evaluation of convolution terms. In the FLEX calculations, we start from a relatively small value of $U$ and gradually increase it during the self-consistency procedure to ensure that the Stoner factor neither becomes larger than 1 nor diverges. Namely, for the Stoner factor reads
\begin{equation}
\alpha_{s}=\text{max}~\text{Re}~\text{eig}[\hat{\chi^{0}}(q)\hat{U}^{S}]
\end{equation}
we ensure that $\alpha_{s}<1$ throughout the FLEX calculations.
We find that, within the parameter regime discussed in this work, the calculations remain stable for $U<1.5~\mathrm{eV}$.

\subsubsection{Gap equation}

After obtaining the converged normal-state quantities from the FLEX self-consistency loop, including the renormalized Green’s function, self-energy, and spin and charge susceptibilities, we construct the linearized Eliashberg equation using the converged Green’s function with the self-energy fully included. The effective pairing interaction is built from the FLEX spin and charge fluctuation propagators in the singlet channel. The gap equation is then treated as an eigenvalue problem of the form $\lambda\Delta=K\Delta$, where $K$ is the Eliashberg kernel acting on the gap function in the combined momentum, Matsubara-frequency, and orbital space. We applying the effective pairing interaction through fast Fourier transformation and intermediate representation basis (irbasis) transformations \cite{Shinaoka2017IRSM,wallerberger2023sparse}, and then imposing the required even-frequency singlet symmetry. The leading eigenvalues and eigenfunctions are obtained using an iterative Arnoldi/ARPACK eigensolver. The largest eigenvalue $\lambda$ characterizes the superconducting pairing strength, while the corresponding eigenfunction at the lowest Matsubara frequency is identified as the superconducting gap structure.

For a given trial gap function $\hat{\Delta}(k)$, the linearized anomalous Green's function in orbital indices is
\begin{equation}
F_{ab}(k) = \sum_{c,d} G_{ac}(k) \Delta_{cd}(k) G^{*}_{db}(-k). 
\end{equation}
For the spin-singlet pairing channel, we define the total effective pairing interaction as 
\begin{equation}
\hat{V}^{\rm pair}(q) = \frac{3}{2} \hat{U}^{S} \hat{\chi}^{S}(q) \hat{U}^{S} - \frac{1}{2} \hat{U}^{C} \hat{\chi}^{C}(q) \hat{U}^{C} + \frac{3\hat{U}^{S}+\hat{U}^{C}}{4}. 
\end{equation}
The full linearized Eliashberg equation is then written as 
\begin{equation}
\lambda \Delta_{ab}(k) = - \frac{T}{N_k} \sum_q \sum_{c,d} \left[ V^{\rm pair}(q) \right]_{ab,cd} F_{cd}(k-q).
\end{equation}
To reveal its physical momentum structure, we transform the orbital-basis gap function into the bare band basis and examine the diagonal band components on the Fermi surface, which is determined from the low-frequency spectral weight of the converged Green’s function. After fixing the arbitrary global sign of each eigenvector for visualization, the sign structure of the real part of the gap projected on the Fermi surface is used to identify its symmetry. In particular, a gap structure that remains invariant under fourfold rotation is assigned to the extended s-wave channel, whereas a gap that changes sign under the corresponding lattice symmetry operations is classified as a d-wave state.

\subsection{C. Orbital selectivity of correlation}

\begin{figure}
\centering
\includegraphics[width=1\textwidth]{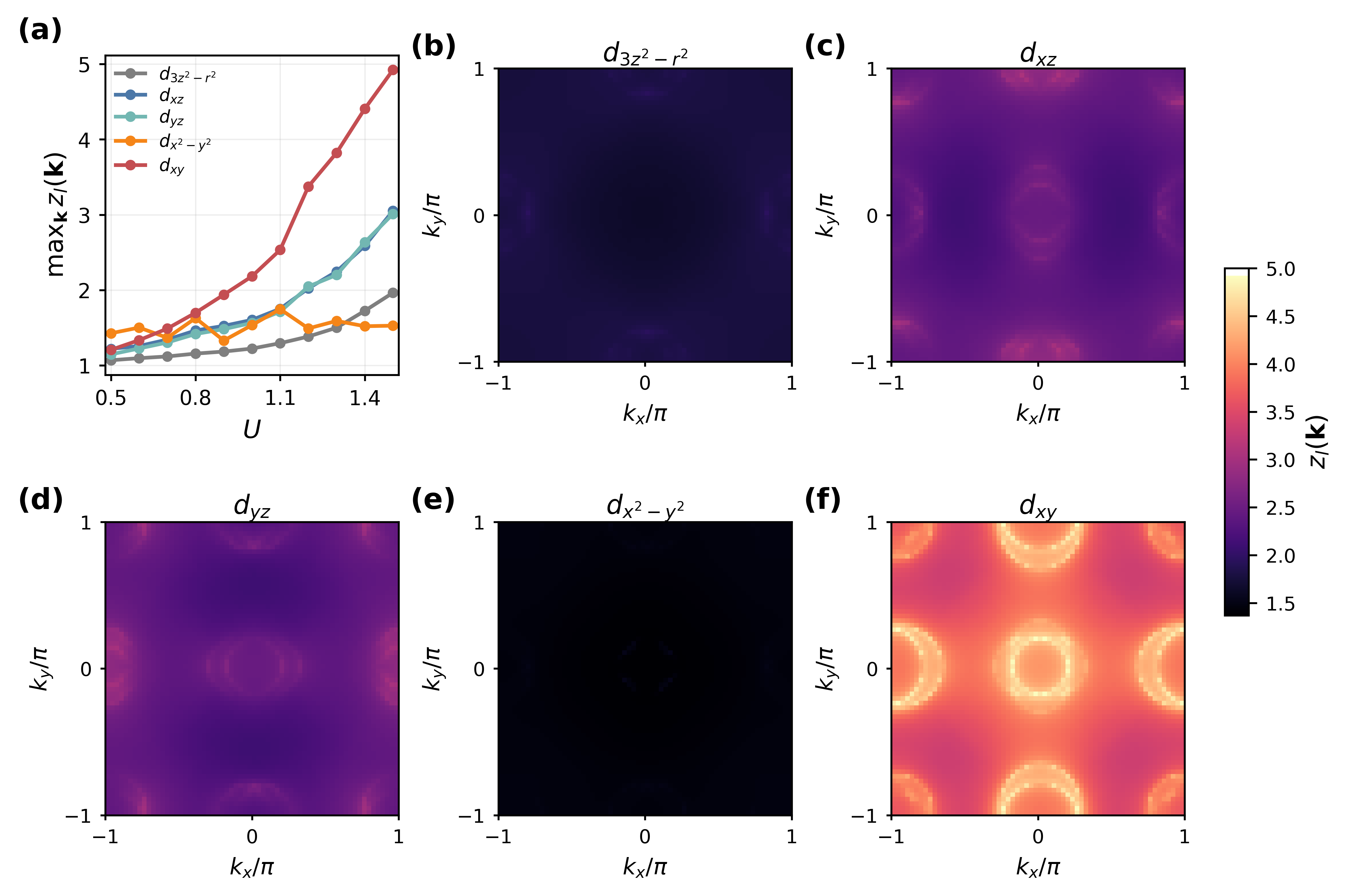}
    \caption{Orbital-resolved renormalization factor in stoichiometric FeTe. (a) Orbital-resolved maximum of the renormalization factor, $\max_{\mathbf{k}}z_l(\mathbf{k})$, as a function of interaction strength U for FeTe at n=1.20 and T=0.001. The five curves correspond to the Fe 3d orbitals $d_{3z^2-r^2}$, $d_{xz}$, $d_{yz}$, $d_{x^2-y^2}$, and $d_{xy}$. The renormalization factor is evaluated from the lowest positive Matsubara self-energy as (b–f) Momentum-resolved diagonal components $z_l(\mathbf{k})$ at U=1.5, n=1.20, and T=0.001, shown separately for the five orbitals. All heat maps in panels (b–f) share the same color scale.
    }
    \label{s3}
\end{figure}
    
\begin{figure}
\centering
\includegraphics[width=1\textwidth]{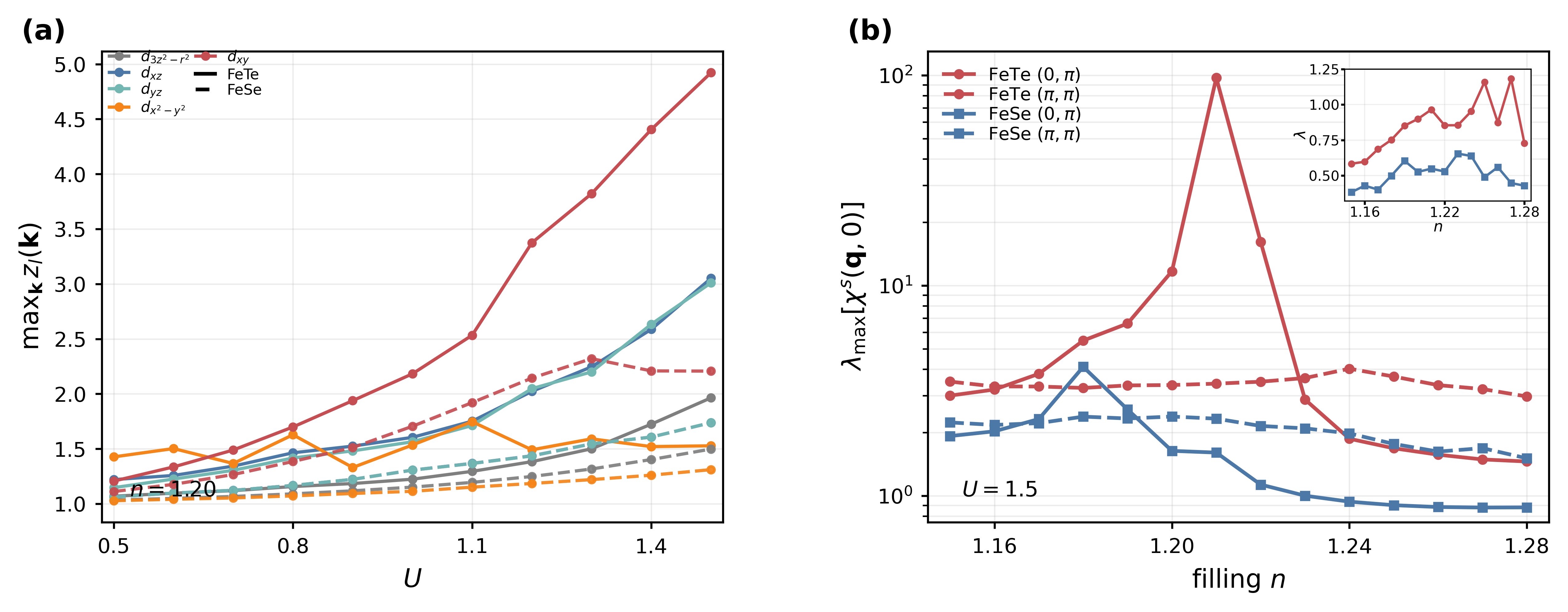}
    \caption{Correlation effects in FeTe and FeSe. (a) Orbital-resolved maximum renormalization factor $\max_{\mathbf{k}} z_l(\mathbf{k})$ as a function of the on-site interaction U at fixed filling n=1.20. Colors label the five Fe \(3d\) orbitals, and solid/dashed lines denote FeTe/FeSe, respectively. FeTe shows a much stronger orbital-selective enhancement, most prominently in the $d_{xy}$ orbital. (b) Maximum eigenvalue of the static spin susceptibility, $\Lambda_{\max}[\chi^s(\mathbf{q},0)]$, as a function of filling n at U=1.5, shown for $\mathbf{q}=(0,\pi)$ and $(\pi,\pi)$. The main panel uses a logarithmic scale, while the inset shows the corresponding superconducting eigenvalue $\lambda$. 
}
    \label{s4}
\end{figure}

The renormalization factor provides a useful measure of the strength of electronic correlations. It is defined as \cite{ikeda2010phase}
\begin{equation}
z_{ab}(\mathbf{k}) =
\delta_{ab}
-
\left.
\frac{\partial \mathrm{Im}\,\Sigma_{ab}(\mathbf{k},i\omega)}
{\partial \omega}
\right|_{\omega\rightarrow 0},
\end{equation}
where $a$ and $b$ denote orbital indices, and $\Sigma_{ab}(\mathbf{k},i\omega)$ is the self-energy matrix in the
orbital basis. At sufficiently low temperature, this quantity can be approximated using the lowest positive Matsubara
frequency as
\begin{equation}
z_{ab}(\mathbf{k})
\simeq
\delta_{ab}
-
\frac{\mathrm{Im}\,\Sigma_{ab}(\mathbf{k},i\pi T)}
{\pi T}.
\end{equation}

In the following, we focus on the diagonal orbital components, $z_l(\mathbf{k})\equiv z_{ll}(\mathbf{k})$.
Fig.~\ref{s3}a shows the orbital-resolved maximum value, $\max_{\mathbf{k}} z_l(\mathbf{k})$, as a function of the on-site interaction strength $U$ for stoichiometric FeTe. In the present filling and interaction range, the $e_g$ orbitals contribute only weakly to the low-energy electronic structure of FeTe, and therefore their renormalization factors remain small even at relatively large $U$. In contrast, the renormalization factors of the $d_{xz}$, $d_{yz}$, and $d_{xy}$ orbitals increase substantially with increasing $U$, consistent with previous theoretical expectations for orbital-selective correlation effects in FeTe. In particular, the $d_{xy}$ orbital is more strongly renormalized than the $d_{xz}$ and $d_{yz}$ orbitals. At $U=1.5$, the maximum value of $z_l$ for the $d_{xy}$ orbital approaches 5, indicating pronounced orbital-selective correlations in FeTe. Fig.~\ref{s3}b-f further show the momentum-resolved distribution of the orbital-dependent renormalization factors at $U=1.5$. The enhanced renormalization is mainly concentrated in the momentum regions carrying low-energy spectral weight, namely the regions associated with the Fermi surface. This observation is consistent with our subsequent analysis of superconducting pairing: within the FLEX framework, the pairing interaction is primarily governed by low-energy quasiparticle states and the spin fluctuations generated by them, which are dominated by the $d_{xz}$, $d_{yz}$, and $d_{xy}$ orbitals. Therefore, the orbital-selective enhancement revealed by the renormalization factor naturally corresponds to the dominant orbital components of the leading gap function and to the spin-fluctuation-mediated pairing structure discussed below. In this sense, Fig.~\ref{s3} provides microscopic support for the orbital origin of the superconducting instability in FeTe. Fig.~\ref{s4} compares the correlation strength and the leading eigenvalues of the spin susceptibility at representative wave vectors between FeTe and FeSe. Across different orbital channels, FeTe exhibits substantially stronger orbital-dependent renormalization than FeSe, indicating a more pronounced orbital-selective correlation effect. Consistently, the leading eigenvalues of the spin susceptibility are also significantly enhanced in FeTe. These results show that the stronger electronic correlations in FeTe promote more robust antiferromagnetic spin fluctuations, providing a natural explanation for the larger superconducting pairing tendency found in FeTe compared with FeSe.

\newpage

\subsection{D. Supplementary result of FeTe}

Fig.~\ref{s5} supplements the main-text results by showing the lowest Matsubara frequency spectral function at different electron fillings, which serves as an approximate representation of the Fermi surface. As the system evolves from hole doping to electron doping, the hole-like low-energy spectral weight around the $\Gamma$ and $M$ points shrinks substantially and eventually disappears, while the electron-like spectral weight around the $X/Y$ points expands slightly.

Fig.~\ref{s6} shows the maximum eigenvalue of the zero-frequency spin susceptibility for the same set of fillings as in Fig.~\ref{s5}. With increasing electron doping, the nesting peak in the $(0,\pi)$ family first increases and then decreases, reaching its maximum near the stoichiometric filling. In contrast, the peak intensity in the $(\pi,\pi)$ family varies much more weakly and remains relatively stable.
This behavior can be naturally understood from the Fermi-surface nesting analysis discussed in the main text. The spin fluctuations near $(0,\pi)$ and $(\pi,0)$ are mainly contributed by scattering between the $X/Y$ electron pockets and the hole pocket around the $M$ point. As electron doping increases, the hole pocket around $M$ gradually shrinks. At a filling close to stoichiometry, its size and shape become best matched to those of the $X/Y$ electron pockets, causing the nesting wave vector to become strongly concentrated near $(0,\pi)$ and $(\pi,0)$, and producing a pronounced enhancement of the spin-susceptibility peak. By contrast, the spin fluctuations near $(\pi,\pi)$ mainly originate from scattering between the $X/Y$ electron pockets themselves. Since the associated low-energy spectral weight changes more moderately with electron doping, the corresponding susceptibility peak remains comparatively stable.

\begin{figure}
\centering
\includegraphics[width=1\textwidth]{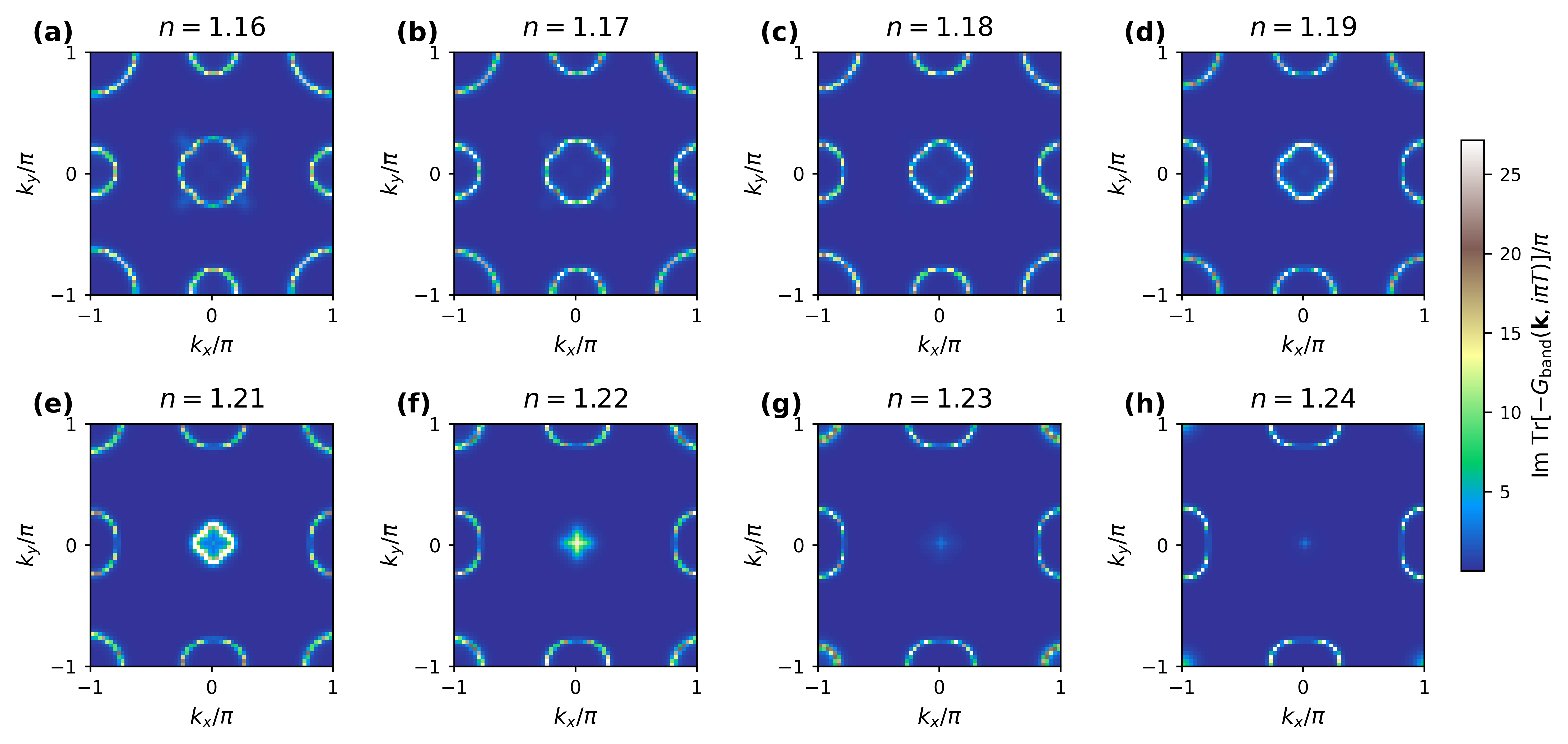}
    \caption{Filling dependence of the Fermi-surface in FeTe. Momentum-resolved spectral-weight proxy for FeTe at U=1.5 and T=0.001, shown for fillings (a) n=1.16, (b) n=1.17, (c) n=1.18, (d) n=1.19, (e) n=1.21, (f) n=1.22, (g) n=1.23, and (h) n=1.24. The plotted quantity is $\frac{1}{\pi}\,\mathrm{Im}\,\mathrm{Tr}\left[-G(\mathbf{k},i\pi T)\right]$. All panels use the same color scale.
    }
    \label{s5}
\end{figure}

\begin{figure}
\centering
\includegraphics[width=1\textwidth]{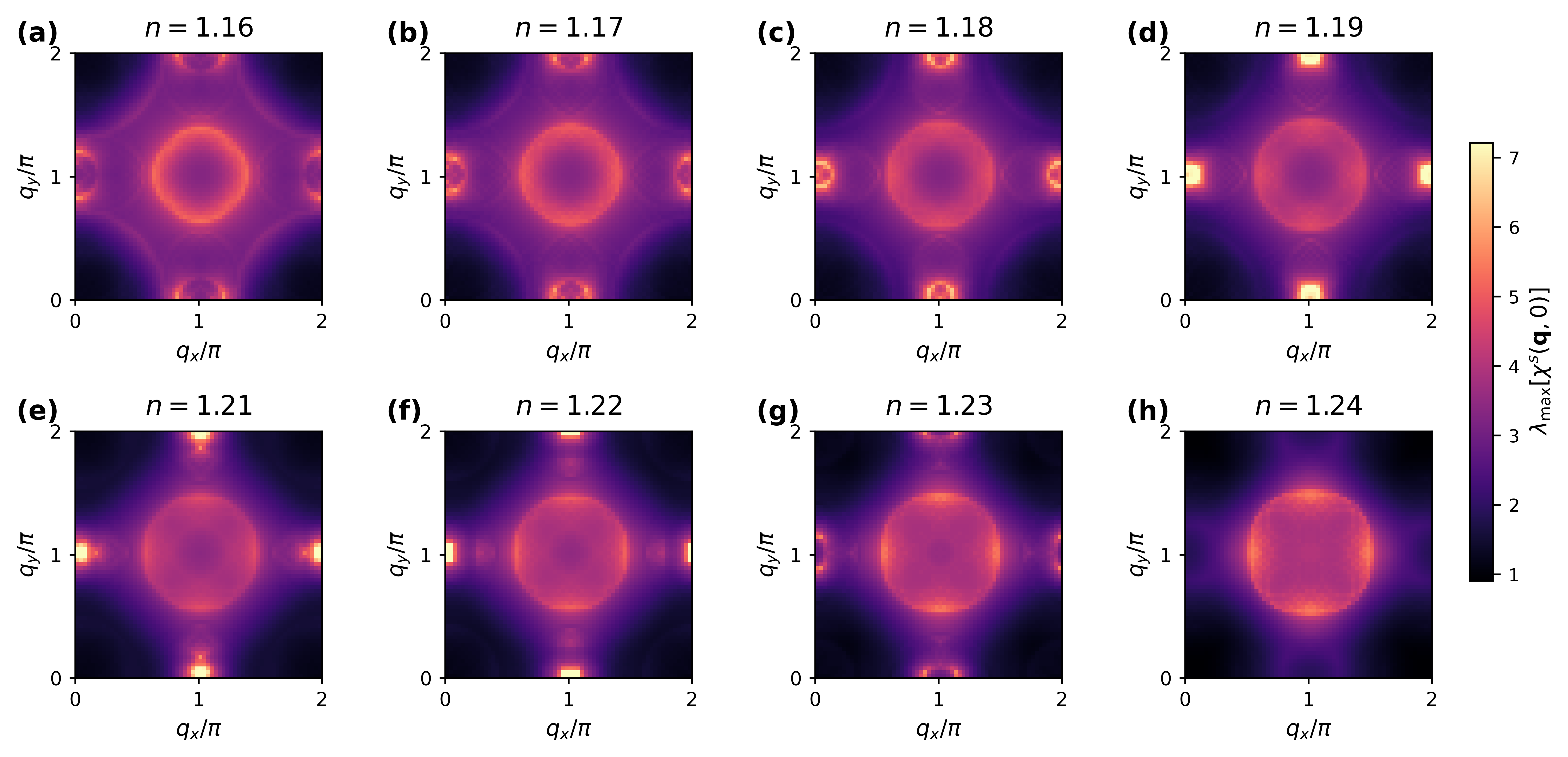}
    \caption{Filling dependence of the spin susceptibility in FeTe.
Momentum-resolved static spin susceptibility for FeTe at U=1.5 and T=0.001, shown for fillings (a) n=1.16, (b) n=1.17, (c) n=1.18, (d) n=1.19, (e) n=1.21, (f) n=1.22, (g) n=1.23, and (h) n=1.24.
The plotted quantity is the largest eigenvalue of the orbital-space spin-susceptibility matrix, $\lambda_{\max}[\chi^s(\mathbf{q},0)]$. All panels use the same color scale, with the upper limit for visualization.
}
    \label{s6}
\end{figure}

\newpage

\subsection{E. Gap function in orbital basis}

Fig.~\ref{s7} and Fig.~\ref{s8} show the orbital-resolved structure of the leading superconducting gap function. Figure S6 corresponds to the leading extended $s$-wave gap function near stoichiometric filling, while Fig. S7 corresponds to the leading $d$-wave gap function at the electron-doped filling $n=1.24$. In both figures, the 25 matrix components of $\mathrm{Re}\,\Delta_{ab}(\mathbf{k},i\pi T)$ are plotted in the orbital basis.
The off-diagonal orbital components of the gap function are generally very small, with the dominant weight concentrated in the diagonal components $\Delta_{aa}$. This indicates that, within the present parameter regime, superconducting pairing occurs predominantly within the same orbital channel. Moreover, the leading gap components mainly reside in the $d_{xz}$, $d_{yz}$, and $d_{xy}$ orbitals, which are also the orbitals showing strong correlation-induced renormalization in Fig. S3 and providing the dominant low-energy spectral weight and spin fluctuations. The orbital structure of the gap function is therefore consistent with the orbital-selective correlations and the spin-fluctuation origin of pairing discussed above.
In momentum space, the extended $s$-wave state in Fig.~\ref{s7} exhibits sign changes between different low-energy spectral-weight regions while preserving the overall $A_{1g}$ symmetry. By contrast, the $d$-wave state in Fig.~\ref{s8} displays a sign structure consistent with \(B_{1g}\) pairing. It should be emphasized that the local appearance of an individual orbital component in the orbital basis should not be directly identified with the conventional $s$-, $p$-, or $d$-wave classification. The pairing symmetry is determined by the transformation property of the full gap function under the crystal symmetry operations, especially after projection onto the low-energy bands and Fermi-surface sheets. In this sense, Fig.~\ref{s7} and Fig.~\ref{s8} do not independently define the pairing symmetry, but rather provide orbital-resolved support for the extended $s$-wave and $d$-wave pairing states identified in the main text.

\begin{figure}
\centering
\includegraphics[width=1\textwidth]{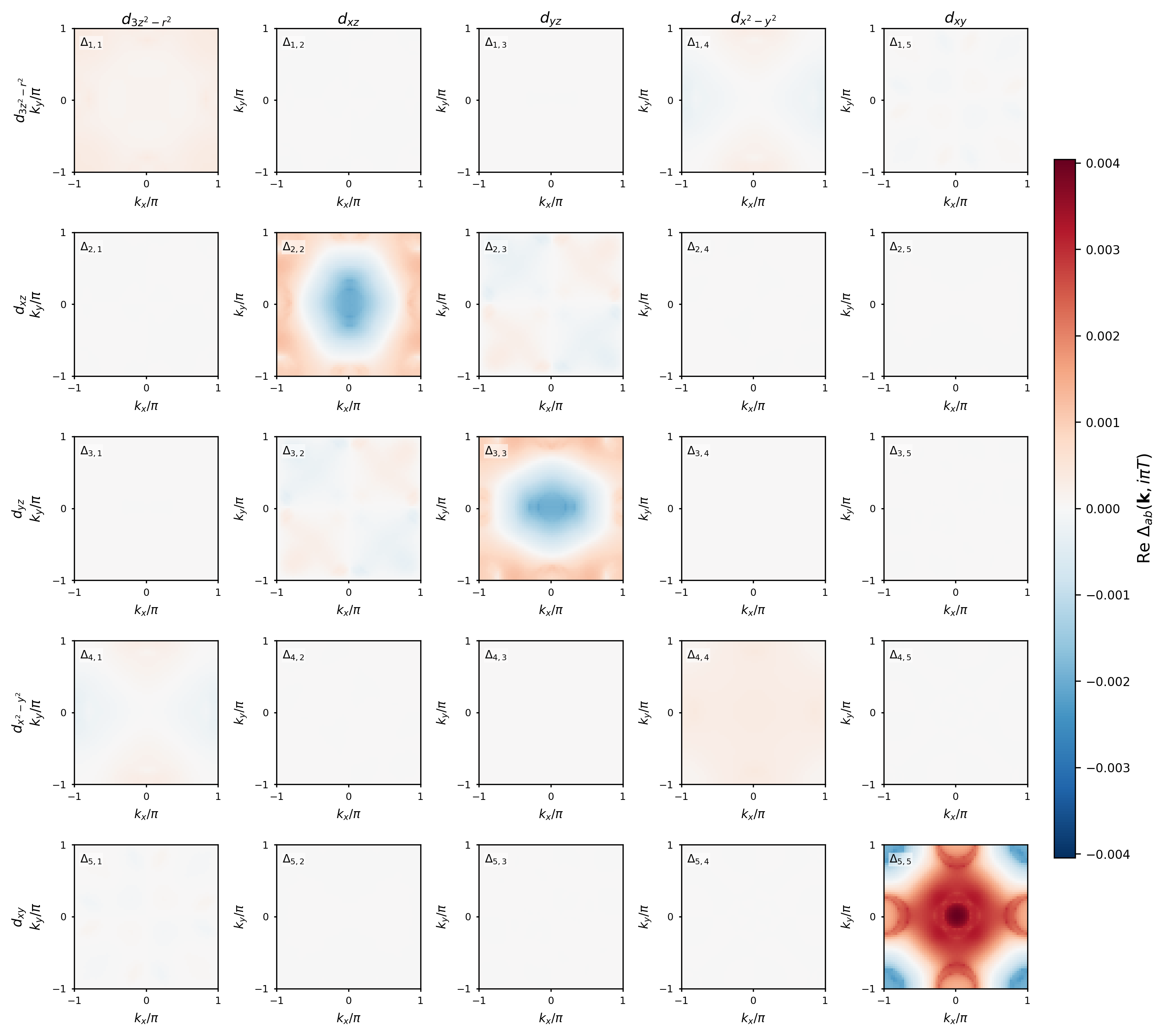}
    \caption{Orbital components of the leading superconducting gap function obtained from the FLEX pairing calculation at filling n=1.20. The momentum dependence is shown in the full Brillouin zone, with color indicating the relative amplitude and sign of the gap on each orbital component. All panels are plotted using the same color scale to facilitate comparison between orbital channels.}
    \label{s7}
\end{figure}

\begin{figure}
\centering
\includegraphics[width=1\textwidth]{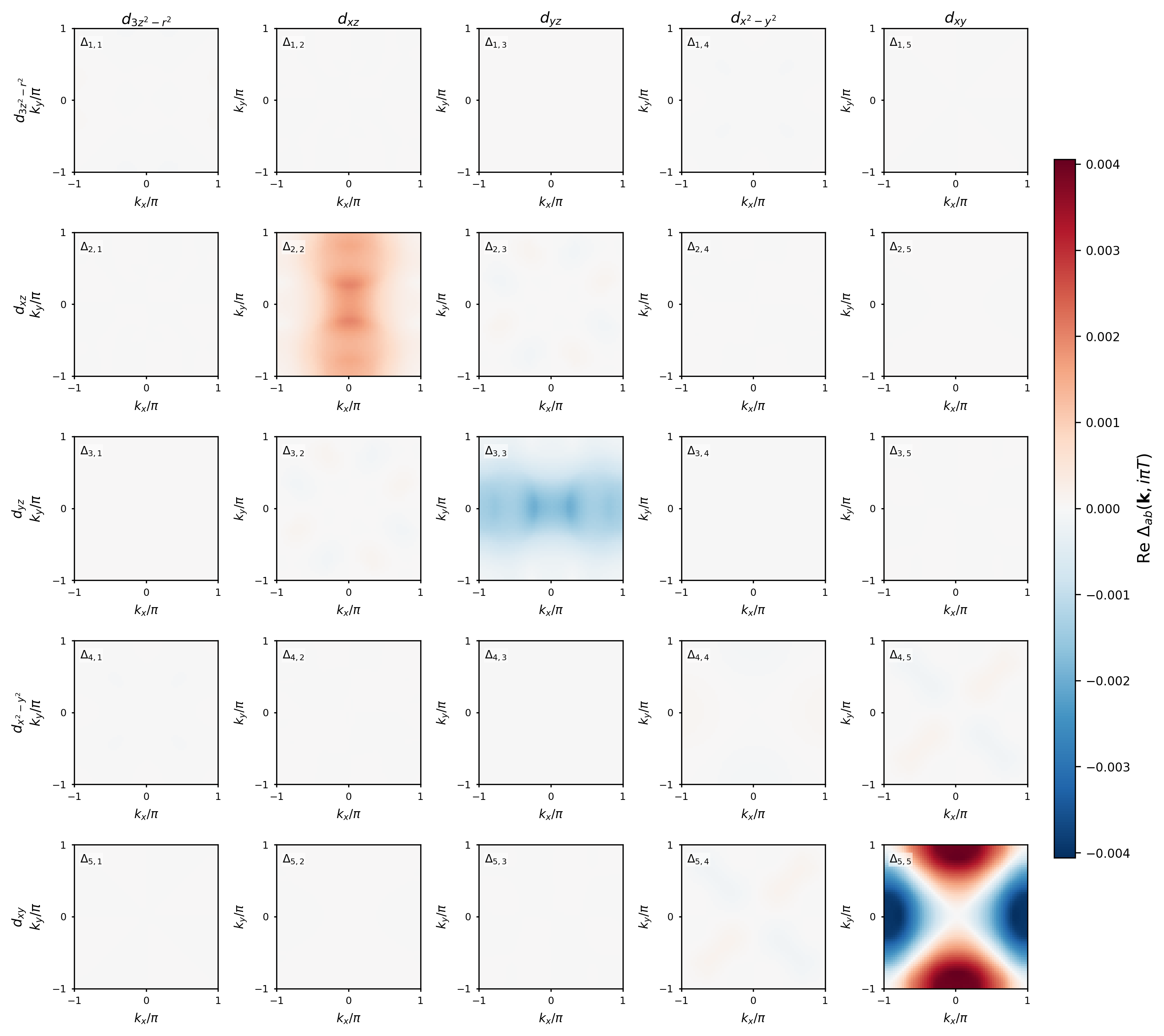}
    \caption{Orbital components of the leading superconducting gap function obtained from the FLEX pairing calculation at filling n=1.24. The momentum dependence is shown in the full Brillouin zone, with color indicating the relative amplitude and sign of the gap on each orbital component. All panels are plotted using the same color scale to facilitate comparison between orbital channels.}
    \label{s8}
\end{figure}

\end{widetext}


\begin{thebibliography}{75}%
\makeatletter
\providecommand \@ifxundefined [1]{%
 \@ifx{#1\undefined}
}%
\providecommand \@ifnum [1]{%
 \ifnum #1\expandafter \@firstoftwo
 \else \expandafter \@secondoftwo
 \fi
}%
\providecommand \@ifx [1]{%
 \ifx #1\expandafter \@firstoftwo
 \else \expandafter \@secondoftwo
 \fi
}%
\providecommand \natexlab [1]{#1}%
\providecommand \enquote  [1]{``#1''}%
\providecommand \bibnamefont  [1]{#1}%
\providecommand \bibfnamefont [1]{#1}%
\providecommand \citenamefont [1]{#1}%
\providecommand \href@noop [0]{\@secondoftwo}%
\providecommand \href [0]{\begingroup \@sanitize@url \@href}%
\providecommand \@href[1]{\@@startlink{#1}\@@href}%
\providecommand \@@href[1]{\endgroup#1\@@endlink}%
\providecommand \@sanitize@url [0]{\catcode `\\12\catcode `\$12\catcode
  `\&12\catcode `\#12\catcode `\^12\catcode `\_12\catcode `\%12\relax}%
\providecommand \@@startlink[1]{}%
\providecommand \@@endlink[0]{}%
\providecommand \url  [0]{\begingroup\@sanitize@url \@url }%
\providecommand \@url [1]{\endgroup\@href {#1}{\urlprefix }}%
\providecommand \urlprefix  [0]{URL }%
\providecommand \Eprint [0]{\href }%
\providecommand \doibase [0]{https://doi.org/}%
\providecommand \selectlanguage [0]{\@gobble}%
\providecommand \bibinfo  [0]{\@secondoftwo}%
\providecommand \bibfield  [0]{\@secondoftwo}%
\providecommand \translation [1]{[#1]}%
\providecommand \BibitemOpen [0]{}%
\providecommand \bibitemStop [0]{}%
\providecommand \bibitemNoStop [0]{.\EOS\space}%
\providecommand \EOS [0]{\spacefactor3000\relax}%
\providecommand \BibitemShut  [1]{\csname bibitem#1\endcsname}%
\let\auto@bib@innerbib\@empty
\bibitem [{\citenamefont {Kamihara}\ \emph {et~al.}(2008)\citenamefont
  {Kamihara}, \citenamefont {Watanabe}, \citenamefont {Hirano},\ and\
  \citenamefont {Hosono}}]{Kamihara2008}%
  \BibitemOpen
  \bibfield  {author} {\bibinfo {author} {\bibfnamefont {Y.}~\bibnamefont
  {Kamihara}}, \bibinfo {author} {\bibfnamefont {T.}~\bibnamefont {Watanabe}},
  \bibinfo {author} {\bibfnamefont {M.}~\bibnamefont {Hirano}},\ and\ \bibinfo
  {author} {\bibfnamefont {H.}~\bibnamefont {Hosono}},\ }\bibfield  {title}
  {\bibinfo {title} {Iron-based layered superconductor {La[O$_{1-x}$F$_x$]FeAs}
  ($x$ = 0.05--0.12) with {$T_c$} = 26 k},\ }\href
  {https://doi.org/10.1021/ja800073m} {\bibfield  {journal} {\bibinfo
  {journal} {Journal of the American Chemical Society}\ }\textbf {\bibinfo
  {volume} {130}},\ \bibinfo {pages} {3296} (\bibinfo {year}
  {2008})}\BibitemShut {NoStop}%
\bibitem [{\citenamefont {Chen}\ \emph {et~al.}(2008)\citenamefont {Chen},
  \citenamefont {Wu}, \citenamefont {Wu}, \citenamefont {Liu}, \citenamefont
  {Chen},\ and\ \citenamefont {Fang}}]{Chen2008SmFeAsO}%
  \BibitemOpen
  \bibfield  {author} {\bibinfo {author} {\bibfnamefont {X.~H.}\ \bibnamefont
  {Chen}}, \bibinfo {author} {\bibfnamefont {T.}~\bibnamefont {Wu}}, \bibinfo
  {author} {\bibfnamefont {G.}~\bibnamefont {Wu}}, \bibinfo {author}
  {\bibfnamefont {R.~H.}\ \bibnamefont {Liu}}, \bibinfo {author} {\bibfnamefont
  {H.}~\bibnamefont {Chen}},\ and\ \bibinfo {author} {\bibfnamefont {D.~F.}\
  \bibnamefont {Fang}},\ }\bibfield  {title} {\bibinfo {title}
  {Superconductivity at 43 k in {SmFeAsO$_{1-x}$F$_x$}},\ }\href
  {https://doi.org/10.1038/nature07045} {\bibfield  {journal} {\bibinfo
  {journal} {Nature}\ }\textbf {\bibinfo {volume} {453}},\ \bibinfo {pages}
  {761} (\bibinfo {year} {2008})}\BibitemShut {NoStop}%
\bibitem [{\citenamefont {Ren}\ \emph {et~al.}(2008)\citenamefont {Ren},
  \citenamefont {Lu}, \citenamefont {Yang}, \citenamefont {Yi}, \citenamefont
  {Shen}, \citenamefont {Li}, \citenamefont {Che}, \citenamefont {Dong},
  \citenamefont {Sun}, \citenamefont {Zhou},\ and\ \citenamefont
  {Zhao}}]{Ren2008SmFeAsO}%
  \BibitemOpen
  \bibfield  {author} {\bibinfo {author} {\bibfnamefont {Z.-A.}\ \bibnamefont
  {Ren}}, \bibinfo {author} {\bibfnamefont {W.}~\bibnamefont {Lu}}, \bibinfo
  {author} {\bibfnamefont {J.}~\bibnamefont {Yang}}, \bibinfo {author}
  {\bibfnamefont {W.}~\bibnamefont {Yi}}, \bibinfo {author} {\bibfnamefont
  {X.-L.}\ \bibnamefont {Shen}}, \bibinfo {author} {\bibfnamefont {Z.-C.}\
  \bibnamefont {Li}}, \bibinfo {author} {\bibfnamefont {G.-C.}\ \bibnamefont
  {Che}}, \bibinfo {author} {\bibfnamefont {X.-L.}\ \bibnamefont {Dong}},
  \bibinfo {author} {\bibfnamefont {L.-L.}\ \bibnamefont {Sun}}, \bibinfo
  {author} {\bibfnamefont {F.}~\bibnamefont {Zhou}},\ and\ \bibinfo {author}
  {\bibfnamefont {Z.-X.}\ \bibnamefont {Zhao}},\ }\bibfield  {title} {\bibinfo
  {title} {Superconductivity at 55 k in iron-based f-doped layered quaternary
  compound {Sm[O$_{1-x}$F$_x$]FeAs}},\ }\href
  {https://doi.org/10.1088/0256-307X/25/6/080} {\bibfield  {journal} {\bibinfo
  {journal} {Chinese Physics Letters}\ }\textbf {\bibinfo {volume} {25}},\
  \bibinfo {pages} {2215} (\bibinfo {year} {2008})}\BibitemShut {NoStop}%
\bibitem [{\citenamefont {Stewart}(2011)}]{Stewart2011}%
  \BibitemOpen
  \bibfield  {author} {\bibinfo {author} {\bibfnamefont {G.~R.}\ \bibnamefont
  {Stewart}},\ }\bibfield  {title} {\bibinfo {title} {Superconductivity in iron
  compounds},\ }\href {https://doi.org/10.1103/RevModPhys.83.1589} {\bibfield
  {journal} {\bibinfo  {journal} {Reviews of Modern Physics}\ }\textbf
  {\bibinfo {volume} {83}},\ \bibinfo {pages} {1589} (\bibinfo {year}
  {2011})}\BibitemShut {NoStop}%
\bibitem [{\citenamefont {Paglione}\ and\ \citenamefont
  {Greene}(2010)}]{Paglione2010}%
  \BibitemOpen
  \bibfield  {author} {\bibinfo {author} {\bibfnamefont {J.}~\bibnamefont
  {Paglione}}\ and\ \bibinfo {author} {\bibfnamefont {R.~L.}\ \bibnamefont
  {Greene}},\ }\bibfield  {title} {\bibinfo {title} {High-temperature
  superconductivity in iron-based materials},\ }\href
  {https://doi.org/10.1038/nphys1759} {\bibfield  {journal} {\bibinfo
  {journal} {Nature Physics}\ }\textbf {\bibinfo {volume} {6}},\ \bibinfo
  {pages} {645} (\bibinfo {year} {2010})}\BibitemShut {NoStop}%
\bibitem [{\citenamefont {Johnston}(2010)}]{Johnston2010}%
  \BibitemOpen
  \bibfield  {author} {\bibinfo {author} {\bibfnamefont {D.~C.}\ \bibnamefont
  {Johnston}},\ }\bibfield  {title} {\bibinfo {title} {The puzzle of high
  temperature superconductivity in layered iron pnictides and chalcogenides},\
  }\href {https://doi.org/10.1080/00018732.2010.513480} {\bibfield  {journal}
  {\bibinfo  {journal} {Advances in Physics}\ }\textbf {\bibinfo {volume}
  {59}},\ \bibinfo {pages} {803} (\bibinfo {year} {2010})}\BibitemShut
  {NoStop}%
\bibitem [{\citenamefont {Fernandes}\ \emph {et~al.}(2022)\citenamefont
  {Fernandes}, \citenamefont {Coldea}, \citenamefont {Ding}, \citenamefont
  {Fisher}, \citenamefont {Hirschfeld},\ and\ \citenamefont
  {Kotliar}}]{Fernandes2022}%
  \BibitemOpen
  \bibfield  {author} {\bibinfo {author} {\bibfnamefont {R.~M.}\ \bibnamefont
  {Fernandes}}, \bibinfo {author} {\bibfnamefont {A.~I.}\ \bibnamefont
  {Coldea}}, \bibinfo {author} {\bibfnamefont {H.}~\bibnamefont {Ding}},
  \bibinfo {author} {\bibfnamefont {I.~R.}\ \bibnamefont {Fisher}}, \bibinfo
  {author} {\bibfnamefont {P.~J.}\ \bibnamefont {Hirschfeld}},\ and\ \bibinfo
  {author} {\bibfnamefont {G.}~\bibnamefont {Kotliar}},\ }\bibfield  {title}
  {\bibinfo {title} {Iron pnictides and chalcogenides: a new paradigm for
  superconductivity},\ }\href {https://doi.org/10.1038/s41586-021-04073-2}
  {\bibfield  {journal} {\bibinfo  {journal} {Nature}\ }\textbf {\bibinfo
  {volume} {601}},\ \bibinfo {pages} {35} (\bibinfo {year} {2022})}\BibitemShut
  {NoStop}%
\bibitem [{\citenamefont {Dai}(2015)}]{Dai2015}%
  \BibitemOpen
  \bibfield  {author} {\bibinfo {author} {\bibfnamefont {P.}~\bibnamefont
  {Dai}},\ }\bibfield  {title} {\bibinfo {title} {Antiferromagnetic order and
  spin dynamics in iron-based superconductors},\ }\href
  {https://doi.org/10.1103/RevModPhys.87.855} {\bibfield  {journal} {\bibinfo
  {journal} {Reviews of Modern Physics}\ }\textbf {\bibinfo {volume} {87}},\
  \bibinfo {pages} {855} (\bibinfo {year} {2015})}\BibitemShut {NoStop}%
\bibitem [{\citenamefont {Hosono}\ and\ \citenamefont
  {Kuroki}(2015)}]{Hosono2015KurokiReview}%
  \BibitemOpen
  \bibfield  {author} {\bibinfo {author} {\bibfnamefont {H.}~\bibnamefont
  {Hosono}}\ and\ \bibinfo {author} {\bibfnamefont {K.}~\bibnamefont
  {Kuroki}},\ }\bibfield  {title} {\bibinfo {title} {Iron-based
  superconductors: Current status of materials and pairing mechanism},\ }\href
  {https://doi.org/10.1016/j.physc.2015.02.020} {\bibfield  {journal} {\bibinfo
   {journal} {Physica C: Superconductivity and its Applications}\ }\textbf
  {\bibinfo {volume} {514}},\ \bibinfo {pages} {399} (\bibinfo {year}
  {2015})}\BibitemShut {NoStop}%
\bibitem [{\citenamefont {Kuroki}\ \emph {et~al.}(2008)\citenamefont {Kuroki},
  \citenamefont {Onari}, \citenamefont {Arita}, \citenamefont {Usui},
  \citenamefont {Tanaka}, \citenamefont {Kontani},\ and\ \citenamefont
  {Aoki}}]{Kuroki2008DisconnectedFS}%
  \BibitemOpen
  \bibfield  {author} {\bibinfo {author} {\bibfnamefont {K.}~\bibnamefont
  {Kuroki}}, \bibinfo {author} {\bibfnamefont {S.}~\bibnamefont {Onari}},
  \bibinfo {author} {\bibfnamefont {R.}~\bibnamefont {Arita}}, \bibinfo
  {author} {\bibfnamefont {H.}~\bibnamefont {Usui}}, \bibinfo {author}
  {\bibfnamefont {Y.}~\bibnamefont {Tanaka}}, \bibinfo {author} {\bibfnamefont
  {H.}~\bibnamefont {Kontani}},\ and\ \bibinfo {author} {\bibfnamefont
  {H.}~\bibnamefont {Aoki}},\ }\bibfield  {title} {\bibinfo {title}
  {Unconventional pairing originating from the disconnected fermi surfaces of
  superconducting {LaFeAsO$_{1-x}$F$_x$}},\ }\href
  {https://doi.org/10.1103/PhysRevLett.101.087004} {\bibfield  {journal}
  {\bibinfo  {journal} {Physical Review Letters}\ }\textbf {\bibinfo {volume}
  {101}},\ \bibinfo {pages} {087004} (\bibinfo {year} {2008})}\BibitemShut
  {NoStop}%
\bibitem [{\citenamefont {Rotter}\ \emph {et~al.}(2008)\citenamefont {Rotter},
  \citenamefont {Tegel},\ and\ \citenamefont {Johrendt}}]{Rotter2008}%
  \BibitemOpen
  \bibfield  {author} {\bibinfo {author} {\bibfnamefont {M.}~\bibnamefont
  {Rotter}}, \bibinfo {author} {\bibfnamefont {M.}~\bibnamefont {Tegel}},\ and\
  \bibinfo {author} {\bibfnamefont {D.}~\bibnamefont {Johrendt}},\ }\bibfield
  {title} {\bibinfo {title} {Superconductivity at 38 k in the iron arsenide
  {($Ba_{1-x}K_x$)Fe$_2$As$_2$}},\ }\href
  {https://doi.org/10.1103/PhysRevLett.101.107006} {\bibfield  {journal}
  {\bibinfo  {journal} {Physical Review Letters}\ }\textbf {\bibinfo {volume}
  {101}},\ \bibinfo {pages} {107006} (\bibinfo {year} {2008})}\BibitemShut
  {NoStop}%
\bibitem [{\citenamefont {Sefat}\ \emph {et~al.}(2008)\citenamefont {Sefat},
  \citenamefont {Jin}, \citenamefont {McGuire}, \citenamefont {Sales},
  \citenamefont {Singh},\ and\ \citenamefont {Mandrus}}]{Sefat2008}%
  \BibitemOpen
  \bibfield  {author} {\bibinfo {author} {\bibfnamefont {A.~S.}\ \bibnamefont
  {Sefat}}, \bibinfo {author} {\bibfnamefont {R.}~\bibnamefont {Jin}}, \bibinfo
  {author} {\bibfnamefont {M.~A.}\ \bibnamefont {McGuire}}, \bibinfo {author}
  {\bibfnamefont {B.~C.}\ \bibnamefont {Sales}}, \bibinfo {author}
  {\bibfnamefont {D.~J.}\ \bibnamefont {Singh}},\ and\ \bibinfo {author}
  {\bibfnamefont {D.}~\bibnamefont {Mandrus}},\ }\bibfield  {title} {\bibinfo
  {title} {Superconductivity at 22 k in co-doped {BaFe$_2$As$_2$} crystals},\
  }\href {https://doi.org/10.1103/PhysRevLett.101.117004} {\bibfield  {journal}
  {\bibinfo  {journal} {Physical Review Letters}\ }\textbf {\bibinfo {volume}
  {101}},\ \bibinfo {pages} {117004} (\bibinfo {year} {2008})}\BibitemShut
  {NoStop}%
\bibitem [{\citenamefont {Canfield}\ and\ \citenamefont
  {Bud'ko}(2010)}]{Canfield2010}%
  \BibitemOpen
  \bibfield  {author} {\bibinfo {author} {\bibfnamefont {P.~C.}\ \bibnamefont
  {Canfield}}\ and\ \bibinfo {author} {\bibfnamefont {S.~L.}\ \bibnamefont
  {Bud'ko}},\ }\bibfield  {title} {\bibinfo {title} {{FeAs}-based
  superconductivity: A case study of the effects of transition metal doping on
  {BaFe$_2$As$_2$}},\ }\href
  {https://doi.org/10.1146/annurev-conmatphys-070909-104041} {\bibfield
  {journal} {\bibinfo  {journal} {Annual Review of Condensed Matter Physics}\
  }\textbf {\bibinfo {volume} {1}},\ \bibinfo {pages} {27} (\bibinfo {year}
  {2010})}\BibitemShut {NoStop}%
\bibitem [{\citenamefont {Hsu}\ \emph {et~al.}(2008)\citenamefont {Hsu},
  \citenamefont {Luo}, \citenamefont {Yeh}, \citenamefont {Chen}, \citenamefont
  {Huang}, \citenamefont {Wu}, \citenamefont {Lee}, \citenamefont {Huang},
  \citenamefont {Chu}, \citenamefont {Yan},\ and\ \citenamefont
  {Wu}}]{Hsu2008}%
  \BibitemOpen
  \bibfield  {author} {\bibinfo {author} {\bibfnamefont {F.-C.}\ \bibnamefont
  {Hsu}}, \bibinfo {author} {\bibfnamefont {J.-Y.}\ \bibnamefont {Luo}},
  \bibinfo {author} {\bibfnamefont {K.-W.}\ \bibnamefont {Yeh}}, \bibinfo
  {author} {\bibfnamefont {T.-K.}\ \bibnamefont {Chen}}, \bibinfo {author}
  {\bibfnamefont {T.-W.}\ \bibnamefont {Huang}}, \bibinfo {author}
  {\bibfnamefont {P.~M.}\ \bibnamefont {Wu}}, \bibinfo {author} {\bibfnamefont
  {Y.-C.}\ \bibnamefont {Lee}}, \bibinfo {author} {\bibfnamefont {Y.-L.}\
  \bibnamefont {Huang}}, \bibinfo {author} {\bibfnamefont {Y.-Y.}\ \bibnamefont
  {Chu}}, \bibinfo {author} {\bibfnamefont {D.-C.}\ \bibnamefont {Yan}},\ and\
  \bibinfo {author} {\bibfnamefont {M.-K.}\ \bibnamefont {Wu}},\ }\bibfield
  {title} {\bibinfo {title} {Superconductivity in the {PbO}-type structure
  {$\alpha$-FeSe}},\ }\href {https://doi.org/10.1073/pnas.0807325105}
  {\bibfield  {journal} {\bibinfo  {journal} {Proceedings of the National
  Academy of Sciences}\ }\textbf {\bibinfo {volume} {105}},\ \bibinfo {pages}
  {14262} (\bibinfo {year} {2008})}\BibitemShut {NoStop}%
\bibitem [{\citenamefont {Fang}\ \emph {et~al.}(2008)\citenamefont {Fang},
  \citenamefont {Pham}, \citenamefont {Qian}, \citenamefont {Liu},
  \citenamefont {Vehstedt}, \citenamefont {Liu}, \citenamefont {Spinu},\ and\
  \citenamefont {Mao}}]{Fang2008}%
  \BibitemOpen
  \bibfield  {author} {\bibinfo {author} {\bibfnamefont {M.~H.}\ \bibnamefont
  {Fang}}, \bibinfo {author} {\bibfnamefont {H.~M.}\ \bibnamefont {Pham}},
  \bibinfo {author} {\bibfnamefont {B.}~\bibnamefont {Qian}}, \bibinfo {author}
  {\bibfnamefont {T.~J.}\ \bibnamefont {Liu}}, \bibinfo {author} {\bibfnamefont
  {E.~K.}\ \bibnamefont {Vehstedt}}, \bibinfo {author} {\bibfnamefont
  {Y.}~\bibnamefont {Liu}}, \bibinfo {author} {\bibfnamefont {L.}~\bibnamefont
  {Spinu}},\ and\ \bibinfo {author} {\bibfnamefont {Z.~Q.}\ \bibnamefont
  {Mao}},\ }\bibfield  {title} {\bibinfo {title} {Superconductivity close to
  magnetic instability in {Fe(Se$_{1-x}$Te$_x$)$_{0.82}$}},\ }\href
  {https://doi.org/10.1103/PhysRevB.78.224503} {\bibfield  {journal} {\bibinfo
  {journal} {Physical Review B}\ }\textbf {\bibinfo {volume} {78}},\ \bibinfo
  {pages} {224503} (\bibinfo {year} {2008})}\BibitemShut {NoStop}%
\bibitem [{\citenamefont {Wang}\ \emph {et~al.}(2012)\citenamefont {Wang},
  \citenamefont {Li}, \citenamefont {Zhang}, \citenamefont {Zhang},
  \citenamefont {Zhang}, \citenamefont {Li}, \citenamefont {Ding},
  \citenamefont {Ou}, \citenamefont {Deng}, \citenamefont {Chang},
  \citenamefont {Wen}, \citenamefont {Song}, \citenamefont {He}, \citenamefont
  {Jia}, \citenamefont {Ji}, \citenamefont {Wang}, \citenamefont {Wang},
  \citenamefont {Chen}, \citenamefont {Ma},\ and\ \citenamefont
  {Xue}}]{Wang2012FeSeSTO}%
  \BibitemOpen
  \bibfield  {author} {\bibinfo {author} {\bibfnamefont {Q.-Y.}\ \bibnamefont
  {Wang}}, \bibinfo {author} {\bibfnamefont {Z.}~\bibnamefont {Li}}, \bibinfo
  {author} {\bibfnamefont {W.-H.}\ \bibnamefont {Zhang}}, \bibinfo {author}
  {\bibfnamefont {Z.-C.}\ \bibnamefont {Zhang}}, \bibinfo {author}
  {\bibfnamefont {J.-S.}\ \bibnamefont {Zhang}}, \bibinfo {author}
  {\bibfnamefont {W.}~\bibnamefont {Li}}, \bibinfo {author} {\bibfnamefont
  {H.}~\bibnamefont {Ding}}, \bibinfo {author} {\bibfnamefont {Y.-B.}\
  \bibnamefont {Ou}}, \bibinfo {author} {\bibfnamefont {P.}~\bibnamefont
  {Deng}}, \bibinfo {author} {\bibfnamefont {K.}~\bibnamefont {Chang}},
  \bibinfo {author} {\bibfnamefont {J.}~\bibnamefont {Wen}}, \bibinfo {author}
  {\bibfnamefont {C.-L.}\ \bibnamefont {Song}}, \bibinfo {author}
  {\bibfnamefont {K.}~\bibnamefont {He}}, \bibinfo {author} {\bibfnamefont
  {J.-F.}\ \bibnamefont {Jia}}, \bibinfo {author} {\bibfnamefont {S.-H.}\
  \bibnamefont {Ji}}, \bibinfo {author} {\bibfnamefont {Y.-Y.}\ \bibnamefont
  {Wang}}, \bibinfo {author} {\bibfnamefont {L.-L.}\ \bibnamefont {Wang}},
  \bibinfo {author} {\bibfnamefont {X.}~\bibnamefont {Chen}}, \bibinfo {author}
  {\bibfnamefont {X.-C.}\ \bibnamefont {Ma}},\ and\ \bibinfo {author}
  {\bibfnamefont {Q.-K.}\ \bibnamefont {Xue}},\ }\bibfield  {title} {\bibinfo
  {title} {Interface-induced high-temperature superconductivity in single
  unit-cell {FeSe} films on {SrTiO$_3$}},\ }\href
  {https://doi.org/10.1088/0256-307X/29/3/037402} {\bibfield  {journal}
  {\bibinfo  {journal} {Chinese Physics Letters}\ }\textbf {\bibinfo {volume}
  {29}},\ \bibinfo {pages} {037402} (\bibinfo {year} {2012})}\BibitemShut
  {NoStop}%
\bibitem [{\citenamefont {Kreisel}\ \emph {et~al.}(2020)\citenamefont
  {Kreisel}, \citenamefont {Hirschfeld},\ and\ \citenamefont
  {Andersen}}]{Kreisel2020}%
  \BibitemOpen
  \bibfield  {author} {\bibinfo {author} {\bibfnamefont {A.}~\bibnamefont
  {Kreisel}}, \bibinfo {author} {\bibfnamefont {P.}~\bibnamefont
  {Hirschfeld}},\ and\ \bibinfo {author} {\bibfnamefont {B.}~\bibnamefont
  {Andersen}},\ }\bibfield  {title} {\bibinfo {title} {On the remarkable
  superconductivity of {FeSe} and its close cousins},\ }\href
  {https://doi.org/10.3390/sym12091402} {\bibfield  {journal} {\bibinfo
  {journal} {Symmetry}\ }\textbf {\bibinfo {volume} {12}},\ \bibinfo {pages}
  {1402} (\bibinfo {year} {2020})}\BibitemShut {NoStop}%
\bibitem [{\citenamefont {Coldea}\ and\ \citenamefont
  {Watson}(2018)}]{Coldea2018}%
  \BibitemOpen
  \bibfield  {author} {\bibinfo {author} {\bibfnamefont {A.~I.}\ \bibnamefont
  {Coldea}}\ and\ \bibinfo {author} {\bibfnamefont {M.~D.}\ \bibnamefont
  {Watson}},\ }\bibfield  {title} {\bibinfo {title} {The key ingredients of the
  electronic structure of {FeSe}},\ }\href
  {https://doi.org/10.1146/annurev-conmatphys-033117-054137} {\bibfield
  {journal} {\bibinfo  {journal} {Annual Review of Condensed Matter Physics}\
  }\textbf {\bibinfo {volume} {9}},\ \bibinfo {pages} {125} (\bibinfo {year}
  {2018})}\BibitemShut {NoStop}%
\bibitem [{\citenamefont {B{\"o}hmer}\ and\ \citenamefont
  {Kreisel}(2018)}]{Bohmer2018}%
  \BibitemOpen
  \bibfield  {author} {\bibinfo {author} {\bibfnamefont {A.~E.}\ \bibnamefont
  {B{\"o}hmer}}\ and\ \bibinfo {author} {\bibfnamefont {A.}~\bibnamefont
  {Kreisel}},\ }\bibfield  {title} {\bibinfo {title} {Nematicity, magnetism and
  superconductivity in {FeSe}},\ }\href
  {https://doi.org/10.1088/1361-648X/aa9caa} {\bibfield  {journal} {\bibinfo
  {journal} {Journal of Physics: Condensed Matter}\ }\textbf {\bibinfo {volume}
  {30}},\ \bibinfo {pages} {023001} (\bibinfo {year} {2018})}\BibitemShut
  {NoStop}%
\bibitem [{\citenamefont {Shibauchi}\ \emph {et~al.}(2020)\citenamefont
  {Shibauchi}, \citenamefont {Hanaguri},\ and\ \citenamefont
  {Matsuda}}]{Shibauchi2020}%
  \BibitemOpen
  \bibfield  {author} {\bibinfo {author} {\bibfnamefont {T.}~\bibnamefont
  {Shibauchi}}, \bibinfo {author} {\bibfnamefont {T.}~\bibnamefont
  {Hanaguri}},\ and\ \bibinfo {author} {\bibfnamefont {Y.}~\bibnamefont
  {Matsuda}},\ }\bibfield  {title} {\bibinfo {title} {Exotic superconducting
  states in {FeSe}-based materials},\ }\href
  {https://doi.org/10.7566/JPSJ.89.102002} {\bibfield  {journal} {\bibinfo
  {journal} {Journal of the Physical Society of Japan}\ }\textbf {\bibinfo
  {volume} {89}},\ \bibinfo {pages} {102002} (\bibinfo {year}
  {2020})}\BibitemShut {NoStop}%
\bibitem [{\citenamefont {McQueen}\ \emph
  {et~al.}(2009{\natexlab{a}})\citenamefont {McQueen}, \citenamefont
  {Williams}, \citenamefont {Stephens}, \citenamefont {Tao}, \citenamefont
  {Zhu}, \citenamefont {Ksenofontov}, \citenamefont {Casper}, \citenamefont
  {Felser},\ and\ \citenamefont {Cava}}]{McQueen2009}%
  \BibitemOpen
  \bibfield  {author} {\bibinfo {author} {\bibfnamefont {T.~M.}\ \bibnamefont
  {McQueen}}, \bibinfo {author} {\bibfnamefont {A.~J.}\ \bibnamefont
  {Williams}}, \bibinfo {author} {\bibfnamefont {P.~W.}\ \bibnamefont
  {Stephens}}, \bibinfo {author} {\bibfnamefont {J.}~\bibnamefont {Tao}},
  \bibinfo {author} {\bibfnamefont {Y.}~\bibnamefont {Zhu}}, \bibinfo {author}
  {\bibfnamefont {V.}~\bibnamefont {Ksenofontov}}, \bibinfo {author}
  {\bibfnamefont {F.}~\bibnamefont {Casper}}, \bibinfo {author} {\bibfnamefont
  {C.}~\bibnamefont {Felser}},\ and\ \bibinfo {author} {\bibfnamefont {R.~J.}\
  \bibnamefont {Cava}},\ }\bibfield  {title} {\bibinfo {title}
  {Tetragonal-to-orthorhombic structural phase transition at 90 k in the
  superconductor {Fe$_{1.01}$Se}},\ }\href
  {https://doi.org/10.1103/PhysRevLett.103.057002} {\bibfield  {journal}
  {\bibinfo  {journal} {Physical Review Letters}\ }\textbf {\bibinfo {volume}
  {103}},\ \bibinfo {pages} {057002} (\bibinfo {year}
  {2009}{\natexlab{a}})}\BibitemShut {NoStop}%
\bibitem [{\citenamefont {Shimojima}\ \emph {et~al.}(2014)\citenamefont
  {Shimojima}, \citenamefont {Suzuki}, \citenamefont {Sonobe}, \citenamefont
  {Nakamura}, \citenamefont {Sakano}, \citenamefont {Omachi}, \citenamefont
  {Yoshioka}, \citenamefont {Kuwata-Gonokami}, \citenamefont {Ono},
  \citenamefont {Kumigashira}, \citenamefont {B{\"o}hmer}, \citenamefont
  {Hardy}, \citenamefont {Wolf}, \citenamefont {Meingast}, \citenamefont
  {L{\"o}hneysen}, \citenamefont {Ikeda},\ and\ \citenamefont
  {Ishizaka}}]{Shimojima2014}%
  \BibitemOpen
  \bibfield  {author} {\bibinfo {author} {\bibfnamefont {T.}~\bibnamefont
  {Shimojima}}, \bibinfo {author} {\bibfnamefont {Y.}~\bibnamefont {Suzuki}},
  \bibinfo {author} {\bibfnamefont {T.}~\bibnamefont {Sonobe}}, \bibinfo
  {author} {\bibfnamefont {A.}~\bibnamefont {Nakamura}}, \bibinfo {author}
  {\bibfnamefont {M.}~\bibnamefont {Sakano}}, \bibinfo {author} {\bibfnamefont
  {J.}~\bibnamefont {Omachi}}, \bibinfo {author} {\bibfnamefont
  {K.}~\bibnamefont {Yoshioka}}, \bibinfo {author} {\bibfnamefont
  {M.}~\bibnamefont {Kuwata-Gonokami}}, \bibinfo {author} {\bibfnamefont
  {K.}~\bibnamefont {Ono}}, \bibinfo {author} {\bibfnamefont {H.}~\bibnamefont
  {Kumigashira}}, \bibinfo {author} {\bibfnamefont {A.~E.}\ \bibnamefont
  {B{\"o}hmer}}, \bibinfo {author} {\bibfnamefont {F.}~\bibnamefont {Hardy}},
  \bibinfo {author} {\bibfnamefont {T.}~\bibnamefont {Wolf}}, \bibinfo {author}
  {\bibfnamefont {C.}~\bibnamefont {Meingast}}, \bibinfo {author}
  {\bibfnamefont {H.~v.}\ \bibnamefont {L{\"o}hneysen}}, \bibinfo {author}
  {\bibfnamefont {H.}~\bibnamefont {Ikeda}},\ and\ \bibinfo {author}
  {\bibfnamefont {K.}~\bibnamefont {Ishizaka}},\ }\bibfield  {title} {\bibinfo
  {title} {Lifting of {$xz$}/{$yz$} orbital degeneracy at the structural
  transition in detwinned {FeSe}},\ }\href
  {https://doi.org/10.1103/PhysRevB.90.121111} {\bibfield  {journal} {\bibinfo
  {journal} {Physical Review B}\ }\textbf {\bibinfo {volume} {90}},\ \bibinfo
  {pages} {121111(R)} (\bibinfo {year} {2014})}\BibitemShut {NoStop}%
\bibitem [{\citenamefont {Nakayama}\ \emph {et~al.}(2014)\citenamefont
  {Nakayama}, \citenamefont {Miyata}, \citenamefont {Phan}, \citenamefont
  {Sato}, \citenamefont {Tanabe}, \citenamefont {Urata}, \citenamefont
  {Tanigaki},\ and\ \citenamefont {Takahashi}}]{Nakayama2014}%
  \BibitemOpen
  \bibfield  {author} {\bibinfo {author} {\bibfnamefont {K.}~\bibnamefont
  {Nakayama}}, \bibinfo {author} {\bibfnamefont {Y.}~\bibnamefont {Miyata}},
  \bibinfo {author} {\bibfnamefont {G.~N.}\ \bibnamefont {Phan}}, \bibinfo
  {author} {\bibfnamefont {T.}~\bibnamefont {Sato}}, \bibinfo {author}
  {\bibfnamefont {Y.}~\bibnamefont {Tanabe}}, \bibinfo {author} {\bibfnamefont
  {T.}~\bibnamefont {Urata}}, \bibinfo {author} {\bibfnamefont
  {K.}~\bibnamefont {Tanigaki}},\ and\ \bibinfo {author} {\bibfnamefont
  {T.}~\bibnamefont {Takahashi}},\ }\bibfield  {title} {\bibinfo {title}
  {Reconstruction of band structure induced by electronic nematicity in an
  {FeSe} superconductor},\ }\href
  {https://doi.org/10.1103/PhysRevLett.113.237001} {\bibfield  {journal}
  {\bibinfo  {journal} {Physical Review Letters}\ }\textbf {\bibinfo {volume}
  {113}},\ \bibinfo {pages} {237001} (\bibinfo {year} {2014})}\BibitemShut
  {NoStop}%
\bibitem [{\citenamefont {Baek}\ \emph {et~al.}(2015)\citenamefont {Baek},
  \citenamefont {Efremov}, \citenamefont {Ok}, \citenamefont {Kim},
  \citenamefont {van~den Brink},\ and\ \citenamefont {B{\"u}chner}}]{Baek2015}%
  \BibitemOpen
  \bibfield  {author} {\bibinfo {author} {\bibfnamefont {S.-H.}\ \bibnamefont
  {Baek}}, \bibinfo {author} {\bibfnamefont {D.~V.}\ \bibnamefont {Efremov}},
  \bibinfo {author} {\bibfnamefont {J.~M.}\ \bibnamefont {Ok}}, \bibinfo
  {author} {\bibfnamefont {J.~S.}\ \bibnamefont {Kim}}, \bibinfo {author}
  {\bibfnamefont {J.}~\bibnamefont {van~den Brink}},\ and\ \bibinfo {author}
  {\bibfnamefont {B.}~\bibnamefont {B{\"u}chner}},\ }\bibfield  {title}
  {\bibinfo {title} {Orbital-driven nematicity in {FeSe}},\ }\href
  {https://doi.org/10.1038/nmat4138} {\bibfield  {journal} {\bibinfo  {journal}
  {Nature Materials}\ }\textbf {\bibinfo {volume} {14}},\ \bibinfo {pages}
  {210} (\bibinfo {year} {2015})}\BibitemShut {NoStop}%
\bibitem [{\citenamefont {B{\"o}hmer}\ \emph {et~al.}(2015)\citenamefont
  {B{\"o}hmer}, \citenamefont {Arai}, \citenamefont {Hardy}, \citenamefont
  {Hattori}, \citenamefont {Iye}, \citenamefont {Wolf}, \citenamefont
  {L{\"o}hneysen}, \citenamefont {Ishida},\ and\ \citenamefont
  {Meingast}}]{Bohmer2015}%
  \BibitemOpen
  \bibfield  {author} {\bibinfo {author} {\bibfnamefont {A.~E.}\ \bibnamefont
  {B{\"o}hmer}}, \bibinfo {author} {\bibfnamefont {T.}~\bibnamefont {Arai}},
  \bibinfo {author} {\bibfnamefont {F.}~\bibnamefont {Hardy}}, \bibinfo
  {author} {\bibfnamefont {T.}~\bibnamefont {Hattori}}, \bibinfo {author}
  {\bibfnamefont {T.}~\bibnamefont {Iye}}, \bibinfo {author} {\bibfnamefont
  {T.}~\bibnamefont {Wolf}}, \bibinfo {author} {\bibfnamefont {H.~v.}\
  \bibnamefont {L{\"o}hneysen}}, \bibinfo {author} {\bibfnamefont
  {K.}~\bibnamefont {Ishida}},\ and\ \bibinfo {author} {\bibfnamefont
  {C.}~\bibnamefont {Meingast}},\ }\bibfield  {title} {\bibinfo {title} {Origin
  of the tetragonal-to-orthorhombic phase transition in {FeSe}: A combined
  thermodynamic and {NMR} study of nematicity},\ }\href
  {https://doi.org/10.1103/PhysRevLett.114.027001} {\bibfield  {journal}
  {\bibinfo  {journal} {Physical Review Letters}\ }\textbf {\bibinfo {volume}
  {114}},\ \bibinfo {pages} {027001} (\bibinfo {year} {2015})}\BibitemShut
  {NoStop}%
\bibitem [{\citenamefont {Watson}\ \emph {et~al.}(2015)\citenamefont {Watson},
  \citenamefont {Kim}, \citenamefont {Haghighirad}, \citenamefont {Davies},
  \citenamefont {McCollam}, \citenamefont {Narayanan}, \citenamefont {Blake},
  \citenamefont {Chen}, \citenamefont {Ghannadzadeh}, \citenamefont
  {Schofield}, \citenamefont {Hoesch}, \citenamefont {Meingast}, \citenamefont
  {Wolf},\ and\ \citenamefont {Coldea}}]{Watson2015}%
  \BibitemOpen
  \bibfield  {author} {\bibinfo {author} {\bibfnamefont {M.~D.}\ \bibnamefont
  {Watson}}, \bibinfo {author} {\bibfnamefont {T.~K.}\ \bibnamefont {Kim}},
  \bibinfo {author} {\bibfnamefont {A.~A.}\ \bibnamefont {Haghighirad}},
  \bibinfo {author} {\bibfnamefont {N.~R.}\ \bibnamefont {Davies}}, \bibinfo
  {author} {\bibfnamefont {A.}~\bibnamefont {McCollam}}, \bibinfo {author}
  {\bibfnamefont {A.}~\bibnamefont {Narayanan}}, \bibinfo {author}
  {\bibfnamefont {S.~F.}\ \bibnamefont {Blake}}, \bibinfo {author}
  {\bibfnamefont {Y.~L.}\ \bibnamefont {Chen}}, \bibinfo {author}
  {\bibfnamefont {S.}~\bibnamefont {Ghannadzadeh}}, \bibinfo {author}
  {\bibfnamefont {A.~J.}\ \bibnamefont {Schofield}}, \bibinfo {author}
  {\bibfnamefont {M.}~\bibnamefont {Hoesch}}, \bibinfo {author} {\bibfnamefont
  {C.}~\bibnamefont {Meingast}}, \bibinfo {author} {\bibfnamefont
  {T.}~\bibnamefont {Wolf}},\ and\ \bibinfo {author} {\bibfnamefont {A.~I.}\
  \bibnamefont {Coldea}},\ }\bibfield  {title} {\bibinfo {title} {Emergence of
  the nematic electronic state in {FeSe}},\ }\href
  {https://doi.org/10.1103/PhysRevB.91.155106} {\bibfield  {journal} {\bibinfo
  {journal} {Physical Review B}\ }\textbf {\bibinfo {volume} {91}},\ \bibinfo
  {pages} {155106} (\bibinfo {year} {2015})}\BibitemShut {NoStop}%
\bibitem [{\citenamefont {Sprau}\ \emph {et~al.}(2017)\citenamefont {Sprau},
  \citenamefont {Kostin}, \citenamefont {Kreisel}, \citenamefont {B{\"o}hmer},
  \citenamefont {Taufour}, \citenamefont {Canfield}, \citenamefont {Mukherjee},
  \citenamefont {Hirschfeld}, \citenamefont {Andersen},\ and\ \citenamefont
  {Davis}}]{Sprau2017}%
  \BibitemOpen
  \bibfield  {author} {\bibinfo {author} {\bibfnamefont {P.~O.}\ \bibnamefont
  {Sprau}}, \bibinfo {author} {\bibfnamefont {A.}~\bibnamefont {Kostin}},
  \bibinfo {author} {\bibfnamefont {A.}~\bibnamefont {Kreisel}}, \bibinfo
  {author} {\bibfnamefont {A.~E.}\ \bibnamefont {B{\"o}hmer}}, \bibinfo
  {author} {\bibfnamefont {V.}~\bibnamefont {Taufour}}, \bibinfo {author}
  {\bibfnamefont {P.~C.}\ \bibnamefont {Canfield}}, \bibinfo {author}
  {\bibfnamefont {S.}~\bibnamefont {Mukherjee}}, \bibinfo {author}
  {\bibfnamefont {P.~J.}\ \bibnamefont {Hirschfeld}}, \bibinfo {author}
  {\bibfnamefont {B.~M.}\ \bibnamefont {Andersen}},\ and\ \bibinfo {author}
  {\bibfnamefont {J.~C.~S.}\ \bibnamefont {Davis}},\ }\bibfield  {title}
  {\bibinfo {title} {Discovery of orbital-selective cooper pairing in {FeSe}},\
  }\href {https://doi.org/10.1126/science.aal1575} {\bibfield  {journal}
  {\bibinfo  {journal} {Science}\ }\textbf {\bibinfo {volume} {357}},\ \bibinfo
  {pages} {75} (\bibinfo {year} {2017})}\BibitemShut {NoStop}%
\bibitem [{\citenamefont {Hashimoto}\ \emph {et~al.}(2018)\citenamefont
  {Hashimoto}, \citenamefont {Ota}, \citenamefont {Yamamoto}, \citenamefont
  {Suzuki}, \citenamefont {Shimojima}, \citenamefont {Watanabe}, \citenamefont
  {Chen}, \citenamefont {Kasahara}, \citenamefont {Matsuda}, \citenamefont
  {Shibauchi}, \citenamefont {Okazaki},\ and\ \citenamefont
  {Shin}}]{Hashimoto2018}%
  \BibitemOpen
  \bibfield  {author} {\bibinfo {author} {\bibfnamefont {T.}~\bibnamefont
  {Hashimoto}}, \bibinfo {author} {\bibfnamefont {Y.}~\bibnamefont {Ota}},
  \bibinfo {author} {\bibfnamefont {H.~Q.}\ \bibnamefont {Yamamoto}}, \bibinfo
  {author} {\bibfnamefont {Y.}~\bibnamefont {Suzuki}}, \bibinfo {author}
  {\bibfnamefont {T.}~\bibnamefont {Shimojima}}, \bibinfo {author}
  {\bibfnamefont {S.}~\bibnamefont {Watanabe}}, \bibinfo {author}
  {\bibfnamefont {C.}~\bibnamefont {Chen}}, \bibinfo {author} {\bibfnamefont
  {S.}~\bibnamefont {Kasahara}}, \bibinfo {author} {\bibfnamefont
  {Y.}~\bibnamefont {Matsuda}}, \bibinfo {author} {\bibfnamefont
  {T.}~\bibnamefont {Shibauchi}}, \bibinfo {author} {\bibfnamefont
  {K.}~\bibnamefont {Okazaki}},\ and\ \bibinfo {author} {\bibfnamefont
  {S.}~\bibnamefont {Shin}},\ }\bibfield  {title} {\bibinfo {title}
  {Superconducting gap anisotropy sensitive to nematic domains in {FeSe}},\
  }\href {https://doi.org/10.1038/s41467-017-02739-y} {\bibfield  {journal}
  {\bibinfo  {journal} {Nature Communications}\ }\textbf {\bibinfo {volume}
  {9}},\ \bibinfo {pages} {282} (\bibinfo {year} {2018})}\BibitemShut {NoStop}%
\bibitem [{\citenamefont {Bao}\ \emph {et~al.}(2009)\citenamefont {Bao},
  \citenamefont {Qiu}, \citenamefont {Huang}, \citenamefont {Green},
  \citenamefont {Zajdel}, \citenamefont {Fitzsimmons}, \citenamefont
  {Zhernenkov}, \citenamefont {Chang}, \citenamefont {Fang}, \citenamefont
  {Qian}, \citenamefont {Vehstedt}, \citenamefont {Yang}, \citenamefont {Pham},
  \citenamefont {Spinu},\ and\ \citenamefont {Mao}}]{Bao2009}%
  \BibitemOpen
  \bibfield  {author} {\bibinfo {author} {\bibfnamefont {W.}~\bibnamefont
  {Bao}}, \bibinfo {author} {\bibfnamefont {Y.}~\bibnamefont {Qiu}}, \bibinfo
  {author} {\bibfnamefont {Q.}~\bibnamefont {Huang}}, \bibinfo {author}
  {\bibfnamefont {M.~A.}\ \bibnamefont {Green}}, \bibinfo {author}
  {\bibfnamefont {P.}~\bibnamefont {Zajdel}}, \bibinfo {author} {\bibfnamefont
  {M.~R.}\ \bibnamefont {Fitzsimmons}}, \bibinfo {author} {\bibfnamefont
  {M.}~\bibnamefont {Zhernenkov}}, \bibinfo {author} {\bibfnamefont
  {S.}~\bibnamefont {Chang}}, \bibinfo {author} {\bibfnamefont
  {M.}~\bibnamefont {Fang}}, \bibinfo {author} {\bibfnamefont {B.}~\bibnamefont
  {Qian}}, \bibinfo {author} {\bibfnamefont {E.~K.}\ \bibnamefont {Vehstedt}},
  \bibinfo {author} {\bibfnamefont {J.}~\bibnamefont {Yang}}, \bibinfo {author}
  {\bibfnamefont {H.~M.}\ \bibnamefont {Pham}}, \bibinfo {author}
  {\bibfnamefont {L.}~\bibnamefont {Spinu}},\ and\ \bibinfo {author}
  {\bibfnamefont {Z.~Q.}\ \bibnamefont {Mao}},\ }\bibfield  {title} {\bibinfo
  {title} {Tunable {($\delta\pi$, $\delta\pi$)}-type antiferromagnetic order in
  {$\alpha$-Fe(Te,Se)} superconductors},\ }\href
  {https://doi.org/10.1103/PhysRevLett.102.247001} {\bibfield  {journal}
  {\bibinfo  {journal} {Physical Review Letters}\ }\textbf {\bibinfo {volume}
  {102}},\ \bibinfo {pages} {247001} (\bibinfo {year} {2009})}\BibitemShut
  {NoStop}%
\bibitem [{\citenamefont {Li}\ \emph {et~al.}(2009)\citenamefont {Li},
  \citenamefont {de~la Cruz}, \citenamefont {Huang}, \citenamefont {Chen},
  \citenamefont {Lynn}, \citenamefont {Hu}, \citenamefont {Huang},
  \citenamefont {Hsu}, \citenamefont {Yeh}, \citenamefont {Wu},\ and\
  \citenamefont {Dai}}]{Li2009}%
  \BibitemOpen
  \bibfield  {author} {\bibinfo {author} {\bibfnamefont {S.}~\bibnamefont
  {Li}}, \bibinfo {author} {\bibfnamefont {C.}~\bibnamefont {de~la Cruz}},
  \bibinfo {author} {\bibfnamefont {Q.}~\bibnamefont {Huang}}, \bibinfo
  {author} {\bibfnamefont {Y.}~\bibnamefont {Chen}}, \bibinfo {author}
  {\bibfnamefont {J.~W.}\ \bibnamefont {Lynn}}, \bibinfo {author}
  {\bibfnamefont {J.}~\bibnamefont {Hu}}, \bibinfo {author} {\bibfnamefont
  {Y.-L.}\ \bibnamefont {Huang}}, \bibinfo {author} {\bibfnamefont {F.-C.}\
  \bibnamefont {Hsu}}, \bibinfo {author} {\bibfnamefont {K.-W.}\ \bibnamefont
  {Yeh}}, \bibinfo {author} {\bibfnamefont {M.-K.}\ \bibnamefont {Wu}},\ and\
  \bibinfo {author} {\bibfnamefont {P.}~\bibnamefont {Dai}},\ }\bibfield
  {title} {\bibinfo {title} {First-order magnetic and structural phase
  transitions in {Fe$_{1+y}$Se$_x$Te$_{1-x}$}},\ }\href
  {https://doi.org/10.1103/PhysRevB.79.054503} {\bibfield  {journal} {\bibinfo
  {journal} {Physical Review B}\ }\textbf {\bibinfo {volume} {79}},\ \bibinfo
  {pages} {054503} (\bibinfo {year} {2009})}\BibitemShut {NoStop}%
\bibitem [{\citenamefont {Ma}\ \emph {et~al.}(2009)\citenamefont {Ma},
  \citenamefont {Ji}, \citenamefont {Hu}, \citenamefont {Lu},\ and\
  \citenamefont {Xiang}}]{Ma2009}%
  \BibitemOpen
  \bibfield  {author} {\bibinfo {author} {\bibfnamefont {F.}~\bibnamefont
  {Ma}}, \bibinfo {author} {\bibfnamefont {W.}~\bibnamefont {Ji}}, \bibinfo
  {author} {\bibfnamefont {J.}~\bibnamefont {Hu}}, \bibinfo {author}
  {\bibfnamefont {Z.-Y.}\ \bibnamefont {Lu}},\ and\ \bibinfo {author}
  {\bibfnamefont {T.}~\bibnamefont {Xiang}},\ }\bibfield  {title} {\bibinfo
  {title} {First-principles calculations of the electronic structure of
  tetragonal {$\alpha$-FeTe} and {$\alpha$-FeSe} crystals: Evidence for a
  bicollinear antiferromagnetic order},\ }\href
  {https://doi.org/10.1103/PhysRevLett.102.177003} {\bibfield  {journal}
  {\bibinfo  {journal} {Physical Review Letters}\ }\textbf {\bibinfo {volume}
  {102}},\ \bibinfo {pages} {177003} (\bibinfo {year} {2009})}\BibitemShut
  {NoStop}%
\bibitem [{\citenamefont {Rodriguez}\ \emph {et~al.}(2011)\citenamefont
  {Rodriguez}, \citenamefont {Stock}, \citenamefont {Zajdel}, \citenamefont
  {Krycka}, \citenamefont {Majkrzak}, \citenamefont {Zavalij},\ and\
  \citenamefont {Green}}]{Rodriguez2011}%
  \BibitemOpen
  \bibfield  {author} {\bibinfo {author} {\bibfnamefont {E.~E.}\ \bibnamefont
  {Rodriguez}}, \bibinfo {author} {\bibfnamefont {C.}~\bibnamefont {Stock}},
  \bibinfo {author} {\bibfnamefont {P.}~\bibnamefont {Zajdel}}, \bibinfo
  {author} {\bibfnamefont {K.~L.}\ \bibnamefont {Krycka}}, \bibinfo {author}
  {\bibfnamefont {C.~F.}\ \bibnamefont {Majkrzak}}, \bibinfo {author}
  {\bibfnamefont {P.}~\bibnamefont {Zavalij}},\ and\ \bibinfo {author}
  {\bibfnamefont {M.~A.}\ \bibnamefont {Green}},\ }\bibfield  {title} {\bibinfo
  {title} {Magnetic-crystallographic phase diagram of the superconducting
  parent compound {Fe$_{1+x}$Te}},\ }\href
  {https://doi.org/10.1103/PhysRevB.84.064403} {\bibfield  {journal} {\bibinfo
  {journal} {Physical Review B}\ }\textbf {\bibinfo {volume} {84}},\ \bibinfo
  {pages} {064403} (\bibinfo {year} {2011})}\BibitemShut {NoStop}%
\bibitem [{\citenamefont {Stock}\ \emph {et~al.}(2011)\citenamefont {Stock},
  \citenamefont {Rodriguez}, \citenamefont {Green}, \citenamefont {Zavalij},\
  and\ \citenamefont {Rodriguez-Rivera}}]{Stock2011}%
  \BibitemOpen
  \bibfield  {author} {\bibinfo {author} {\bibfnamefont {C.}~\bibnamefont
  {Stock}}, \bibinfo {author} {\bibfnamefont {E.~E.}\ \bibnamefont
  {Rodriguez}}, \bibinfo {author} {\bibfnamefont {M.~A.}\ \bibnamefont
  {Green}}, \bibinfo {author} {\bibfnamefont {P.}~\bibnamefont {Zavalij}},\
  and\ \bibinfo {author} {\bibfnamefont {J.~A.}\ \bibnamefont
  {Rodriguez-Rivera}},\ }\bibfield  {title} {\bibinfo {title} {Interstitial
  iron tuning of the spin fluctuations in the nonsuperconducting parent phase
  {Fe$_{1+x}$Te}},\ }\href {https://doi.org/10.1103/PhysRevB.84.045124}
  {\bibfield  {journal} {\bibinfo  {journal} {Physical Review B}\ }\textbf
  {\bibinfo {volume} {84}},\ \bibinfo {pages} {045124} (\bibinfo {year}
  {2011})}\BibitemShut {NoStop}%
\bibitem [{\citenamefont {Zaliznyak}\ \emph {et~al.}(2012)\citenamefont
  {Zaliznyak}, \citenamefont {Xu}, \citenamefont {Wen}, \citenamefont
  {Tranquada}, \citenamefont {Gu}, \citenamefont {Solovyov}, \citenamefont
  {Glazkov}, \citenamefont {Zheludev}, \citenamefont {Garlea},\ and\
  \citenamefont {Stone}}]{Zaliznyak2012}%
  \BibitemOpen
  \bibfield  {author} {\bibinfo {author} {\bibfnamefont {I.~A.}\ \bibnamefont
  {Zaliznyak}}, \bibinfo {author} {\bibfnamefont {Z.~J.}\ \bibnamefont {Xu}},
  \bibinfo {author} {\bibfnamefont {J.~S.}\ \bibnamefont {Wen}}, \bibinfo
  {author} {\bibfnamefont {J.~M.}\ \bibnamefont {Tranquada}}, \bibinfo {author}
  {\bibfnamefont {G.~D.}\ \bibnamefont {Gu}}, \bibinfo {author} {\bibfnamefont
  {V.}~\bibnamefont {Solovyov}}, \bibinfo {author} {\bibfnamefont {V.~N.}\
  \bibnamefont {Glazkov}}, \bibinfo {author} {\bibfnamefont {A.~I.}\
  \bibnamefont {Zheludev}}, \bibinfo {author} {\bibfnamefont {V.~O.}\
  \bibnamefont {Garlea}},\ and\ \bibinfo {author} {\bibfnamefont {M.~B.}\
  \bibnamefont {Stone}},\ }\bibfield  {title} {\bibinfo {title} {Continuous
  magnetic and structural phase transitions in {Fe$_{1+y}$Te}},\ }\href
  {https://doi.org/10.1103/PhysRevB.85.085105} {\bibfield  {journal} {\bibinfo
  {journal} {Physical Review B}\ }\textbf {\bibinfo {volume} {85}},\ \bibinfo
  {pages} {085105} (\bibinfo {year} {2012})}\BibitemShut {NoStop}%
\bibitem [{\citenamefont {Gr{\o}nvold}\ \emph {et~al.}(1954)\citenamefont
  {Gr{\o}nvold}, \citenamefont {Haraldsen}, \citenamefont {Vihovde},\ and\
  \citenamefont {S{\"o}rensen}}]{Gronvold1954}%
  \BibitemOpen
  \bibfield  {author} {\bibinfo {author} {\bibfnamefont {F.}~\bibnamefont
  {Gr{\o}nvold}}, \bibinfo {author} {\bibfnamefont {H.}~\bibnamefont
  {Haraldsen}}, \bibinfo {author} {\bibfnamefont {J.}~\bibnamefont {Vihovde}},\
  and\ \bibinfo {author} {\bibfnamefont {N.~A.}\ \bibnamefont {S{\"o}rensen}},\
  }\bibfield  {title} {\bibinfo {title} {Phase and structural relations in the
  system iron tellurium},\ }\href
  {https://doi.org/10.3891/acta.chem.scand.08-1927} {\bibfield  {journal}
  {\bibinfo  {journal} {Acta Chemica Scandinavica}\ }\textbf {\bibinfo {volume}
  {8}},\ \bibinfo {pages} {1927} (\bibinfo {year} {1954})}\BibitemShut
  {NoStop}%
\bibitem [{\citenamefont {Ward}\ and\ \citenamefont {McCann}(1979)}]{Ward1979}%
  \BibitemOpen
  \bibfield  {author} {\bibinfo {author} {\bibfnamefont {J.~B.}\ \bibnamefont
  {Ward}}\ and\ \bibinfo {author} {\bibfnamefont {V.~H.}\ \bibnamefont
  {McCann}},\ }\bibfield  {title} {\bibinfo {title} {On the {$^{57}$Fe}
  m{\"o}ssbauer spectra of {FeTe} and {Fe$_2$Te$_3$}},\ }\href
  {https://doi.org/10.1088/0022-3719/12/5/016} {\bibfield  {journal} {\bibinfo
  {journal} {Journal of Physics C: Solid State Physics}\ }\textbf {\bibinfo
  {volume} {12}},\ \bibinfo {pages} {873} (\bibinfo {year} {1979})}\BibitemShut
  {NoStop}%
\bibitem [{\citenamefont {Han}\ \emph {et~al.}(2010)\citenamefont {Han},
  \citenamefont {Li}, \citenamefont {Cao}, \citenamefont {Wang}, \citenamefont
  {Xu}, \citenamefont {Zhao}, \citenamefont {Guo},\ and\ \citenamefont
  {Yang}}]{Han2010}%
  \BibitemOpen
  \bibfield  {author} {\bibinfo {author} {\bibfnamefont {Y.}~\bibnamefont
  {Han}}, \bibinfo {author} {\bibfnamefont {W.~Y.}\ \bibnamefont {Li}},
  \bibinfo {author} {\bibfnamefont {L.~X.}\ \bibnamefont {Cao}}, \bibinfo
  {author} {\bibfnamefont {X.~Y.}\ \bibnamefont {Wang}}, \bibinfo {author}
  {\bibfnamefont {B.}~\bibnamefont {Xu}}, \bibinfo {author} {\bibfnamefont
  {B.~R.}\ \bibnamefont {Zhao}}, \bibinfo {author} {\bibfnamefont {Y.~Q.}\
  \bibnamefont {Guo}},\ and\ \bibinfo {author} {\bibfnamefont {J.~L.}\
  \bibnamefont {Yang}},\ }\bibfield  {title} {\bibinfo {title}
  {Superconductivity in iron telluride thin films under tensile stress},\
  }\href {https://doi.org/10.1103/PhysRevLett.104.017003} {\bibfield  {journal}
  {\bibinfo  {journal} {Physical Review Letters}\ }\textbf {\bibinfo {volume}
  {104}},\ \bibinfo {pages} {017003} (\bibinfo {year} {2010})}\BibitemShut
  {NoStop}%
\bibitem [{\citenamefont {Yan}\ \emph {et~al.}(2026)\citenamefont {Yan},
  \citenamefont {Wang}, \citenamefont {Xia}, \citenamefont {Paolini},
  \citenamefont {Chan}, \citenamefont {Dihingia}, \citenamefont {Rong},
  \citenamefont {Xiao}, \citenamefont {Halanayake}, \citenamefont {Song},
  \citenamefont {Gowda}, \citenamefont {Hickey}, \citenamefont {Wu},
  \citenamefont {Yu}, \citenamefont {Hirschfeld},\ and\ \citenamefont
  {Chang}}]{Yan2026FeTe}%
  \BibitemOpen
  \bibfield  {author} {\bibinfo {author} {\bibfnamefont {Z.-J.}\ \bibnamefont
  {Yan}}, \bibinfo {author} {\bibfnamefont {Z.}~\bibnamefont {Wang}}, \bibinfo
  {author} {\bibfnamefont {B.}~\bibnamefont {Xia}}, \bibinfo {author}
  {\bibfnamefont {S.}~\bibnamefont {Paolini}}, \bibinfo {author} {\bibfnamefont
  {Y.-T.}\ \bibnamefont {Chan}}, \bibinfo {author} {\bibfnamefont
  {N.}~\bibnamefont {Dihingia}}, \bibinfo {author} {\bibfnamefont
  {H.}~\bibnamefont {Rong}}, \bibinfo {author} {\bibfnamefont {P.}~\bibnamefont
  {Xiao}}, \bibinfo {author} {\bibfnamefont {K.~D.}\ \bibnamefont
  {Halanayake}}, \bibinfo {author} {\bibfnamefont {J.}~\bibnamefont {Song}},
  \bibinfo {author} {\bibfnamefont {V.}~\bibnamefont {Gowda}}, \bibinfo
  {author} {\bibfnamefont {D.~R.}\ \bibnamefont {Hickey}}, \bibinfo {author}
  {\bibfnamefont {W.}~\bibnamefont {Wu}}, \bibinfo {author} {\bibfnamefont
  {J.}~\bibnamefont {Yu}}, \bibinfo {author} {\bibfnamefont {P.~J.}\
  \bibnamefont {Hirschfeld}},\ and\ \bibinfo {author} {\bibfnamefont {C.-Z.}\
  \bibnamefont {Chang}},\ }\bibfield  {title} {\bibinfo {title} {Stoichiometric
  {FeTe} is a superconductor},\ }\href
  {https://doi.org/10.1038/s41586-026-10321-0} {\bibfield  {journal} {\bibinfo
  {journal} {Nature}\ }\textbf {\bibinfo {volume} {652}},\ \bibinfo {pages}
  {342} (\bibinfo {year} {2026})}\BibitemShut {NoStop}%
\bibitem [{\citenamefont {Subedi}\ \emph {et~al.}(2008)\citenamefont {Subedi},
  \citenamefont {Zhang}, \citenamefont {Singh},\ and\ \citenamefont
  {Du}}]{Subedi2008FeChalcogenides}%
  \BibitemOpen
  \bibfield  {author} {\bibinfo {author} {\bibfnamefont {A.}~\bibnamefont
  {Subedi}}, \bibinfo {author} {\bibfnamefont {L.}~\bibnamefont {Zhang}},
  \bibinfo {author} {\bibfnamefont {D.~J.}\ \bibnamefont {Singh}},\ and\
  \bibinfo {author} {\bibfnamefont {M.~H.}\ \bibnamefont {Du}},\ }\bibfield
  {title} {\bibinfo {title} {Density functional study of {FeS}, {FeSe}, and
  {FeTe}: Electronic structure, magnetism, phonons, and superconductivity},\
  }\href {https://doi.org/10.1103/PhysRevB.78.134514} {\bibfield  {journal}
  {\bibinfo  {journal} {Physical Review B}\ }\textbf {\bibinfo {volume} {78}},\
  \bibinfo {pages} {134514} (\bibinfo {year} {2008})}\BibitemShut {NoStop}%
\bibitem [{\citenamefont {Fang}\ \emph {et~al.}(2009)\citenamefont {Fang},
  \citenamefont {Bernevig},\ and\ \citenamefont {Hu}}]{Fang2009MagneticOrder}%
  \BibitemOpen
  \bibfield  {author} {\bibinfo {author} {\bibfnamefont {C.}~\bibnamefont
  {Fang}}, \bibinfo {author} {\bibfnamefont {B.~A.}\ \bibnamefont {Bernevig}},\
  and\ \bibinfo {author} {\bibfnamefont {J.}~\bibnamefont {Hu}},\ }\bibfield
  {title} {\bibinfo {title} {Theory of magnetic order in
  {Fe$_{1+y}$Te$_{1-x}$Se$_x$}},\ }\href
  {https://doi.org/10.1209/0295-5075/86/67005} {\bibfield  {journal} {\bibinfo
  {journal} {EPL (Europhysics Letters)}\ }\textbf {\bibinfo {volume} {86}},\
  \bibinfo {pages} {67005} (\bibinfo {year} {2009})}\BibitemShut {NoStop}%
\bibitem [{\citenamefont {Turner}\ \emph {et~al.}(2009)\citenamefont {Turner},
  \citenamefont {Wang},\ and\ \citenamefont {Vishwanath}}]{Turner2009FeTe}%
  \BibitemOpen
  \bibfield  {author} {\bibinfo {author} {\bibfnamefont {A.~M.}\ \bibnamefont
  {Turner}}, \bibinfo {author} {\bibfnamefont {F.}~\bibnamefont {Wang}},\ and\
  \bibinfo {author} {\bibfnamefont {A.}~\bibnamefont {Vishwanath}},\ }\bibfield
   {title} {\bibinfo {title} {Kinetic magnetism and orbital order in iron
  telluride},\ }\href {https://doi.org/10.1103/PhysRevB.80.224504} {\bibfield
  {journal} {\bibinfo  {journal} {Physical Review B}\ }\textbf {\bibinfo
  {volume} {80}},\ \bibinfo {pages} {224504} (\bibinfo {year}
  {2009})}\BibitemShut {NoStop}%
\bibitem [{\citenamefont {Moon}\ and\ \citenamefont
  {Choi}(2010)}]{Moon2010ChalcogenHeight}%
  \BibitemOpen
  \bibfield  {author} {\bibinfo {author} {\bibfnamefont {C.-Y.}\ \bibnamefont
  {Moon}}\ and\ \bibinfo {author} {\bibfnamefont {H.~J.}\ \bibnamefont
  {Choi}},\ }\bibfield  {title} {\bibinfo {title} {Chalcogen-height dependent
  magnetic interactions and magnetic order switching in {FeSe$_x$Te$_{1-x}$}},\
  }\href {https://doi.org/10.1103/PhysRevLett.104.057003} {\bibfield  {journal}
  {\bibinfo  {journal} {Physical Review Letters}\ }\textbf {\bibinfo {volume}
  {104}},\ \bibinfo {pages} {057003} (\bibinfo {year} {2010})}\BibitemShut
  {NoStop}%
\bibitem [{\citenamefont {Ducatman}\ \emph {et~al.}(2012)\citenamefont
  {Ducatman}, \citenamefont {Perkins},\ and\ \citenamefont
  {Chubukov}}]{Ducatman2012Chalcogenides}%
  \BibitemOpen
  \bibfield  {author} {\bibinfo {author} {\bibfnamefont {S.}~\bibnamefont
  {Ducatman}}, \bibinfo {author} {\bibfnamefont {N.~B.}\ \bibnamefont
  {Perkins}},\ and\ \bibinfo {author} {\bibfnamefont {A.}~\bibnamefont
  {Chubukov}},\ }\bibfield  {title} {\bibinfo {title} {Magnetism in parent iron
  chalcogenides: Quantum fluctuations select plaquette order},\ }\href
  {https://doi.org/10.1103/PhysRevLett.109.157206} {\bibfield  {journal}
  {\bibinfo  {journal} {Physical Review Letters}\ }\textbf {\bibinfo {volume}
  {109}},\ \bibinfo {pages} {157206} (\bibinfo {year} {2012})}\BibitemShut
  {NoStop}%
\bibitem [{\citenamefont {Ducatman}\ \emph {et~al.}(2014)\citenamefont
  {Ducatman}, \citenamefont {Fernandes},\ and\ \citenamefont
  {Perkins}}]{Ducatman2014FeTe}%
  \BibitemOpen
  \bibfield  {author} {\bibinfo {author} {\bibfnamefont {S.}~\bibnamefont
  {Ducatman}}, \bibinfo {author} {\bibfnamefont {R.~M.}\ \bibnamefont
  {Fernandes}},\ and\ \bibinfo {author} {\bibfnamefont {N.~B.}\ \bibnamefont
  {Perkins}},\ }\bibfield  {title} {\bibinfo {title} {Theory of the evolution
  of magnetic order in {Fe$_{1+y}$Te} compounds with increasing interstitial
  iron},\ }\href {https://doi.org/10.1103/PhysRevB.90.165123} {\bibfield
  {journal} {\bibinfo  {journal} {Physical Review B}\ }\textbf {\bibinfo
  {volume} {90}},\ \bibinfo {pages} {165123} (\bibinfo {year}
  {2014})}\BibitemShut {NoStop}%
\bibitem [{\citenamefont {Glasbrenner}\ \emph {et~al.}(2015)\citenamefont
  {Glasbrenner}, \citenamefont {Mazin}, \citenamefont {Jeschke}, \citenamefont
  {Hirschfeld}, \citenamefont {Fernandes},\ and\ \citenamefont
  {Valenti}}]{Glasbrenner2015IronChalcogenides}%
  \BibitemOpen
  \bibfield  {author} {\bibinfo {author} {\bibfnamefont {J.~K.}\ \bibnamefont
  {Glasbrenner}}, \bibinfo {author} {\bibfnamefont {I.~I.}\ \bibnamefont
  {Mazin}}, \bibinfo {author} {\bibfnamefont {H.~O.}\ \bibnamefont {Jeschke}},
  \bibinfo {author} {\bibfnamefont {P.~J.}\ \bibnamefont {Hirschfeld}},
  \bibinfo {author} {\bibfnamefont {R.~M.}\ \bibnamefont {Fernandes}},\ and\
  \bibinfo {author} {\bibfnamefont {R.}~\bibnamefont {Valenti}},\ }\bibfield
  {title} {\bibinfo {title} {Effect of magnetic frustration on nematicity and
  superconductivity in iron chalcogenides},\ }\href
  {https://doi.org/10.1038/nphys3434} {\bibfield  {journal} {\bibinfo
  {journal} {Nature Physics}\ }\textbf {\bibinfo {volume} {11}},\ \bibinfo
  {pages} {953} (\bibinfo {year} {2015})}\BibitemShut {NoStop}%
\bibitem [{\citenamefont {Ciechan}\ \emph {et~al.}(2014)\citenamefont
  {Ciechan}, \citenamefont {Winiarski},\ and\ \citenamefont
  {Samsel-Czekala}}]{Ciechan2014StrainedFeTe}%
  \BibitemOpen
  \bibfield  {author} {\bibinfo {author} {\bibfnamefont {A.}~\bibnamefont
  {Ciechan}}, \bibinfo {author} {\bibfnamefont {M.~J.}\ \bibnamefont
  {Winiarski}},\ and\ \bibinfo {author} {\bibfnamefont {M.}~\bibnamefont
  {Samsel-Czekala}},\ }\bibfield  {title} {\bibinfo {title} {Magnetic phase
  transitions and superconductivity in strained {FeTe}},\ }\href
  {https://doi.org/10.1088/0953-8984/26/2/025702} {\bibfield  {journal}
  {\bibinfo  {journal} {Journal of Physics: Condensed Matter}\ }\textbf
  {\bibinfo {volume} {26}},\ \bibinfo {pages} {025702} (\bibinfo {year}
  {2014})}\BibitemShut {NoStop}%
\bibitem [{\citenamefont {Xu}\ \emph {et~al.}(2026)\citenamefont {Xu},
  \citenamefont {Jiang}, \citenamefont {Gai}, \citenamefont {Cao},
  \citenamefont {Chen}, \citenamefont {Man}, \citenamefont {Lin}, \citenamefont
  {Deng}, \citenamefont {He}, \citenamefont {Liu}, \citenamefont {Zhao},
  \citenamefont {Lu}, \citenamefont {Chang},\ and\ \citenamefont
  {Liu}}]{Xu2026StrainedFeTe}%
  \BibitemOpen
  \bibfield  {author} {\bibinfo {author} {\bibfnamefont {H.}~\bibnamefont
  {Xu}}, \bibinfo {author} {\bibfnamefont {J.}~\bibnamefont {Jiang}}, \bibinfo
  {author} {\bibfnamefont {X.}~\bibnamefont {Gai}}, \bibinfo {author}
  {\bibfnamefont {R.-Q.}\ \bibnamefont {Cao}}, \bibinfo {author} {\bibfnamefont
  {K.}~\bibnamefont {Chen}}, \bibinfo {author} {\bibfnamefont {X.-X.}\
  \bibnamefont {Man}}, \bibinfo {author} {\bibfnamefont {H.}~\bibnamefont
  {Lin}}, \bibinfo {author} {\bibfnamefont {P.}~\bibnamefont {Deng}}, \bibinfo
  {author} {\bibfnamefont {K.}~\bibnamefont {He}}, \bibinfo {author}
  {\bibfnamefont {K.}~\bibnamefont {Liu}}, \bibinfo {author} {\bibfnamefont
  {D.}~\bibnamefont {Zhao}}, \bibinfo {author} {\bibfnamefont {Z.-Y.}\
  \bibnamefont {Lu}}, \bibinfo {author} {\bibfnamefont {K.}~\bibnamefont
  {Chang}},\ and\ \bibinfo {author} {\bibfnamefont {C.}~\bibnamefont {Liu}},\
  }\bibfield  {title} {\bibinfo {title} {Reversible tuning of magnetic order
  and intrinsic superconductivity in strained {FeTe} films via stoichiometry
  control},\ }\href {https://doi.org/10.1021/acsnano.6c05058} {\bibfield
  {journal} {\bibinfo  {journal} {ACS Nano}\ }\textbf {\bibinfo {volume}
  {20}},\ \bibinfo {pages} {16426} (\bibinfo {year} {2026})}\BibitemShut
  {NoStop}%
\bibitem [{\citenamefont {Song}\ \emph {et~al.}(2026)\citenamefont {Song},
  \citenamefont {Lee}, \citenamefont {Park}, \citenamefont {Park},
  \citenamefont {Lee}, \citenamefont {Lee}, \citenamefont {Kim}, \citenamefont
  {Lee}, \citenamefont {Chang}, \citenamefont {Kim},\ and\ \citenamefont
  {Kim}}]{Changyoung-PRB-2026}%
  \BibitemOpen
  \bibfield  {author} {\bibinfo {author} {\bibfnamefont {H.}~\bibnamefont
  {Song}}, \bibinfo {author} {\bibfnamefont {S.}~\bibnamefont {Lee}}, \bibinfo
  {author} {\bibfnamefont {K.-Y.}\ \bibnamefont {Park}}, \bibinfo {author}
  {\bibfnamefont {J.}~\bibnamefont {Park}}, \bibinfo {author} {\bibfnamefont
  {S.}~\bibnamefont {Lee}}, \bibinfo {author} {\bibfnamefont {Y.}~\bibnamefont
  {Lee}}, \bibinfo {author} {\bibfnamefont {J.}~\bibnamefont {Kim}}, \bibinfo
  {author} {\bibfnamefont {J.}~\bibnamefont {Lee}}, \bibinfo {author}
  {\bibfnamefont {C.~S.}\ \bibnamefont {Chang}}, \bibinfo {author}
  {\bibfnamefont {Y.}~\bibnamefont {Kim}},\ and\ \bibinfo {author}
  {\bibfnamefont {C.}~\bibnamefont {Kim}},\ }\bibfield  {title} {\bibinfo
  {title} {Strain-tuned orbital-dependent electronic correlations in fete thin
  films},\ }\href {https://doi.org/10.1103/xffv-kpjp} {\bibfield  {journal}
  {\bibinfo  {journal} {Phys. Rev. B}\ }\textbf {\bibinfo {volume} {114}},\
  \bibinfo {pages} {045103} (\bibinfo {year} {2026})}\BibitemShut {NoStop}%
\bibitem [{\citenamefont {Kresse}\ and\ \citenamefont
  {Hafner}(1993)}]{Kresse1993VASP}%
  \BibitemOpen
  \bibfield  {author} {\bibinfo {author} {\bibfnamefont {G.}~\bibnamefont
  {Kresse}}\ and\ \bibinfo {author} {\bibfnamefont {J.}~\bibnamefont
  {Hafner}},\ }\bibfield  {title} {\bibinfo {title} {Ab initio molecular
  dynamics for liquid metals},\ }\href
  {https://doi.org/10.1103/PhysRevB.47.558} {\bibfield  {journal} {\bibinfo
  {journal} {Physical Review B}\ }\textbf {\bibinfo {volume} {47}},\ \bibinfo
  {pages} {558} (\bibinfo {year} {1993})}\BibitemShut {NoStop}%
\bibitem [{\citenamefont {Kresse}\ and\ \citenamefont
  {Furthm{\"u}ller}(1996)}]{Kresse1996VASP}%
  \BibitemOpen
  \bibfield  {author} {\bibinfo {author} {\bibfnamefont {G.}~\bibnamefont
  {Kresse}}\ and\ \bibinfo {author} {\bibfnamefont {J.}~\bibnamefont
  {Furthm{\"u}ller}},\ }\bibfield  {title} {\bibinfo {title} {Efficient
  iterative schemes for ab initio total-energy calculations using a plane-wave
  basis set},\ }\href {https://doi.org/10.1103/PhysRevB.54.11169} {\bibfield
  {journal} {\bibinfo  {journal} {Physical Review B}\ }\textbf {\bibinfo
  {volume} {54}},\ \bibinfo {pages} {11169} (\bibinfo {year}
  {1996})}\BibitemShut {NoStop}%
\bibitem [{\citenamefont {Bl\"ochl}(1994)}]{PhysRevB.50.17953}%
  \BibitemOpen
  \bibfield  {author} {\bibinfo {author} {\bibfnamefont {P.~E.}\ \bibnamefont
  {Bl\"ochl}},\ }\bibfield  {title} {\bibinfo {title} {Projector augmented-wave
  method},\ }\href {https://doi.org/10.1103/PhysRevB.50.17953} {\bibfield
  {journal} {\bibinfo  {journal} {Phys. Rev. B}\ }\textbf {\bibinfo {volume}
  {50}},\ \bibinfo {pages} {17953} (\bibinfo {year} {1994})}\BibitemShut
  {NoStop}%
\bibitem [{\citenamefont {Kresse}\ and\ \citenamefont
  {Joubert}(1999)}]{Kresse1999PAW}%
  \BibitemOpen
  \bibfield  {author} {\bibinfo {author} {\bibfnamefont {G.}~\bibnamefont
  {Kresse}}\ and\ \bibinfo {author} {\bibfnamefont {D.}~\bibnamefont
  {Joubert}},\ }\bibfield  {title} {\bibinfo {title} {From ultrasoft
  pseudopotentials to the projector augmented-wave method},\ }\href
  {https://doi.org/10.1103/PhysRevB.59.1758} {\bibfield  {journal} {\bibinfo
  {journal} {Physical Review B}\ }\textbf {\bibinfo {volume} {59}},\ \bibinfo
  {pages} {1758} (\bibinfo {year} {1999})}\BibitemShut {NoStop}%
\bibitem [{\citenamefont {Perdew}\ \emph {et~al.}(1996)\citenamefont {Perdew},
  \citenamefont {Burke},\ and\ \citenamefont
  {Ernzerhof}}]{PhysRevLett.77.3865}%
  \BibitemOpen
  \bibfield  {author} {\bibinfo {author} {\bibfnamefont {J.~P.}\ \bibnamefont
  {Perdew}}, \bibinfo {author} {\bibfnamefont {K.}~\bibnamefont {Burke}},\ and\
  \bibinfo {author} {\bibfnamefont {M.}~\bibnamefont {Ernzerhof}},\ }\bibfield
  {title} {\bibinfo {title} {Generalized gradient approximation made simple},\
  }\href {https://doi.org/10.1103/PhysRevLett.77.3865} {\bibfield  {journal}
  {\bibinfo  {journal} {Phys. Rev. Lett.}\ }\textbf {\bibinfo {volume} {77}},\
  \bibinfo {pages} {3865} (\bibinfo {year} {1996})}\BibitemShut {NoStop}%
\bibitem [{\citenamefont {Mostofi}\ \emph {et~al.}(2008)\citenamefont
  {Mostofi}, \citenamefont {Yates}, \citenamefont {Lee}, \citenamefont {Souza},
  \citenamefont {Vanderbilt},\ and\ \citenamefont {Marzari}}]{MOSTOFI2008685}%
  \BibitemOpen
  \bibfield  {author} {\bibinfo {author} {\bibfnamefont {A.~A.}\ \bibnamefont
  {Mostofi}}, \bibinfo {author} {\bibfnamefont {J.~R.}\ \bibnamefont {Yates}},
  \bibinfo {author} {\bibfnamefont {Y.-S.}\ \bibnamefont {Lee}}, \bibinfo
  {author} {\bibfnamefont {I.}~\bibnamefont {Souza}}, \bibinfo {author}
  {\bibfnamefont {D.}~\bibnamefont {Vanderbilt}},\ and\ \bibinfo {author}
  {\bibfnamefont {N.}~\bibnamefont {Marzari}},\ }\bibfield  {title} {\bibinfo
  {title} {wannier90: A tool for obtaining maximally-localised wannier
  functions},\ }\href
  {https://doi.org/https://doi.org/10.1016/j.cpc.2007.11.016} {\bibfield
  {journal} {\bibinfo  {journal} {Computer Physics Communications}\ }\textbf
  {\bibinfo {volume} {178}},\ \bibinfo {pages} {685} (\bibinfo {year}
  {2008})}\BibitemShut {NoStop}%
\bibitem [{\citenamefont {Marzari}\ \emph {et~al.}(2012)\citenamefont
  {Marzari}, \citenamefont {Mostofi}, \citenamefont {Yates}, \citenamefont
  {Souza},\ and\ \citenamefont {Vanderbilt}}]{RevModPhys.84.1419}%
  \BibitemOpen
  \bibfield  {author} {\bibinfo {author} {\bibfnamefont {N.}~\bibnamefont
  {Marzari}}, \bibinfo {author} {\bibfnamefont {A.~A.}\ \bibnamefont
  {Mostofi}}, \bibinfo {author} {\bibfnamefont {J.~R.}\ \bibnamefont {Yates}},
  \bibinfo {author} {\bibfnamefont {I.}~\bibnamefont {Souza}},\ and\ \bibinfo
  {author} {\bibfnamefont {D.}~\bibnamefont {Vanderbilt}},\ }\bibfield  {title}
  {\bibinfo {title} {Maximally localized wannier functions: Theory and
  applications},\ }\href {https://doi.org/10.1103/RevModPhys.84.1419}
  {\bibfield  {journal} {\bibinfo  {journal} {Rev. Mod. Phys.}\ }\textbf
  {\bibinfo {volume} {84}},\ \bibinfo {pages} {1419} (\bibinfo {year}
  {2012})}\BibitemShut {NoStop}%
\bibitem [{\citenamefont {Kanamori}(1963)}]{Kanamori1963TransitionMetals}%
  \BibitemOpen
  \bibfield  {author} {\bibinfo {author} {\bibfnamefont {J.}~\bibnamefont
  {Kanamori}},\ }\bibfield  {title} {\bibinfo {title} {Electron correlation and
  ferromagnetism of transition metals},\ }\href
  {https://doi.org/10.1143/PTP.30.275} {\bibfield  {journal} {\bibinfo
  {journal} {Progress of Theoretical Physics}\ }\textbf {\bibinfo {volume}
  {30}},\ \bibinfo {pages} {275} (\bibinfo {year} {1963})}\BibitemShut
  {NoStop}%
\bibitem [{\citenamefont {Bickers}\ and\ \citenamefont
  {Scalapino}(1989)}]{Bickers1989FLEX}%
  \BibitemOpen
  \bibfield  {author} {\bibinfo {author} {\bibfnamefont {N.~E.}\ \bibnamefont
  {Bickers}}\ and\ \bibinfo {author} {\bibfnamefont {D.~J.}\ \bibnamefont
  {Scalapino}},\ }\bibfield  {title} {\bibinfo {title} {Conserving
  approximations for strongly fluctuating electron systems. {I}. formalism and
  calculational approach},\ }\href
  {https://doi.org/10.1016/0003-4916(89)90359-X} {\bibfield  {journal}
  {\bibinfo  {journal} {Annals of Physics}\ }\textbf {\bibinfo {volume}
  {193}},\ \bibinfo {pages} {206} (\bibinfo {year} {1989})}\BibitemShut
  {NoStop}%
\bibitem [{\citenamefont {Bickers}\ and\ \citenamefont
  {White}(1991)}]{PhysRevB.43.8044}%
  \BibitemOpen
  \bibfield  {author} {\bibinfo {author} {\bibfnamefont {N.~E.}\ \bibnamefont
  {Bickers}}\ and\ \bibinfo {author} {\bibfnamefont {S.~R.}\ \bibnamefont
  {White}},\ }\bibfield  {title} {\bibinfo {title} {Conserving approximations
  for strongly fluctuating electron systems. ii. numerical results and parquet
  extension},\ }\href {https://doi.org/10.1103/PhysRevB.43.8044} {\bibfield
  {journal} {\bibinfo  {journal} {Phys. Rev. B}\ }\textbf {\bibinfo {volume}
  {43}},\ \bibinfo {pages} {8044} (\bibinfo {year} {1991})}\BibitemShut
  {NoStop}%
\bibitem [{\citenamefont {Esirgen}\ and\ \citenamefont
  {Bickers}(1997)}]{PhysRevB.55.2122}%
  \BibitemOpen
  \bibfield  {author} {\bibinfo {author} {\bibfnamefont {G.}~\bibnamefont
  {Esirgen}}\ and\ \bibinfo {author} {\bibfnamefont {N.~E.}\ \bibnamefont
  {Bickers}},\ }\bibfield  {title} {\bibinfo {title} {Fluctuation-exchange
  theory for general lattice hamiltonians},\ }\href
  {https://doi.org/10.1103/PhysRevB.55.2122} {\bibfield  {journal} {\bibinfo
  {journal} {Phys. Rev. B}\ }\textbf {\bibinfo {volume} {55}},\ \bibinfo
  {pages} {2122} (\bibinfo {year} {1997})}\BibitemShut {NoStop}%
\bibitem [{\citenamefont {Ikeda}\ \emph {et~al.}(2010)\citenamefont {Ikeda},
  \citenamefont {Arita},\ and\ \citenamefont
  {Kune{\v{s}}}}]{Ikeda2010GapAnisotropy}%
  \BibitemOpen
  \bibfield  {author} {\bibinfo {author} {\bibfnamefont {H.}~\bibnamefont
  {Ikeda}}, \bibinfo {author} {\bibfnamefont {R.}~\bibnamefont {Arita}},\ and\
  \bibinfo {author} {\bibfnamefont {J.}~\bibnamefont {Kune{\v{s}}}},\
  }\bibfield  {title} {\bibinfo {title} {Phase diagram and gap anisotropy in
  iron-pnictide superconductors},\ }\href
  {https://doi.org/10.1103/PhysRevB.81.054502} {\bibfield  {journal} {\bibinfo
  {journal} {Physical Review B}\ }\textbf {\bibinfo {volume} {81}},\ \bibinfo
  {pages} {054502} (\bibinfo {year} {2010})}\BibitemShut {NoStop}%
\bibitem [{\citenamefont {Onari}\ and\ \citenamefont
  {Kontani}(2012)}]{PhysRevB.85.134507}%
  \BibitemOpen
  \bibfield  {author} {\bibinfo {author} {\bibfnamefont {S.}~\bibnamefont
  {Onari}}\ and\ \bibinfo {author} {\bibfnamefont {H.}~\bibnamefont
  {Kontani}},\ }\bibfield  {title} {\bibinfo {title} {Non-fermi-liquid
  transport phenomena and superconductivity driven by orbital fluctuations in
  iron pnictides: Analysis by fluctuation-exchange approximation},\ }\href
  {https://doi.org/10.1103/PhysRevB.85.134507} {\bibfield  {journal} {\bibinfo
  {journal} {Phys. Rev. B}\ }\textbf {\bibinfo {volume} {85}},\ \bibinfo
  {pages} {134507} (\bibinfo {year} {2012})}\BibitemShut {NoStop}%
\bibitem [{\citenamefont {Usui}\ \emph {et~al.}(2012)\citenamefont {Usui},
  \citenamefont {Suzuki},\ and\ \citenamefont {Kuroki}}]{Usui2012SweetSpot}%
  \BibitemOpen
  \bibfield  {author} {\bibinfo {author} {\bibfnamefont {H.}~\bibnamefont
  {Usui}}, \bibinfo {author} {\bibfnamefont {K.}~\bibnamefont {Suzuki}},\ and\
  \bibinfo {author} {\bibfnamefont {K.}~\bibnamefont {Kuroki}},\ }\bibfield
  {title} {\bibinfo {title} {Least momentum space frustration as a condition
  for a high-{$T_c$} sweet spot in iron-based superconductors},\ }\href
  {https://doi.org/10.1088/0953-2048/25/8/084004} {\bibfield  {journal}
  {\bibinfo  {journal} {Superconductor Science and Technology}\ }\textbf
  {\bibinfo {volume} {25}},\ \bibinfo {pages} {084004} (\bibinfo {year}
  {2012})}\BibitemShut {NoStop}%
\bibitem [{\citenamefont {Rademaker}\ \emph {et~al.}(2021)\citenamefont
  {Rademaker}, \citenamefont {Alvarez-Suchini}, \citenamefont {Nakatsukasa},
  \citenamefont {Wang},\ and\ \citenamefont {Johnston}}]{Rademaker2021FeSeSTO}%
  \BibitemOpen
  \bibfield  {author} {\bibinfo {author} {\bibfnamefont {L.}~\bibnamefont
  {Rademaker}}, \bibinfo {author} {\bibfnamefont {G.}~\bibnamefont
  {Alvarez-Suchini}}, \bibinfo {author} {\bibfnamefont {K.}~\bibnamefont
  {Nakatsukasa}}, \bibinfo {author} {\bibfnamefont {Y.}~\bibnamefont {Wang}},\
  and\ \bibinfo {author} {\bibfnamefont {S.}~\bibnamefont {Johnston}},\
  }\bibfield  {title} {\bibinfo {title} {Enhanced superconductivity in
  {FeSe/SrTiO$_3$} from the combination of forward scattering phonons and spin
  fluctuations},\ }\href {https://doi.org/10.1103/PhysRevB.103.144504}
  {\bibfield  {journal} {\bibinfo  {journal} {Physical Review B}\ }\textbf
  {\bibinfo {volume} {103}},\ \bibinfo {pages} {144504} (\bibinfo {year}
  {2021})}\BibitemShut {NoStop}%
\bibitem [{\citenamefont {Sakakibara}\ \emph {et~al.}(2024)\citenamefont
  {Sakakibara}, \citenamefont {Kitamine}, \citenamefont {Ochi},\ and\
  \citenamefont {Kuroki}}]{Kuroki-2024-PRL-327}%
  \BibitemOpen
  \bibfield  {author} {\bibinfo {author} {\bibfnamefont {H.}~\bibnamefont
  {Sakakibara}}, \bibinfo {author} {\bibfnamefont {N.}~\bibnamefont
  {Kitamine}}, \bibinfo {author} {\bibfnamefont {M.}~\bibnamefont {Ochi}},\
  and\ \bibinfo {author} {\bibfnamefont {K.}~\bibnamefont {Kuroki}},\
  }\bibfield  {title} {\bibinfo {title} {Possible high ${T}_{c}$
  superconductivity in ${\mathrm{la}}_{3}{\mathrm{ni}}_{2}{\mathrm{o}}_{7}$
  under high pressure through manifestation of a nearly half-filled bilayer
  hubbard model},\ }\href {https://doi.org/10.1103/PhysRevLett.132.106002}
  {\bibfield  {journal} {\bibinfo  {journal} {Phys. Rev. Lett.}\ }\textbf
  {\bibinfo {volume} {132}},\ \bibinfo {pages} {106002} (\bibinfo {year}
  {2024})}\BibitemShut {NoStop}%
\bibitem [{\citenamefont {Witt}\ \emph {et~al.}(2021)\citenamefont {Witt},
  \citenamefont {van Loon}, \citenamefont {Nomoto}, \citenamefont {Arita},\
  and\ \citenamefont {Wehling}}]{Witt2021EfficientFLEX}%
  \BibitemOpen
  \bibfield  {author} {\bibinfo {author} {\bibfnamefont {N.}~\bibnamefont
  {Witt}}, \bibinfo {author} {\bibfnamefont {E.~G. C.~P.}\ \bibnamefont {van
  Loon}}, \bibinfo {author} {\bibfnamefont {T.}~\bibnamefont {Nomoto}},
  \bibinfo {author} {\bibfnamefont {R.}~\bibnamefont {Arita}},\ and\ \bibinfo
  {author} {\bibfnamefont {T.~O.}\ \bibnamefont {Wehling}},\ }\bibfield
  {title} {\bibinfo {title} {Efficient fluctuation-exchange approach to
  low-temperature spin fluctuations and superconductivity: From the hubbard
  model to {Na$_x$CoO$_2\cdot y$H$_2$O}},\ }\href
  {https://doi.org/10.1103/PhysRevB.103.205148} {\bibfield  {journal} {\bibinfo
   {journal} {Physical Review B}\ }\textbf {\bibinfo {volume} {103}},\ \bibinfo
  {pages} {205148} (\bibinfo {year} {2021})}\BibitemShut {NoStop}%
\bibitem [{\citenamefont {Shinaoka}\ \emph {et~al.}(2017)\citenamefont
  {Shinaoka}, \citenamefont {Otsuki}, \citenamefont {Ohzeki},\ and\
  \citenamefont {Yoshimi}}]{Shinaoka2017IR}%
  \BibitemOpen
  \bibfield  {author} {\bibinfo {author} {\bibfnamefont {H.}~\bibnamefont
  {Shinaoka}}, \bibinfo {author} {\bibfnamefont {J.}~\bibnamefont {Otsuki}},
  \bibinfo {author} {\bibfnamefont {M.}~\bibnamefont {Ohzeki}},\ and\ \bibinfo
  {author} {\bibfnamefont {K.}~\bibnamefont {Yoshimi}},\ }\bibfield  {title}
  {\bibinfo {title} {Compressing green's function using intermediate
  representation between imaginary-time and real-frequency domains},\ }\href
  {https://doi.org/10.1103/PhysRevB.96.035147} {\bibfield  {journal} {\bibinfo
  {journal} {Physical Review B}\ }\textbf {\bibinfo {volume} {96}},\ \bibinfo
  {pages} {035147} (\bibinfo {year} {2017})}\BibitemShut {NoStop}%
\bibitem [{\citenamefont {Wallerberger}\ \emph {et~al.}(2023)\citenamefont
  {Wallerberger}, \citenamefont {Badr}, \citenamefont {Hoshino}, \citenamefont
  {Huber}, \citenamefont {Kakizawa}, \citenamefont {Koretsune}, \citenamefont
  {Nagai}, \citenamefont {Nogaki}, \citenamefont {Nomoto}, \citenamefont
  {Mori}, \citenamefont {Otsuki}, \citenamefont {Ozaki}, \citenamefont
  {Plaikner}, \citenamefont {Sakurai}, \citenamefont {Vogel}, \citenamefont
  {Witt}, \citenamefont {Yoshimi},\ and\ \citenamefont
  {Shinaoka}}]{Wallerberger2023SparseIR}%
  \BibitemOpen
  \bibfield  {author} {\bibinfo {author} {\bibfnamefont {M.}~\bibnamefont
  {Wallerberger}}, \bibinfo {author} {\bibfnamefont {S.}~\bibnamefont {Badr}},
  \bibinfo {author} {\bibfnamefont {S.}~\bibnamefont {Hoshino}}, \bibinfo
  {author} {\bibfnamefont {S.}~\bibnamefont {Huber}}, \bibinfo {author}
  {\bibfnamefont {F.}~\bibnamefont {Kakizawa}}, \bibinfo {author}
  {\bibfnamefont {T.}~\bibnamefont {Koretsune}}, \bibinfo {author}
  {\bibfnamefont {Y.}~\bibnamefont {Nagai}}, \bibinfo {author} {\bibfnamefont
  {K.}~\bibnamefont {Nogaki}}, \bibinfo {author} {\bibfnamefont
  {T.}~\bibnamefont {Nomoto}}, \bibinfo {author} {\bibfnamefont
  {H.}~\bibnamefont {Mori}}, \bibinfo {author} {\bibfnamefont {J.}~\bibnamefont
  {Otsuki}}, \bibinfo {author} {\bibfnamefont {S.}~\bibnamefont {Ozaki}},
  \bibinfo {author} {\bibfnamefont {T.}~\bibnamefont {Plaikner}}, \bibinfo
  {author} {\bibfnamefont {R.}~\bibnamefont {Sakurai}}, \bibinfo {author}
  {\bibfnamefont {C.}~\bibnamefont {Vogel}}, \bibinfo {author} {\bibfnamefont
  {N.}~\bibnamefont {Witt}}, \bibinfo {author} {\bibfnamefont {K.}~\bibnamefont
  {Yoshimi}},\ and\ \bibinfo {author} {\bibfnamefont {H.}~\bibnamefont
  {Shinaoka}},\ }\bibfield  {title} {\bibinfo {title} {{sparse-ir}: Optimal
  compression and sparse sampling of many-body propagators},\ }\href
  {https://doi.org/10.1016/j.softx.2022.101266} {\bibfield  {journal} {\bibinfo
   {journal} {SoftwareX}\ }\textbf {\bibinfo {volume} {21}},\ \bibinfo {pages}
  {101266} (\bibinfo {year} {2023})}\BibitemShut {NoStop}%
\bibitem [{\citenamefont {Okabe}\ \emph {et~al.}(2010)\citenamefont {Okabe},
  \citenamefont {Takeshita}, \citenamefont {Horigane}, \citenamefont
  {Muranaka},\ and\ \citenamefont {Akimitsu}}]{PhysRevB.81.205119}%
  \BibitemOpen
  \bibfield  {author} {\bibinfo {author} {\bibfnamefont {H.}~\bibnamefont
  {Okabe}}, \bibinfo {author} {\bibfnamefont {N.}~\bibnamefont {Takeshita}},
  \bibinfo {author} {\bibfnamefont {K.}~\bibnamefont {Horigane}}, \bibinfo
  {author} {\bibfnamefont {T.}~\bibnamefont {Muranaka}},\ and\ \bibinfo
  {author} {\bibfnamefont {J.}~\bibnamefont {Akimitsu}},\ }\bibfield  {title}
  {\bibinfo {title} {Pressure-induced high-${T}_{c}$ superconducting phase in
  fese: Correlation between anion height and ${T}_{c}$},\ }\href
  {https://doi.org/10.1103/PhysRevB.81.205119} {\bibfield  {journal} {\bibinfo
  {journal} {Phys. Rev. B}\ }\textbf {\bibinfo {volume} {81}},\ \bibinfo
  {pages} {205119} (\bibinfo {year} {2010})}\BibitemShut {NoStop}%
\bibitem [{\citenamefont {Mizuguchi}\ \emph {et~al.}(2010)\citenamefont
  {Mizuguchi}, \citenamefont {Hara}, \citenamefont {Deguchi}, \citenamefont
  {Tsuda}, \citenamefont {Yamaguchi}, \citenamefont {Takeda}, \citenamefont
  {Kotegawa}, \citenamefont {Tou},\ and\ \citenamefont
  {Takano}}]{Mizuguchi_2010}%
  \BibitemOpen
  \bibfield  {author} {\bibinfo {author} {\bibfnamefont {Y.}~\bibnamefont
  {Mizuguchi}}, \bibinfo {author} {\bibfnamefont {Y.}~\bibnamefont {Hara}},
  \bibinfo {author} {\bibfnamefont {K.}~\bibnamefont {Deguchi}}, \bibinfo
  {author} {\bibfnamefont {S.}~\bibnamefont {Tsuda}}, \bibinfo {author}
  {\bibfnamefont {T.}~\bibnamefont {Yamaguchi}}, \bibinfo {author}
  {\bibfnamefont {K.}~\bibnamefont {Takeda}}, \bibinfo {author} {\bibfnamefont
  {H.}~\bibnamefont {Kotegawa}}, \bibinfo {author} {\bibfnamefont
  {H.}~\bibnamefont {Tou}},\ and\ \bibinfo {author} {\bibfnamefont
  {Y.}~\bibnamefont {Takano}},\ }\bibfield  {title} {\bibinfo {title} {Anion
  height dependence of tc for the fe-based superconductor},\ }\href
  {https://doi.org/10.1088/0953-2048/23/5/054013} {\bibfield  {journal}
  {\bibinfo  {journal} {Superconductor Science and Technology}\ }\textbf
  {\bibinfo {volume} {23}},\ \bibinfo {pages} {054013} (\bibinfo {year}
  {2010})}\BibitemShut {NoStop}%
\bibitem [{\citenamefont {Kuroki}(2011)}]{KUROKI2011307}%
  \BibitemOpen
  \bibfield  {author} {\bibinfo {author} {\bibfnamefont {K.}~\bibnamefont
  {Kuroki}},\ }\bibfield  {title} {\bibinfo {title} {Anion height as a
  controlling parameter for the superconductivity in iron pnictides and
  cuprates},\ }\href
  {https://doi.org/https://doi.org/10.1016/j.jpcs.2010.10.011} {\bibfield
  {journal} {\bibinfo  {journal} {Journal of Physics and Chemistry of Solids}\
  }\textbf {\bibinfo {volume} {72}},\ \bibinfo {pages} {307} (\bibinfo {year}
  {2011})},\ \bibinfo {note} {spectroscopies in Novel Superconductors
  2010}\BibitemShut {NoStop}%
\bibitem [{\citenamefont {McQueen}\ \emph
  {et~al.}(2009{\natexlab{b}})\citenamefont {McQueen}, \citenamefont {Huang},
  \citenamefont {Ksenofontov}, \citenamefont {Felser}, \citenamefont {Xu},
  \citenamefont {Zandbergen}, \citenamefont {Hor}, \citenamefont {Allred},
  \citenamefont {Williams}, \citenamefont {Qu}, \citenamefont {Checkelsky},
  \citenamefont {Ong},\ and\ \citenamefont {Cava}}]{PhysRevB.79.014522}%
  \BibitemOpen
  \bibfield  {author} {\bibinfo {author} {\bibfnamefont {T.~M.}\ \bibnamefont
  {McQueen}}, \bibinfo {author} {\bibfnamefont {Q.}~\bibnamefont {Huang}},
  \bibinfo {author} {\bibfnamefont {V.}~\bibnamefont {Ksenofontov}}, \bibinfo
  {author} {\bibfnamefont {C.}~\bibnamefont {Felser}}, \bibinfo {author}
  {\bibfnamefont {Q.}~\bibnamefont {Xu}}, \bibinfo {author} {\bibfnamefont
  {H.}~\bibnamefont {Zandbergen}}, \bibinfo {author} {\bibfnamefont {Y.~S.}\
  \bibnamefont {Hor}}, \bibinfo {author} {\bibfnamefont {J.}~\bibnamefont
  {Allred}}, \bibinfo {author} {\bibfnamefont {A.~J.}\ \bibnamefont
  {Williams}}, \bibinfo {author} {\bibfnamefont {D.}~\bibnamefont {Qu}},
  \bibinfo {author} {\bibfnamefont {J.}~\bibnamefont {Checkelsky}}, \bibinfo
  {author} {\bibfnamefont {N.~P.}\ \bibnamefont {Ong}},\ and\ \bibinfo {author}
  {\bibfnamefont {R.~J.}\ \bibnamefont {Cava}},\ }\bibfield  {title} {\bibinfo
  {title} {Extreme sensitivity of superconductivity to stoichiometry in
  ${\text{fe}}_{1+\ensuremath{\delta}}\text{Se}$},\ }\href
  {https://doi.org/10.1103/PhysRevB.79.014522} {\bibfield  {journal} {\bibinfo
  {journal} {Phys. Rev. B}\ }\textbf {\bibinfo {volume} {79}},\ \bibinfo
  {pages} {014522} (\bibinfo {year} {2009}{\natexlab{b}})}\BibitemShut
  {NoStop}%
\bibitem [{\citenamefont {Kang}\ \emph {et~al.}(2018)\citenamefont {Kang},
  \citenamefont {Fernandes},\ and\ \citenamefont {Chubukov}}]{Kang2018Nematic}%
  \BibitemOpen
  \bibfield  {author} {\bibinfo {author} {\bibfnamefont {J.}~\bibnamefont
  {Kang}}, \bibinfo {author} {\bibfnamefont {R.~M.}\ \bibnamefont
  {Fernandes}},\ and\ \bibinfo {author} {\bibfnamefont {A.}~\bibnamefont
  {Chubukov}},\ }\bibfield  {title} {\bibinfo {title} {Superconductivity in
  {FeSe}: The role of nematic order},\ }\href
  {https://doi.org/10.1103/PhysRevLett.120.267001} {\bibfield  {journal}
  {\bibinfo  {journal} {Physical Review Letters}\ }\textbf {\bibinfo {volume}
  {120}},\ \bibinfo {pages} {267001} (\bibinfo {year} {2018})}\BibitemShut
  {NoStop}%
\bibitem [{\citenamefont {Kreisel}\ \emph {et~al.}(2017)\citenamefont
  {Kreisel}, \citenamefont {Andersen}, \citenamefont {Sprau}, \citenamefont
  {Kostin}, \citenamefont {Davis},\ and\ \citenamefont
  {Hirschfeld}}]{Kreisel2017Orbital}%
  \BibitemOpen
  \bibfield  {author} {\bibinfo {author} {\bibfnamefont {A.}~\bibnamefont
  {Kreisel}}, \bibinfo {author} {\bibfnamefont {B.~M.}\ \bibnamefont
  {Andersen}}, \bibinfo {author} {\bibfnamefont {P.~O.}\ \bibnamefont {Sprau}},
  \bibinfo {author} {\bibfnamefont {A.}~\bibnamefont {Kostin}}, \bibinfo
  {author} {\bibfnamefont {J.~C.~S.}\ \bibnamefont {Davis}},\ and\ \bibinfo
  {author} {\bibfnamefont {P.~J.}\ \bibnamefont {Hirschfeld}},\ }\bibfield
  {title} {\bibinfo {title} {Orbital selective pairing and gap structures of
  iron-based superconductors},\ }\href
  {https://doi.org/10.1103/PhysRevB.95.174504} {\bibfield  {journal} {\bibinfo
  {journal} {Physical Review B}\ }\textbf {\bibinfo {volume} {95}},\ \bibinfo
  {pages} {174504} (\bibinfo {year} {2017})}\BibitemShut {NoStop}%
\bibitem [{\citenamefont {Linscheid}\ \emph {et~al.}(2016)\citenamefont
  {Linscheid}, \citenamefont {Maiti}, \citenamefont {Wang}, \citenamefont
  {Johnston},\ and\ \citenamefont {Hirschfeld}}]{Linscheid2016Incipient}%
  \BibitemOpen
  \bibfield  {author} {\bibinfo {author} {\bibfnamefont {A.}~\bibnamefont
  {Linscheid}}, \bibinfo {author} {\bibfnamefont {S.}~\bibnamefont {Maiti}},
  \bibinfo {author} {\bibfnamefont {Y.}~\bibnamefont {Wang}}, \bibinfo {author}
  {\bibfnamefont {S.}~\bibnamefont {Johnston}},\ and\ \bibinfo {author}
  {\bibfnamefont {P.~J.}\ \bibnamefont {Hirschfeld}},\ }\bibfield  {title}
  {\bibinfo {title} {High-{$T_c$} via spin fluctuations from incipient bands:
  Application to monolayers and intercalates of {FeSe}},\ }\href
  {https://doi.org/10.1103/PhysRevLett.117.077003} {\bibfield  {journal}
  {\bibinfo  {journal} {Physical Review Letters}\ }\textbf {\bibinfo {volume}
  {117}},\ \bibinfo {pages} {077003} (\bibinfo {year} {2016})}\BibitemShut
  {NoStop}%
\bibitem [{\citenamefont {Gao}\ \emph {et~al.}(2018)\citenamefont {Gao},
  \citenamefont {Wang}, \citenamefont {Zhou}, \citenamefont {Huang},\ and\
  \citenamefont {Wang}}]{Gao2018QPI}%
  \BibitemOpen
  \bibfield  {author} {\bibinfo {author} {\bibfnamefont {Y.}~\bibnamefont
  {Gao}}, \bibinfo {author} {\bibfnamefont {Y.}~\bibnamefont {Wang}}, \bibinfo
  {author} {\bibfnamefont {T.}~\bibnamefont {Zhou}}, \bibinfo {author}
  {\bibfnamefont {H.}~\bibnamefont {Huang}},\ and\ \bibinfo {author}
  {\bibfnamefont {Q.-H.}\ \bibnamefont {Wang}},\ }\bibfield  {title} {\bibinfo
  {title} {Possible pairing symmetry in the {FeSe}-based superconductors
  determined by quasiparticle interference},\ }\href
  {https://doi.org/10.1103/PhysRevLett.121.267005} {\bibfield  {journal}
  {\bibinfo  {journal} {Physical Review Letters}\ }\textbf {\bibinfo {volume}
  {121}},\ \bibinfo {pages} {267005} (\bibinfo {year} {2018})}\BibitemShut
  {NoStop}%
\end{thebibliography}

\clearpage

\clearpage

\begin{thebibliography}{99}
\bibitem{blochl1994projector} P. E. Bl"ochl, Projector augmented-wave method, Phys. Rev. B \textbf{50}, 17953 (1994).
\bibitem{kresse1996efficient} G. Kresse and J. Furthm"uller, Efficient iterative schemes for ab initio total-energy calculations using a plane-wave basis set, Phys. Rev. B \textbf{54}, 11169 (1996).
\bibitem{perdew1996generalized} J. P. Perdew, K. Burke, and M. Ernzerhof, Generalized Gradient Approximation Made Simple, Phys. Rev. Lett. \textbf{77}, 3865 (1996).
\bibitem{mostofi2008wannier90} A. A. Mostofi, J. R. Yates, Y.-S. Lee, I. Souza, D. Vanderbilt, and N. Marzari, wannier90: A tool for obtaining maximally-localised Wannier functions, Comput. Phys. Commun. \textbf{178}, 685 (2008).
\bibitem{kuroki2008unconventional} K. Kuroki, S. Onari, R. Arita, H. Usui, Y. Tanaka, H. Kontani, and H. Aoki, Unconventional Pairing Originating from the Disconnected Fermi Surfaces of Superconducting $\mathrm{LaFeAsO}_{1-x}\mathrm{F}_x$, Phys. Rev. Lett. \textbf{101}, 087004 (2008).
\bibitem{bickers1989conserving} N. E. Bickers and D. J. Scalapino, 
Conserving approximations for strongly fluctuating electron systems. I. Formalism and calculational approach, Ann. Phys. \textbf{193}, 206--251 (1989).
\bibitem{bickers1991conserving} N. E. Bickers and S. R. White, Conserving approximations for strongly fluctuating electron systems. II. Numerical results and parquet extension, Phys. Rev. B \textbf{43}, 8044 (1991).
\bibitem{yanase2003theory} Y. Yanase, T. Jujo, T. Nomura, H. Ikeda, T. Hotta, and K. Yamada, 
Theory of superconductivity in strongly correlated electron systems, Phys. Rep. \textbf{387}, 1--149 (2003).
\bibitem{witt2021efficient} N. Witt, E. G. C. P. Van Loon, T. Nomoto, R. Arita, and T. O. Wehling, 
Efficient fluctuation-exchange approach to low-temperature spin fluctuations and superconductivity: From the Hubbard model to Na$_x$CoO$_2${\textperiodcentered}$y$H$_2$O, Phys. Rev. B \textbf{103}, 205148 (2021).
\bibitem{Shinaoka2017IRSM} H. Shinaoka, J. Otsuki, M. Ohzeki, and K. Yoshimi,
Compressing Green's function using intermediate representation between imaginary-time and real-frequency domains, Phys. Rev. B \textbf{96}, 035147 (2017).
\bibitem{wallerberger2023sparse} M. Wallerberger, S. Badr, S. Hoshino, S. Huber, F. Kakizawa, T. Koretsune, Y. Nagai, K. Nogaki, T. Nomoto, H. Mori, \textit{et al.}, sparse-ir: Optimal compression and sparse sampling of many-body propagators, SoftwareX \textbf{21}, 101266 (2023).
\bibitem{ikeda2010phase} H. Ikeda, R. Arita, and J. Kune{\v{s}},
Phase diagram and gap anisotropy in iron-pnictide superconductors, Phys. Rev. B \textbf{81}, 054502 (2010).
\end{thebibliography}
\end{document}